\documentclass[graybox, secnum]{svmult}

\usepackage{mathptmx}       
\usepackage{helvet}         
\usepackage{courier}        
\usepackage{type1cm}        
\usepackage{makeidx}         
\usepackage{graphicx}        
\usepackage{multicol}        
\usepackage[bottom]{footmisc}
\usepackage{hyperref}        
\usepackage{soul}            
\hypersetup{colorlinks=true,urlcolor=blue}
\usepackage[square,numbers]{natbib}
\usepackage{amsmath,amssymb}
\UseRawInputEncoding

\def\pref#1{(\ref{#1})}
\def\exd{{\rm d}}

\def\gsim{\mbox{~{\protect\raisebox{0.3ex}{$>$}}\hspace{-0.9em}
        {\protect\raisebox{-0.6ex}{$\sim$}}~}}
\def\lsim{\mbox{~{\protect\raisebox{0.3ex}{$<$}}\hspace{-0.9em}
        {\protect\raisebox{-0.6ex}{$\sim$}}~}}

\def\cG{{\cal G}}
\def\cH{{\cal H}}

\def\cK{{\cal K}}
\def\cL{{\cal L}}

\def\cO{{\cal O}}
\def\cP{{\cal P}}
\def\cQ{{\cal Q}}
\def\cR{{\cal R}}

\def\cT{{\cal T}}

\def\cW{{\cal W}}

\def\bfh{{\bf h}}
\def\bfk{{\bf k}}

\def\bfp{{\bf p}}
\def\bfq{{\bf q}}

\def\bfx{{\bf x}}
\def\bfy{{\bf y}}

\def\ssA{{\scriptscriptstyle A}}
\def\ssB{{\scriptscriptstyle B}}

\def\ssF{{\scriptscriptstyle F}}

\def\ssH{{\scriptscriptstyle H}}
\def\ssI{{\scriptscriptstyle I}}

\def\ssM{{\scriptscriptstyle M}}

\def\ssP{{\scriptscriptstyle P}}

\def\ssR{{\scriptscriptstyle R}}
\def\ssS{{\scriptscriptstyle S}}
\def\ssT{{\scriptscriptstyle T}}

\def\IR{{\scriptscriptstyle IR}}
\def\GH{{\scriptscriptstyle GH}}

\def\IF{{\scriptscriptstyle IF}}

\def\mfa{\mathfrak{a}}
\def\mfp{\mathfrak{p}}

\def\llangle{\langle\hspace{-2mm}\langle}
\def\rrangle{\rangle\hspace{-2mm}\rangle}

\def\nn{\nonumber}
\def\mathi{i}
\def\tr{\hbox{ tr }}
\def\Tr{\hbox{ Tr }}
\def\trB{\hbox{ tr}_\ssB}
\def\trA{\hbox{ tr}_\ssA}
\def\slrl{\varepsilon_1}
\def\Mp{M_p}

\newcommand{\be}{\begin{equation}}
\newcommand{\bee}{\begin{equation}}
\newcommand{\ee}{\end{equation}}
\newcommand{\beea}{\begin{eqnarray}}
\newcommand{\eea}{\end{eqnarray}}
\newcommand{\bea}{\begin{eqnarray}}

\def\napprox{\hbox{\;\;\big/\kern-.7000em$\simeq$\;\;}}

\makeindex             

\begin{document}
\title*{Gravity, Horizons and Open EFTs}
\author{C.P.~Burgess 
and Greg Kaplanek}
\institute{C.P.~Burgess (\email{cburgess@perimeterinstitute.ca}) \at McMaster University, 1280 Main St.~W, Hamilton ON, Canada, L8S 4M1\\
Perimeter Institute for Theoretical Physics, 31 Caroline St.~N, Waterloo ON, Canada, N2L 2Y5\\ 
Theoretical Physics Department, CERN, Gen\`eve 23, Switzerland\\
School of Theoretical Physics, 
DIAS, 10 Burlington Rd., Dublin, 
Ireland
\and Greg Kaplanek (\email{g.kaplanek@imperial.ac.uk}) \at  Theoretical Physics, Blackett Laboratory, Imperial College, London, SW7 2AZ, UK}

%
%
\maketitle
\abstract{Wilsonian effective theories exploit hierarchies of scale to simplify the description of low-energy behaviour and play as central a role for gravity as for the rest of physics. They are useful both when hierarchies of scale are explicit in a gravitating system and more generally for understanding precisely what controls the size of quantum corrections in gravitational systems. But effective descriptions are also relevant for open systems ({\it e.g.}~fluid mechanics as a long-distance description of statistical systems) for which the `integrating out' of unobserved {\it low-energy} degrees of freedom complicate a straightforward application of Wilsonian methods. Observations performed only on one side of an apparent horizon provide examples where open system descriptions also arise in gravitational physics. This chapter describes some early adaptations of Open Effective Theories ({\it i.e.}~techniques for exploiting hierarchies of scale in open systems) in gravitational settings. Besides allowing the description of new types of phenomena (such as decoherence) these techniques also have an additional benefit: they sometimes can be used to resum perturbative expansions at late times and thereby to obtain controlled predictions in a regime where perturbative predictions otherwise generically fail. }

\section*{Keywords} 
effective field theory, open systems, secular growth, late-time resummation\\
CERN-TH-2022-175

\section{Open systems and gravity as a medium}

Nature comes to us with a riot of scales: from the smallest known subatomic particles out to the observable universe as a whole. Because we do not yet know how Nature works at {\it all} scales a central question asks how badly unknown physics can interfere with predictions for physics we think we do understand. Wilsonian effective field theories (EFTs) are important because (at least so far) they provide the only known answer to this question.

They do so by quantifying how physics at a high energy $M$ contributes to observables at a low energy $E$. A central result shows that most of the high-energy details only enter into low-energy observables suppressed by powers of the small ratio $E/M$, and furthermore shows why these can be captured order-by-order in $E/M$ -- for {\it all} low-energy observables -- as effective couplings within some form of a local Wilsonian action, or Hamiltonian
\begin{equation}
   H_{\rm eff} = \int {\rm d}^3x \sum_s c_s \; {\cal O}_s(x) \,.
\end{equation}
Here the effective couplings $c_s$ are often proportional to $M_s^{4-d_s}$ where $d_s$ is the scaling dimension of the local effective interaction operator ${\cal O}_s(x)$ that is built using only fields $\phi(x)$ that represent the low-energy states. 

In the normal treatment much discussion goes into how to compute the effective couplings $c_s$ and interactions ${\cal O}_s(x)$ that are generated by particular types of known high-energy physics, taking for granted the {\it existence} of an appropriate $H_{\rm eff}$. This is usually reasonable to do because the rules of quantum mechanics guarantee some sort of a Hamiltonian must exist, basically because energy conservation ensures that states that start at low energy remain at low energy (and so their evolution must be describable using the low-energy basis of field operators $\phi$). The uncertainty principle then ensures that the low-energy effects of virtual high-energy states are local in time -- at least order-by-order in $E/M$ -- from which (at least for relativistic systems) locality in space usually also follows. (See {\it e.g.}~\cite{EFTBook} for an extensive review).

The theory of Open Quantum Systems (see for example \cite{OpenSystems}) provides the general framework for describing situations where only a subset of states are observed within a wider quantum system, and because of this one might expect EFTs to closely resemble open systems. This turns out not to be true. In particular it is in general {\it not} true that an effective Hamiltonian can capture how unobserved degrees of freedom influence others that we do measure. 

EFTs turn out to be a very special case where measured and unmeasured degrees of freedom are distinguishable from one another by the eigenvalues of a conserved charge (the value of their energy, plus possibly other conserved quantities), and this plays a crucial role when inferring the existence of an effective Hamiltonian. For general open systems the measured and unmeasured degrees of freedom can both have similar energies and can exchange information and mutually entangle, making the evolution of any observed subsector usually more complicated. This difference is why open systems can exhibit rich phenomena like thermalization and decoherence not normally seen in a simple Wilsonian low-energy/high-energy split. 

For quantum gravity the relative richness of open systems compared to Wilsonian evolution is not just an academic point. This is because the existence of horizons can famously prevent low-energy degrees of freedom from interacting (and being measured) by other low-energy degrees of freedom. The description appropriate for such situations is necessarily open and this is partly why horizons can have surprising quantum implications. 

Because near-horizon systems are open, their effective description can differ from standard Wilsonian reasoning and it is worth exploring what this can mean. This involves exploring how predictions are made for open systems and identifying the differences from the standard Wilsonian approach. Since some steps differ from standard Wilsonian practice, special attention must be paid to the domain of validity for the approximations that are used. Of particular interest are features of Wilsonian intuition (such as locality) that might not generalize to open systems more broadly.

This chapter summarizes some of those issues, but also describes how the systematic study of open systems brings a welcome bonus that is interesting in its own right: it provides a method for dealing with  `secular-growth' problems often faced by quantum systems in gravitational fields. Secular growth is the gravitational version of something that is also generic elsewhere in physics:\footnote{Scattering of wave packets of particles is one of the few physical questions where late-time interactions need not be a problem because packet separation itself turns off late-time interactions.} perturbative predictions break down at sufficiently late times because there is always a time for which 
\begin{equation}
   U(t) = e^{-i(H_0 + H_{\rm int} )t}  \napprox e^{-i H_0 t}(1 -i H_{\rm int} t)
\end{equation}
no matter how small $H_{\rm int}$ is compared with $H_0$. As we review below, one of the reasons for using open-system tools is they can sometimes allow reliable late-time predictions to be made even when $H_{\rm int}t$ is order unity, even if the only available calculations rely on perturbing in $H_{\rm int}$. 

Such tools are particularly important when drawing late-time inferences about quantum systems in gravitational fields because intuition about these systems is often based on simple examples that interact only with the background gravitational field and do not self-interact. It is tempting to think that this intuition should remain reliable in the presence of self-interactions because these self-interactions might be chosen as small as one wants. But smaller couplings only really delay {\it when} perturbation theory fails, rather than remove its failure altogether.  So more robust means for making prediction are required when late-time behaviour is important. Quantum-gravitational problems for which late-time evolution {\it is} crucial include black-hole information loss and some issues of eternal inflation.

This chapter summarizes some early steps that have been taken applying open-system techniques to gravity (for other preliminary applications in particle physics and gravity respectively, see \cite{OpenPP} and \cite{OpenQG}), usually with late-time evolution specifically in mind. We do so from a purely personal perspective with the goal of laying out the conceptual framework as clearly as we can, rather than trying to survey exhaustively everything that has been done. We inevitably will have missed describing some gems within the literature along the way, and apologize in advance -- both to the authors and to you, the reader -- for doing so.  

\subsection{Open quantum systems}
\label{ssec:OpenQS}

The generic open-system problem restricts attention to a subset of observables lying within some sector $A$ within the full system's Hilbert space. The goal is to compute how these observables evolve in time, including in particular how this is influenced by their interactions with all of the other unmeasured degrees of freedom. 

For the reasons alluded to above, one usually does not seek to construct a universal $H_{\rm eff}$ that captures the influence of all unmeasured degrees of freedom. One instead sets up and solves a {\it master equation} that describes the time-evolution of the state of the measured subsystem, and uses this to predict how measurements evolve. 

To see how this works, suppose the Hilbert space for the quantum system of interest contains two sectors, $A$ and $B$, and measurements are only performed in sector $A$ (the `system') while sector $B$ (the `environment') remains unmeasured. For simplicity consider the case where the full system's Hilbert space can be written as a product of states in sector $A$ and those in $B$, so a basis of states in the full system can be decomposed as 
\be
  |a,b\rangle = | a \rangle \otimes | b\rangle \,. 
\ee
Observables involving only sector $A$ are hermitian operators that can be written 
\begin{equation} \label{eq:AObsDef}
   \cO_\ssA = O_\ssA \otimes I_\ssB \quad\hbox{with matrix elements} \quad 
   \langle a, b | \cO_\ssA | a', b' \rangle = \langle a | O_\ssA | a' \rangle \, \delta_{bb'} \,.
\end{equation}

The focus in what follows is on describing the evolution of the system's state, as described by its density matrix 
\be
  \hat\rho := \sum_I p_\ssI \, | I \rangle \langle I | := \sum_{ab} p_{ab} \, | a,b \rangle \langle  a,b|  \,,
\ee
where $p_\ssI = p_{ab}$ is the probability to find the system in state $|I \rangle = |a,b \rangle$. As usual, the sum over mutually exclusive probabilities must give one so~$\Tr \hat\rho = \sum_\ssI p_\ssI = 1$. A `pure' state corresponds to the special case where the system's state $|\psi\rangle$ is precisely known and so 
\be\label{PureStatePsi}
  \hat\rho = | \psi \rangle \, \langle \psi | \quad \hbox{(pure state)} \,. 
\ee
Condition \pref{PureStatePsi} being true for some (normalized) state $|\psi\rangle$ is equivalent to the condition $\hat\rho^2 = \hat\rho$ and because all probabilities satisfy $0 \leq p_{ab} \leq 1$ this is also equivalent to the condition $\Tr \hat \rho^2 = \Tr \hat \rho = 1$. When $\hat\rho^2 \neq \hat\rho$ the state is said to be `mixed' and there is no $|\psi \rangle$ for which \pref{PureStatePsi} is true. 

Expectation values for observables are computed using $\hat \rho$ by evaluating\footnote{The `hat' symbol is meant to emphasize that the operator is written in Schr\"odinger picture.} 
\be \label{ObservableOexample}
  \langle \hat\cO \rangle = \sum_\ssI p_\ssI \langle I | \hat\cO | I \rangle = \hbox{Tr}\,(\hat\rho \, \hat\cO) \,. 
\ee
Within the Schr\"odinger picture this acquires a time-dependence through the time-dependence of $\hat\rho(t)$, which is described by the evolution equation 
\be \label{eq:Liouvilleeq}
  \mathi \,\partial_t \hat\rho = \Bigl[ H \,, \hat\rho \,\Bigr] \,,
\ee
where $H$ is the system's Hamiltonian. Eq.~\pref{eq:Liouvilleeq} is equivalent to the usual Schr\"odinger evolution $\mathi \, \partial_t | \psi(t) \rangle = H | \psi(t) \rangle$ for a pure state satisfying \pref{PureStatePsi}. 

Assuming the initial condition $\hat \rho(t=t_0) = \hat \rho_0$, eq.~\pref{eq:Liouvilleeq} is formally solved by
\be \label{rhoevolution}
  \hat\rho(t) = \hat U(t,t_0) \, \hat\rho_0 \, \hat U^*(t, t_0) \,,
\ee
where hermiticity of $H$ ensures $\hat U(t,t_0) = \exp[ - \mathi H(t-t_0) ]$ is unitary, and so satisfies $\hat U(t,t_0) \hat U^*(t,t_0) = 1$. Furthermore  $\hat U(t,t_1) \hat U(t_1,t_0) = \hat U(t,t_0)$ while
\be \label{unitaryU}
  \mathi \, \partial_t \hat U(t,t_0) = H \, \hat U(t,t_0) \quad \hbox{with initial condition} \quad \hat U(t_0,t_0) = I \,.
\ee
So far this is all bog-standard quantum mechanics. The `open-system' part of the story starts once we assume the only observables of interest have the form \pref{eq:AObsDef}, expressing that they probe only sector $A$.

For open systems the goal is to understand how measurements restricted to sector $A$ evolve in a way that refers as much as possible only to the measured sector $A$. A key tool to this end is the `reduced' density matrix, defined by tracing the full density matrix over the unmeasured sector $B$: 
\be \label{reducedrhodef}
   \hat\rho_\ssA := \trB\, \hat\rho \quad \hbox{so} \quad
   \langle a |\, \hat\rho_\ssA \,| a' \rangle = \sum_b \langle a, b | \,\hat\rho\, | a', b \rangle \,.
\ee
This is useful because it defines an operator that acts only in sector $A$ whose time evolution completely determines the time-dependence of measurements of type \pref{eq:AObsDef} that also act only in sector $A$, as may be seen from expressions like
\be
 \langle \hat\cO_\ssA \rangle := \Tr \bigl[ \hat\rho(t) \, \hat\cO_\ssA \bigr] = \sum_{aa'} \sum_b \langle a,b | \,\hat\rho\, | a', b \rangle \, \langle a'|\, \hat O_\ssA \,| a \rangle  = \trA \bigl[ \hat\rho_\ssA(t) \, \hat O_\ssA \bigr] \,.
\ee

In principle the evolution of $\hat \rho_\ssA$ is obtained by taking the trace of \pref{eq:Liouvilleeq} over sector $B$ and integrating the result. In practice this is not so useful because the right-hand side of the resulting equation depends on the entire density matrix $\hat \rho$ (including what the environmental sector is doing) rather than just being a function of $\hat\rho_\ssA$. Happily, this is a problem that has been solved in some generality, as is possible because both the Liouville evolution \pref{eq:Liouvilleeq} and the projection from $\hat\rho$ onto $\hat \rho_\ssA$ are linear operations on the space of density matrices. One must set up and solve the evolution equation for the state of the environment and use this to eliminate the environment from the expression for $\partial_t \hat \rho$.  

To this end define the super-operator $\cP$ to act on a general operator $\cO$ by 
\be \label{projPdef}
   \cP(\cO) := \trB(\cO) \otimes \rho_\ssB  \,.
\ee 
Here $\rho_\ssB$ is a density matrix that characterizes the state of sector $B$ (the environment) that is specified in more detail later. Because $\trB \rho_\ssB = 1$ it follows that $\cP$ is a projection operator: $\cP^2 = \cP$. It also satisfies $\cP(\rho_\ssA \otimes \rho_\ssB) = \rho_\ssA \otimes \rho_\ssB$ and so $\cP$ does nothing if it acts on a density matrix for which sectors $A$ and $B$ are uncorrelated. Finally, $\cP[\hat\rho(t)] = \hat\rho_\ssA(t) \otimes \rho_\ssB$, where $\hat\rho_\ssA(t) = \trB \hat\rho$ is the reduced density matrix whose time-evolution we seek. 

Because $\cP$ is a projection super-operator, the same is also true for its complement $\cQ := 1 - \cP$: {\it i.e.}~ $\cQ$ satisfies $\cQ^2 = \cQ$ and $\cP \cQ = \cQ \,\cP = 0$. In this same language time evolution is also described by a linear super-operator, since \pref{eq:Liouvilleeq} can be written $\partial_t \, \hat\rho = \cL_t(\hat\rho)$ where 
\be
   \cL_t(\cO) := -\mathi \Bigl[H ,  \cO \Bigr] \,.
\ee
The evolution of $\cP(\hat \rho)$ and $\cQ(\hat \rho)$ with time is easily found by projecting the evolution equation \pref{eq:Liouvilleeq} using $\cP$ and $\cQ$. Using $\cP( \partial_t \,\hat\rho ) = \partial_t  \cP( \hat\rho )$ and $\cP + \cQ = 1$ allows \pref{eq:Liouvilleeq} to be rewritten as the pair of equations
\bea\label{NZintro}
  &&\partial_t \cP(\hat\rho) = \cP ( \partial_t \hat\rho) = \cP \cL_t (\hat\rho) = \cP \cL_t  \cP (\hat\rho) + \cP \cL_t \cQ (\hat\rho) \\
  \hbox{and}
  &&\partial_t \cQ(\hat\rho) = \cQ (\partial_t \hat\rho) = \cQ \cL_t (\hat\rho) = \cQ \cL_t \cP (\hat\rho) + \cQ \cL_t \cQ (\hat\rho) \,.\nn
\eea

Now comes the main point: the second of these equations can be used to eliminate $\cQ(\hat\rho)$ from the right-hand side of the first equation, thereby obtaining an evolution equation that involves only $\cP(\hat\rho)$. This solves our problem of setting up an evolution equation for the reduced matrix $\hat \rho_\ssA$ to the extent that we can use $\cP(\hat\rho)$ as a proxy for $\hat\rho_\ssA$, and these indeed would agree at an initial time $t_0$ if the initial state were assumed to be uncorrelated:
\be \label{intial_uncorr}
   \hat\rho(t_0) = \hat\rho_\ssA(t_0) \otimes \rho_\ssB \,. 
\ee
In general the evolution of $\cP(\hat\rho(t)) = \hat\rho_\ssA(t) \otimes \rho_\ssB$ does not agree with that of $\hat \rho(t)$ because for $\cP(\hat\rho)$ the environment (sector $B$) does not evolve. But $\hat\rho$ and $\cP(\hat\rho)$ nonetheless by construction agree with one another once sector $B$ is traced out and so agree on time dependence for observables acting only in $A$. 

Because eqs.~\pref{NZintro} are linear they can be solved fairly explicitly. If we define the super-operator $\cG(t,s)$ as the solution to $\partial_t \cG(t,s) = \cQ \cL_t \cG(t,s)$ with initial condition $\cG(t,t) = 1$ then $\cG(t,s)$ is given explicitly by
\bea \label{cGasSoln}
 \cG(t,s) &=& 1+ \sum_{n=1}^\infty \int_s^t \exd s_1 \cdots  \int_s^{s_{n-1}} \exd s_n \, \cQ \cL_{s_1}\cdots  \cQ \cL_{s_n}  \nn\\
 &=& 1+ \sum_{n=1}^\infty\frac{1}{n!} \int_s^t \exd s_1 \cdots  \int_s^{t} \exd s_n \, {\cal T} \Bigl[ \cQ \cL_{s_1}\cdots  \cQ \cL_{s_n} \Bigr] \,,
\eea
where ${\cal T}$ denotes time-ordering of the $\cQ \cL_{s_i}$. This allows the solution for $\cQ[\hat\rho(t)]$ with initial condition $\cQ[\hat\rho(t_0)] = \cQ(\hat\rho_0)$ to be written
\be
 \cQ[\hat\rho(t)] = \cG(t,t_0) \cQ(\hat\rho_0) + \int_{t_0}^t \exd s \, \cG(t,s) \, \cQ \cL_s \cP[ \hat\rho(s)] \,,
\ee
as can be verified by explicit differentiation, using $\partial_t \cG(t,s) = \cQ \cL_t \cG(t,s)$. Inserting this into the first of eqs.~\pref{NZintro} gives the desired evolution equation for $\cP(\hat\rho)$
\be \label{eq:NakajimaZwanzig}
 \partial_t \cP[ \hat\rho(t)] = \cP \cL_t \cP [\hat\rho(t)] + \cP \cL_t \cG(t,t_0) \cQ(\rho_0) +  \int_{t_0}^t \exd s \, \cK(t,s) [\hat\rho(s)] \,,
\ee
where $\cK(t,s) = \cP \cL_t \cG(t,s) \cQ \cL_s \cP$. The second term on the right-hand side vanishes for the uncorrelated initial condition $\rho_0 = \hat\rho(t_0) = \hat\rho_\ssA(t_0) \otimes \rho_\ssB$ since this satisfies $\cP(\rho_0) = \rho_0$ and so $\cQ(\rho_0) = 0$. 

Eq.~\pref{eq:NakajimaZwanzig} is an integro-differential {\it master} equation -- initially due to Nakajima and Zwanzig \cite{Nakajima, Zwanzig} -- that contains the same information for sector $A$ as does the original evolution equation \pref{eq:Liouvilleeq}. It is also typically no easier to solve. Its main virtue is that $\hat\rho(t)$ only appears in it through the combination $\cP[\hat\rho(t)]$ and so it expresses $\partial_t \hat\rho_\ssA$ directly in terms of $\hat\rho_\ssA$ itself. 

The final result is most useful when the interaction between sectors $A$ and $B$ can be treated perturbatively. To this end we write
\be
   H = H_0 + \hat V \quad \hbox{with} \quad H_0 = H_\ssA + H_\ssB 
\ee
and switch to the interaction picture so $\cO(t) = e^{\mathi H_0 (t-t_0)} \hat \cO \, e^{-\mathi H_0(t-t_0)}$, so that state evolution is controlled only by $V(t)  = e^{\mathi H_0 (t-t_0)} \hat V \, e^{-\mathi H_0(t-t_0)}$, which we expand in a basis of operators 
\be
   V(t) = \sum_n A_{n} (t) \otimes B_{n}(t) \,.
\ee
In interaction picture the evolution superoperator becomes $\cL_t(\cO) = -i[ V(t) \,, \cO]$ and \pref{eq:NakajimaZwanzig} can be fruitfully expanded in powers of $V(t)$. 

For our later purposes it suffices to work only to second order in $V$, in which case the kernel becomes $\cK \simeq \cK_2 = \cP \cL_t \cQ \cL_s \cP$. Choosing an uncorrelated initial condition for the interaction-picture density matrix, $\rho(t_0) = \rho_\ssA(t_0) \otimes \rho_\ssB$, eq.~\pref{eq:NakajimaZwanzig} reduces to the following approximate expression
\bea \label{NakaZwanExplicit}
  \partial_t \,\rho_\ssA(t) &=& - \mathi \sum_n \Bigl[ A_n(t) , \rho_\ssA(t) \Bigr] \, \llangle \; B_n(t) \; \rrangle \nn\\
  && \quad 
  + (-\mathi)^2 \sum_{mn}  \int_{t_0}^t \exd s \Biggl\{ \Bigl[ A_m(t) , A_n(s) \,\rho_\ssA(s) \Bigr] \, \llangle \, \delta B_m(t) \, \delta B_n(s) \, \rrangle   \\
  && \qquad\qquad\qquad  -  \Bigl[ A_m(t) , \rho_\ssA(s)\,A_n(s)  \Bigr] \, \llangle \, \delta B_n(s) \,\delta B_m(t) \, \rrangle \,\Biggr\} + \cO(V^3)\,,\nn
\eea
which introduces the notation $\llangle \,(\cdots)\, \rrangle = \trB[(\cdots) \,\rho_\ssB]$ for averages over sector $B$ and $\delta B := B - \llangle \; B \; \rrangle$. The implications of this equation when applied to gravitating systems are explored at some length below.

So far the discussion has been general, but perhaps puzzling to find in a review volume on effective field theory. We next introduce a hierarchy of scales since this lies at the root of the simplicity that EFT methods exploit. In the present instance it is more useful to express this hierarchy in the time domain (rather than energy) and so we now assume sector $B$ includes `fast' degrees of freedom relative to a set of slower variables that are of interest in sector $A$. In particular, we assume the correlation functions $\llangle\, \delta B_n (t) \, \delta B_m(s) \, \rrangle\,$ fall to zero once $t - s$ is much larger than a characteristic time-scale, $\tau_c$. A useful hierarchy arises if $\tau_c$ is much smaller than the time scale $t_\ssA$ over which the evolution of $\rho_\ssA(t)$ is sought. 

Under these assumptions eq.~\pref{NakaZwanExplicit} becomes approximately local in time because the rest of the integrand varies much more slowly than $\llangle\, \delta B_n (t) \, \delta B_m(s) \, \rrangle\,$ near $s = t$ and so can be Taylor expanded about $s = t$, with the logic that contributions of successive terms to the integral should be suppressed by powers of $\tau_c/t_\ssA$. Once this is done $\rho_\ssA(t)$ (and its derivatives) can be factored out of the integral. In this case the evolution equation for $\rho_\ssA$ simplifies to
\be \label{LindbladH}
  \partial_t \,\rho_\ssA \simeq - \mathi  \Bigl[ \sum_n A_n \, \llangle \, B_n \, \rrangle + \sum_{mn} h_{mn} A_{m} A_{n} , \rho_\ssA \Bigr] + \sum_{mn} \gamma_{mn} \Bigl[ A_n\, \rho_\ssA A_m - \frac{1}{2}\left\{ A_{m} A_{n}, \rho_{\ssA} \right\} \Bigr] \,,
\ee 
where the coefficients and operators on the right-hand side are all evaluated at a the same time, $t$, as for the left-hand side and
\begin{equation}\label{CDefEq}
\begin{cases}
\ \gamma_{mn} & := C_{mn} + C_{nm}^{\ast} \\
\ h_{mn} & := \frac{1}{2i} \left( C_{mn} - C_{nm}^{\ast} \right) 
\end{cases} \qquad \mathrm{with} \ \ C_{mn}(t) := \int_{t_0}^t \exd s \  \llangle \, \delta B_m(t)\, \delta B_n(s) \, \rrangle \ ,
\end{equation}
where hermiticity of the $B_n$'s implies $\gamma_{nm}^* = \gamma_{mn}$ and $h_{nm}^* = h_{mn}$. 

This is precisely the equation we would have obtained if the sector-$B$ correlation function were approximately local in time,
\be \label{BBcorrCmn}
  \llangle \, \delta B_m(t)\, \delta B_n(s) \, \rrangle \simeq C_{mn}(t) \, \delta(t-s) \,,
\ee
An approximate master equation of this type is called a {\em Lindblad} --- or GKSL (Gorini, Kossakowski, Sudarshan, Lindblad) --- equation \cite{Lindblad, Gorini}. The hermiticity and positivity of $\gamma_{mn}$ is crucial for ensuring that $\rho_\ssA$ remain hermitian and non-negative for all $t$, as is required for its eigenvalues to carry a probability interpretation. 

Eq.~\pref{LindbladH} is much easier to work with than \pref{NakaZwanExplicit} because it is Markovian in the sense that $\partial_t\, \rho_\ssA(t)$ depends only on other variables at time $t$ and not on the entire history of evolution prior to this time. We next expand on why this property also can be useful for extending to later times the predictions of perturbation theory.

\subsection{Late-time failure of perturbative methods}
\label{LateTimeBreakdown}

As mentioned earlier, whenever a system interacts with a persistent environment it is generic that perturbative methods eventually fail to accurately capture time-evolution: $\exp[- \mathi \int V \exd t]$ eventually differs significantly from $1 - \mathi \int V \exd t$. Yet we are often able to make reliable late-time predictions anyway, even when $\int V \exd t$ is order unity. In \S\ref{ssec:Secular} we provide a few examples where this kind of `secular growth' of perturbative corrections actually arise in gravitational settings, but for now we just briefly review when and why this is possible, and argue why an evolution like eq.~\pref{LindbladH} can sometimes give reliable late-time predictions despite being derived from \pref{NakaZwanExplicit} (whose utility in the far future generically breaks down).

\subsubsection{Radioactive decay}
\label{ssec:RadDecay}

Radioactive decay is instructive in this context because it provides a well-understood example where the apparent breakdown of perturbation theory at late times can be circumvented. Consider therefore an unstable parent particle that spontaneously decays (perhaps through the weak interactions) into a collection of daughter particles: $P \to D_1+ D_2+ \cdots$. In the simplest situations such decays can be computed perturbatively and arise -- in the absence of a conservation law that forbids the decay -- due to the existence of a nonzero matrix element $\langle D_1, D_2, \cdots | V | P \rangle \sim O(g)$ for some coupling $g \ll 1$. Standard expressions give the differential decay rate 
\be
    \exd \Gamma \propto |\langle D_1, D_2, \cdots | V | P \rangle|^2 = O(g^2)\,, 
\ee
showing that decays first arise at second order in the interaction responsible.

More subtle is the justification for the (survival) probability for a parent particle not to decay, that follows the well-known exponential decay law, 
\be \label{ExpDecay}
 p(t) = e^{- \Gamma (t-t_0)} \,.
\ee
Eq.~\pref{ExpDecay} is experimentally verified to hold for times much longer than the decay's mean lifetime, $\tau = 1/\Gamma$, and so for times $\Gamma (t-t_0)  \gg 1$. Given that $\Gamma$ is computed only to order $g^2$ how can \pref{ExpDecay} be regarded as more accurate than the strictly perturbative result
\be \label{ExpDecayLin}
   p(t) \simeq 1 - \Gamma(t - t_0)+ \cdots \,, 
\ee
once $t-t_0 \gsim \tau$? 

\begin{figure}[h]
\centering
\includegraphics[scale=0.26]{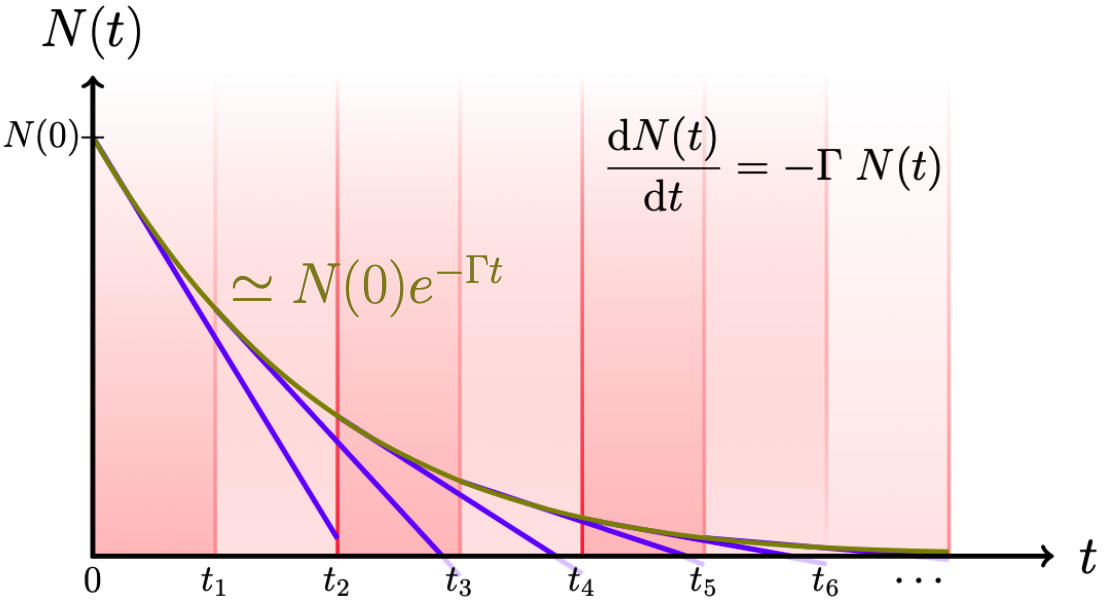} \\
\caption{Perturbation theory gives a survival probability that is linear in $t$, but does so over a tiny interval $t \in [0,t_1]$ for which $\Gamma t \ll 1$. Because the evolution equation for $N(t) = p(t) N_0$  implies $\exd N/\exd t$ is local in time it applies equally well for {\it any} interval $(t_n, t_{n+1})$ using perturbative methods, making its solutions valid over the union of all possible such intervals. The evolution equation stitches together the perturbative expressions to give the resummed solution $N(t) \simeq N(0) e^{- \Gamma t}$ out to very late times.}
\label{fig:resummation}
\end{figure}

Exponential decays arise whenever the survival probability $p(t)$ is a solution to 
\be \label{ExpDecayDE}
   \frac{\exd p}{\exd t} = - \Gamma  p \,.
\ee
Although this evolution equation is consistent with \pref{ExpDecayLin} it has a broader domain of validity because it relies only on the likelihood of decays in any short interval $\exd t$ being independent of decays in any other time windows. The value of $\Gamma$ appearing in the exponential can be extracted from \pref{ExpDecayLin} because it agrees with \pref{ExpDecayDE} for small times, but once this is done the {\it solutions} to \pref{ExpDecayDE} can be trusted for much longer times (see Fig.~\ref{fig:resummation}).

This is a very powerful line of argument, and when it works it allows working to all orders in $g^2 t$ without having to understand all observables at all orders in $g$. Use of the leading-order expression $\Gamma \sim O(g^2)$ when integrating \pref{ExpDecayDE} amounts to resumming all orders in $g^2 t$ while dropping terms like $g^4t$ that involve extra powers of the interaction $V$ without the corresponding extra powers of $t$. Integrating \pref{ExpDecayDE} using an order $g^4$ expression for $\Gamma$ similarly gives a result valid to all orders in $g^4 t$ while dropping terms like $g^6 t$ and so on.

The key to this argument is the derivation of an evolution equation like \pref{ExpDecayDE} that can have broader validity than a straight-up perturbative approach. This is possible because the evolution equation itself does not make explicit reference to the initial time, and so could apply equally well for {\it any} initial starting point $t_0$. The same argument would not be expected to be possible using \pref{NakaZwanExplicit}, for instance, since although this equation is derived on very general grounds its right-hand side makes explicit reference to both $t_0$ and $t$. It {\it could} apply to the evolution given in \pref{LindbladH} however -- such as if $C_{mn}$ defined in \pref{CDefEq} were independent of $t_0$ -- because this also makes explicit reference only to physics at time $t$ (and not to quantities like $t_0$).

\subsubsection{Qubit thermalization}
\label{ssec:ApprToEq}

\medskip\noindent
To make the above discussion concrete it is useful to examine an explicit example for which a Lindblad-type equation describes late-time behaviour. We also build in two features that are useful in some of the examples to follow: they involve thermal environments and reduced evolution is only sought for a simple two-level system (a qubit) for which all calculations can be made very explicit. 

To this end consider a two-level system (with levels split by energy $\omega$) coupled to an environment (taken here to be a relativistic real massless scalar field) prepared in a thermal state with temperature $T$. The unperturbed Hamiltonian is therefore taken to be $H_0 = H_\ssA \otimes I_\ssB + I_\ssA \otimes H_\ssB$ where  
\be \label{SampleLindblad}
   H_\ssA = \frac{\omega}{2} \left(\begin{array}{cc} 1 & 0 \\ 0 & -1  \end{array} \right) 
   \quad \hbox{and} \quad
 H_\ssB = \frac12 \int \exd^3x \; \Bigl[ (\partial_t \phi)^2 + (\nabla \phi)^2   \Bigr] \,.
\ee
Massive fields can be treated equally explicitly though involve somewhat more cumbersome expressions. 

The interaction-picture interaction linking these two systems is assumed to be
\be \label{SampleLindbladInt}
  V(t) = g \alpha(t) \otimes \phi [\bfx_0 , t] 
\ee
with the field evaluated at the qubit's (static) position $\bfx(t) = \bfx_0$ and 
\be
  \alpha(t) = \left(\begin{array}{cc} 0 & 0 \\ 1 & 0  \end{array} \right) e^{-\mathi \omega t} +
   \left(\begin{array}{cc} 0 & 1 \\ 0 & 0  \end{array} \right) e^{\mathi \omega t} 
\ee
involves the qubit's raising and lowering operators in the interaction picture. The dimensionless coupling $g \ll 1$ is assumed small enough to justify perturbative methods. Putting $\omega$ into $H_0$ rather than $V$ and working with non-degenerate perturbation theory assumes $\omega$ to be much larger than any perturbative field-induced shift in qubit energies, a condition made more explicit in \pref{largeomegacond} below. 

The field $\phi$ is prepared in an initial thermal state
\be \label{ThermalFieldState}
 \rho_\ssB = \frac{1}{Z} \; \exp\bigl[ - \beta H_\ssB \bigr] \,,
\ee
with temperature $T = 1/\beta$ and $Z = \trB[ \exp(-\beta H_\ssB)]$ and so its Wightman function $\,W(\bfx, t; \bfx',t') :=  \llangle \,\phi(\bfx,t) \phi(\bfx',t') \,\rrangle\,$  can be explicitly evaluated at coincident spatial points $\bfx = \bfx' = \bfx_0$. For a massless scalar field this gives
\be \label{masslessWlims}
  W(s)  := W(\bfx_0, t+s; \bfx_0, t) =  -\, \frac{1 }{4 \beta^2 \big[ \sinh({\pi s}/{\beta} ) - \mathi \epsilon \big]^2}   \,,
\ee
where $\varepsilon \to 0^+$ at the end of the calculation. Notice that $W(s)$ falls off exponentially once $s \gg \beta/\pi$ and diverges like $1/s^2$ as $s \to 0$.

Substituting these expressions into the Nakajima-Zwanzig equation \pref{NakaZwanExplicit} and eliminating $\rho_{\downarrow\downarrow}$ and $\rho_{\downarrow\uparrow}$ using$\tr\rho = 1$ and $\rho = \rho^\dagger$ shows the diagonal and off-diagonal components of $\rho_\ssA$ evolve independent of one another at $O(g^2)$,
\be  \label{NZQubit1}
\frac{\partial \rho_{\uparrow\uparrow}}{\partial t}   =   g^2 \int_{-t}^{t} \exd s\ W(s) \,e^{- i \omega s} - 4 g^2 \int_0^{t} \exd s\ \mathrm{Re}[W(s) ]\, \cos(\omega s) \,\rho_{\uparrow\uparrow}(t - s) \,, 
\ee 
and
\bea \label{NZQubit2}
\frac{\partial \rho_{\uparrow\downarrow}}{\partial t} & = &  - 2 g^2 \int_0^t \exd s \ \mathrm{Re}[W(s)]\, e^{\mathi \omega s} \rho_{\uparrow\downarrow}(t - s) \\
&& \qquad \qquad \qquad + 2 g^2 \, e^{2\mathi \omega t} \int_0^t \exd s \ \mathrm{Re}[W(s)]\, e^{-\mathi \omega s} \rho^*_{\uparrow\downarrow}(t - s) \,,\nn
\eea
where we assume uncorrelated initial conditions $\rho(t=0) = \rho_0 \otimes \rho_\ssB$ with $\rho_0$ (so far) unspecified.

If we choose the initial condition $\rho_0 = |\hspace{-0.5mm} \downarrow \rangle \, \langle \downarrow\hspace{-0.5mm} |$ then both $\rho_{\uparrow\uparrow}$ and $\rho_{\uparrow\downarrow}$ are at most $O(g)$ and so can be dropped on the right-hand side if we strictly work only to $O(g^2)$.
The resulting expressions reduce to $\partial_t \rho_{\uparrow\downarrow} \lsim O(g^3)$ and 
\be  \label{NZQubit1a}
\frac{\partial \rho_{\uparrow\uparrow}}{\partial t}  \simeq  g^2 \int_{-t}^{t} \exd s\ W(s) \,e^{- i \omega s} \,, 
\ee 
showing how the interaction with the thermal field starts to occupy the qubit's excited state. Because $W(s)$ is exponentially peaked around $s = 0$ with width $\beta$ the right-hand side of this equation quickly approaches the $t$-independent constant $g^2 R(\omega)$ for $t \gg \beta$, where
\be \label{QLRdef}
 R(\omega)  = \int_{-\infty}^{\infty} \exd s \, W(s) \, e^{- i \omega s} = \frac{1}{2 \pi} \;\frac{\omega}{e^{\beta\omega} - 1}   \,.
\ee
Integrating \pref{NZQubit1a} to obtain $\rho_{\uparrow\uparrow}(t)$ then gives the characteristic linear dependence on $t$ that signals a breakdown of perturbation theory at times $t \gsim \tau_p := [g^2R(\omega)]^{-1}$. Notice that this breakdown occurs at a time depending sensitively on $\beta \omega$, with $\tau_p \simeq 2\pi \beta/g^2$ when $\beta\omega \ll 1$ and $\tau_p \simeq (2\pi/g^2 \omega) e^{\beta\omega}$ when $\beta \omega \gg 1$. 

To learn the behaviour of $\rho_\ssA(t)$ for $t \sim \tau_p$ we return to equations \pref{NZQubit1} or \pref{NZQubit2}. Progress can be made if the rest of the integrand varies only over times very long compared with the support of $W(\tau)$, in which case we can expand terms like
\be\label{MarkovianSExp}
  \alpha(t-s) \, \rho_{ab}(t-s) \simeq \alpha(t) \, \rho_{ab}(t) - s \, \left[\partial_t \Bigl(\alpha \, \rho_{ab} \Bigr) \right]_{s=0} + \cdots \,,
\ee
and integrate the result term by term. Here $\alpha(t)$ is the interaction-picture matrix appearing in \pref{SampleLindbladInt} whose presence is responsible for oscillatory factors involving $e^{\pm\mathi \omega t}$. In particular, the combination $\alpha \, \rho_{ab}$ can only vary slowly over times of order $\beta$ if $\beta \omega \ll 1$, which we henceforth assume.

Dropping all but the first term of the expansion \pref{MarkovianSExp} in \pref{NZQubit1} and \pref{NZQubit2} leads to Markovian evolution of the form of \pref{LindbladH}, which for $\rho_{\uparrow\uparrow}$ and $\rho_{\uparrow\downarrow}$ becomes
\be \label{markovianQubit0}
\frac{\partial \rho_{\uparrow\uparrow}}{\partial t}  \simeq  g^2 R_0 \Bigl[ 1 - 2 \,\rho_{\uparrow\uparrow}(t) \Bigr]\,,
\ee
and
\be \label{markovianQubit20}
\frac{\partial \rho_{\uparrow\downarrow}}{\partial t}  \simeq   - g^2 R_0 \Bigl[ \rho_{\uparrow\downarrow}(t)  + e^{2\mathi \omega t}  \rho^*_{\uparrow\downarrow}(t) \Bigr] \,,
\ee
with $R_0 = (2\pi \beta)^{-1}$ the $\beta\omega \ll 1$ limit of the function $R(\omega)$ given in \pref{QLRdef}. Because \pref{markovianQubit0} makes no reference to the initial time, its domain of validity is broader than straight-up perturbation theory would be, allowing its solutions to be trusted at later times through the same reasoning as used above for exponential decays. 

Integrating leads to the solutions
\be \label{rho11relax0}
  \rho_{\uparrow\uparrow}(t) = \frac12 + \left[ \rho_{\uparrow\uparrow}(0) -  \frac12 \right] e^{ - {t}/{\xi_{d0}} } ,,
\ee
and
\be \label{rho12relax0}
\rho_{\uparrow\downarrow}(t) = \left[ \rho_{\uparrow\downarrow}(0) + \frac{\mathi g^2 R_0}{2\omega}  \Bigl(1 - e^{ 2 \mathi \omega t}\Bigr) \,  \rho_{\uparrow\downarrow}^*(0)\right] \, e^{ - t/\xi_{c0}} \,,
\ee
with\footnote{This result turns out to saturate a general upper bound $\xi_{c0} < 2 \xi_{d0}$.}
\be \label{xiDxiTvsC0}
 \xi_{c0} = 2\, \xi_{d0} = \frac{1}{g^2 R_0} = \frac{2\pi \beta}{g^2} \,.
\ee
For small times eq.~\pref{rho11relax0} agrees with the $\beta\omega \to 0$ limit of \pref{NZQubit1a} and describes initial qubit excitation due to the presence of the field. But eqs.~\pref{rho11relax0} and \pref{rho12relax0} also apply for $t \sim \xi_{d0}$ and describe late-time relaxation towards a steady state in which the qubit becomes completely mixed (more about which below): $\rho_\ssA \to \rho_\infty = \hbox{diag}(\frac12,\frac12)$.  Notice that the late-time relaxation rate $\xi_{d0}$ differs from the timescale $1/(g^2 R_0)$ that describes the early-time perturbative excitation rate out of the ground state -- given by \pref{NZQubit1a} and \pref{QLRdef} -- even when $\beta \omega \to 0$.

We can now better quantify the size of the contributions to \pref{NZQubit1} and \pref{NZQubit2} by the subdominant terms in the expansion \pref{MarkovianSExp}. These should be suppressed by powers of either $\beta \omega/\pi$ or $\beta/(\pi \xi_{d0}) = 2g^2\beta R_0/\pi \simeq  (g/\pi)^2$ and so are indeed negligible for perturbatively small $g$ provided $\beta\omega \ll 1$ as well. We may also better quantify the lower limit on $\omega$ alluded to earlier that is required by our use of nongenerate perturbation theory (which assumes the effects of $V(t)$ are perturbatively small relative to $H_\ssA$). This requires $\omega \gg g^2 R_0 \simeq {g^2}/({2\pi \beta})$ and so combining all requirements we find that late-time evolution becomes Markovian and reliably computable within the regime 
\be \label{largeomegacond}
  1 \gg \omega \beta \gg \frac{g^2}{2\pi}\,.
\ee

In the above treatment predictions acquire corrections as powers of $\beta\omega$ once the higher-order contributions in \pref{MarkovianSExp} are included. We note in passing that some of this $\omega$-dependence can be explored if only the density matrix is Taylor expanded: $\rho_{ab}(t-s) \simeq \rho_{ab}(t) - s \partial_t \rho_{ab}(t) + \cdots$ without also expanding $\alpha(t-s)$ as in \pref{MarkovianSExp}. Doing so in \pref{NZQubit1} modifies \pref{markovianQubit0} to
\be \label{markovianQubit}
\frac{\partial \rho_{\uparrow\uparrow}}{\partial t}  \simeq  g^2 R - 2 g^2  C  \,\rho_{\uparrow\uparrow}(t) \,,
\ee
where $R(\omega)$ is given by \pref{QLRdef} and the new function $C(\omega)$ is defined by
\be\label{QLCdef}
  C(\omega) := \int_{-\infty}^\infty \exd s \; \Bigl[ \hbox{Re } W(s) \Bigr] \cos (\omega s)   = \frac{\omega}{4\pi} \, \coth \left( \frac{\beta \omega}{2} \right) \,. 
\ee
This does not mean we get to drop the requirement that $\beta\omega$ must be small since this is still required to ensure that subdominant terms of the expansion of $\rho_{ab}(t-s)$ are small relative to the leading contribution \cite{Kaplanek:2019dqu}. 

The solution to \pref{markovianQubit} with $R$ and $C$ given by \pref{QLRdef} and \pref{QLCdef} is 
\be \label{rho11relax}
  \rho_{\uparrow\uparrow}(t) = \frac{1}{e^{\beta\omega} + 1} + \left[ \rho_{\uparrow\uparrow}(0) -  \frac{1}{e^{\beta\omega} + 1} \right] e^{ - {t}/{\xi_d} } \,,
\ee
with
\be \label{xiDxiTvsC}
  \xi_d = \frac{1}{2g^2 C(\omega)} = \frac{2\pi}{g^2 \omega} \, \tanh\left( \frac{\beta \omega}{2} \right)   \,.
\ee
We see from this that the late-time evolution of $\rho_\ssA$ describes thermalization; the qubit relaxes towards the thermal state
\be \label{ThermalStateQubit}
  \rho_{\rm th} :=  \left[ {\begin{array}{cc} \frac{1}{1+e^{\beta\omega}} & 0 \\ 0 & \frac{1}{1+e^{-\beta\omega}} \end{array}} \right] =  \left[ {\begin{array}{cc} e^{-\beta\omega} & 0 \\ 0 & 1 \end{array}} \right] \frac{1}{1+e^{-\beta\omega}}\,,
\ee
that shares the field's temperature. $\rho_{\rm th}$ is the unique static solution to \pref{markovianQubit} -- and to \pref{markovianQubit2} below -- and so is the state to which solutions relax. 

A similar story goes through for $\rho_{\uparrow\downarrow}$, with Taylor expansion of $\rho_{\uparrow\downarrow}(t-s)$ again removing the history-dependence of eq.~\pref{NZQubit2}, but with a few complications. The complications arise because the equation obtained after Taylor expansion involves a new function
\be
  \Delta(\omega) := 2 \int_0^\infty \exd s\, \mathrm{Re}[W(s)] \sin(\omega s) \,.
\ee
as well as the function $C(\omega)$ encountered in \pref{QLCdef}. This new function causes problems partly because it diverges in the $s \to 0$ part of the integration region. (The function $C$ does {\it not} similarly diverge because of the Wightman function's $\mathi \epsilon$ factor seen in eq.~\pref{masslessWlims}.) This is an ultraviolet divergence, and because it appears together with the qubit frequency $\omega$ it can be renormalized into the physical frequency: $\omega_\ssR = \omega + g^2 \Delta$. 

The second complication arises because the finite part of $\Delta$ is proportional to $\omega$ in the limit $\beta \omega \ll 1$ and as a result is systematically smaller in this limit than is $C \propto \beta^{-1}$, requiring $\Delta$ to be neglected relative to $C$. If these $\Delta$-dependent terms are mistakenly kept then comparing \pref{markovianQubit} and \pref{markovianQubit2} with the general Lindblad form \pref{LindbladH} shows that the matrix $\gamma_{mn}$ that results is not positive.\footnote{Similar issues also arise in optics where a laser plays the role of the environment, and is tuned to a frequency close to $\omega$, which is not small. In these applications a sensible Lindblad form is instead obtained only after performing a ‘rotating wave’ approximation that averages over the fast oscillations. Such steps are not required in the applications considered here.} Careful treatment shows that the apparent negative eigenvalues are always spurious if one religiously restricts to the domain of validity of all approximations (as must be the case). 

Keeping these points in mind, the resulting leading evolution equation again has the Lindblad form, \pref{LindbladH}, 
\be \label{markovianQubit2}
\frac{\partial \rho_{\uparrow\downarrow}}{\partial t}  \simeq   - g^2 C\, \rho_{\uparrow\downarrow}(t)  + g^2 C\,e^{2\mathi \omega t}  \rho^*_{\uparrow\downarrow}(t) \,,
\ee
with solution 
\be \label{rho12relax}
\rho_{\uparrow\downarrow}(t) =   \left[ \rho_{\uparrow\downarrow}(0) + \frac{\mathi g^2 C}{2\omega}  \Bigl(1 - e^{ 2 \mathi \omega t}\Bigr) \,  \rho_{\uparrow\downarrow}^*(0)\right] \,  e^{ - t/\xi_c}\,,
\ee
where $\xi_c = 2 \xi_d$ and we drop the subscript `$R$' on $\omega_\ssR$.

We learn two general things from this example. First, open systems can exhibit phenomena not seen in isolated quantum systems, such as the evolution from pure to mixed states that underlies the processes of decoherence and thermalization. Second, although straight-up perturbation in $g$ fails to reliably capture evolution at late times where $g^2 t$ cannot be neglected, this failure can under some circumstances be resummed to give reliable results that are valid to all orders in $g^2 t$. In this example the late-time evolution can be inferred because the full Nakajima-Zwanzig evolution becomes well-described by an approximate Lindblad equation that expresses how very slow evolution compared with the environment's typical correlation time can become Markovian (and so simpler). Solutions to the resulting Lindblad equation can be trusted at late times if its perturbative derivation works equally well in any small time interval.

\subsubsection{Secular growth for thermal fields}
\label{seculargrowth}

Although it is generic that secular growth {\it can} arise for open quantum systems, {\it does} it actually arise for gravitating systems with horizons? The complete answer to this is not known because in many systems (such as black holes) the required calculations have not yet been completely performed. But the oft-remarked similarity between systems with horizons and thermal systems provides strong circumstantial evidence that secular growth is as ubiquitous as it is for thermal systems. 

This section describes a simple example of this that is useful for the purposes of later comparison. To this end consider the same real scalar field prepared in a thermal state that was examined above as the environment with which a qubit interacted. But this time we ignore the qubit sector completely and instead study scalar-field self-interactions as expressed by the action
\be \label{lambdaphi4}
 S = - \int \exd^4 x \, \sqrt{-g} \; \left[ \frac12 \, g^{\mu\nu} \partial_\mu \phi \, \partial_\nu \phi +  \frac{\lambda}4 \, \phi^4 \right] \,,
\ee
with coupling $\lambda$ chosen to be small and our interest lies in the validity of perturbative calculations in powers of $\lambda$. The gravitational field is included for future purposes though the appearance of the metric $g_{\mu\nu}$, but in this section we work purely in flat Minkowski space. In a Hamiltonian formalism this amounts to replacing the operator $H_\ssB$ of \pref{SampleLindblad} with $H_{\rm tot} = H_\ssB + H_{\rm int}$ where the self-interaction term is $H_{\rm int} = \frac{1}{4!} \int \exd^3x\, \lambda \phi^4$. In terms of this the field's thermal state is -- {\it c.f.}~eq.~\pref{ThermalFieldState}:
\be \label{ThermalFieldStateInt}
 \rho = \frac{1}{Z} \; \exp\bigl[ - \beta H_{\rm tot} \bigr] \,,
\ee
with $\beta = 1/T$ the inverse temperature and $Z$ chosen as usual to ensure Tr $\rho = 1$. 

\vspace{-1.4cm}
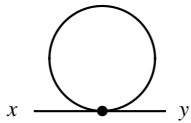
\begin{figure}
\centerline{
\begin{picture}(260,100)
\put(100,0){\begin{picture}(100,100)
    \thicklines
   \put(36,40){\circle{40}}
   \put(10,20){\line(1,0){50}}
   \put(36,20){\circle*{4}}
   \put(0,18){$x$}
   \put(65,18){$y$}
    \end{picture}}
\end{picture}
}
\caption{Feynman graph giving the leading one-loop `tadpole' correction to the scalar propagator. The result from this graph must be summed with the tree-level two-point counter-term graph.\label{figure:Tadpole}}
\end{figure}

We seek an example of how secular growth arises for this scalar-field system and so identify a quantity whose $O(\lambda)$ correction is a growing function of time. We follow \cite{Burgess:2018sou} and use the Feynman correlation function $G(x; y) = \langle \cT \phi(x) \phi(y) \rangle$ as our example, where $\cT$ denotes time-ordering and the average is taken using the thermal state \pref{ThermalFieldStateInt}. To study secular evolution we compute order-$\lambda$ corrections to $G(x,y)$, and because we wish time-dependence to be explicit we work in the real-time formalism computed within the Schwinger-Keldysh (or ``in-in'') framework \cite{Schwinger:1960qe, Keldysh:1964ud}. 

The leading correction in this case comes from the tadpole graph of Fig.~\ref{figure:Tadpole}. The loop part of this graph turns out to be position-independent and diverges in the ultraviolet (UV) in a temperature-independent way. Because it is temperature independent the UV divergence can be renormalized in the same way as at zero temperature, by choosing a mass counterterm to ensure that the renormalized zero-temperature mass remains zero. Once this is done, evaluation of the loop subgraph within Fig.~\ref{figure:Tadpole} using the thermal state gives the finite $O(\lambda)$ temperature-dependent mass shift
\be \label{dmsqT}
  \delta m_\ssT^2 = \frac{ \lambda T^2}{4} = \frac{ \lambda}{4\beta^2} \,.
\ee

Using this self-energy in the remainder of the graph and evaluating the result at zero spatial separation, $\bfy = \bfx$ leads to the following form in the limit of large time difference $y^0 - x^0 = t$:
\be \label{Tsect}
  G_{\rm tad}(x^0, \bfx; x^0+t, \bfx) \simeq \frac{\delta m_\ssT^2 \,T t}{8\pi} + \cdots = \frac{\lambda T^3 t}{32 \pi} + \cdots\,,
\ee
where the ellipses denote terms that grow more slowly than linearly for large $t$. Behold secular growth: for sufficiently large $t$ the correction to $G(x,y)$ is not small no matter how small $\lambda$ is chosen to be.

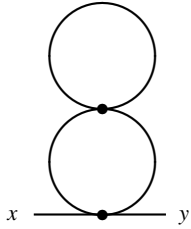
\begin{figure}
\centerline{
\begin{picture}(260,100)
\put(100,0){\begin{picture}(100,100)
    \thicklines
   \put(36,40){\circle{40}}
   \put(36,80){\circle{40}}
   \put(10,20){\line(1,0){50}}
   \put(36,20){\circle*{4}}
   \put(36,60){\circle*{4}}
   \put(0,18){$x$}
   \put(65,18){$y$}
    \end{picture}}
\end{picture}
}
\caption{Feynman graph giving a subleading two-loop `cactus' correction to the scalar propagator. Although this is not the only two-loop contribution, it is noteworthy because of the power-law IR divergences it acquires due to the singularity of the Bose-Einstein distribution at small momenta. \label{figure:Cactus}}
\end{figure}

Secular growth is related to (but not identical with) the infrared divergences that can arise when performing loops because both involve intermediate states with arbitrarily small frequency. This connection schematically arises because unusual growth at large times in a correlation function is usually related to singular behaviour for the small-frequency part of its Fourier transform, and such singularities can cause loop integrals to diverge because of contributions near $\omega = 0$. It is useful to further explore this connection since the resummation method used to handle IR divergences suggests how secular effects might also be resummed (at least in this particular example).

Although the graph of Fig.~\ref{figure:Tadpole} itself is infrared finite when the particles in the loop are massless, the same is not true for the higher loop `cactus' graph of Fig.~\ref{figure:Cactus}. The dangerous part of this graph for small $k$ comes from the two propagators in the bottom loop which make it diverge logarithmically $\propto \int \exd^4k/(k^2)^2$ even at zero temperature. By contrast, the bottom loop instead diverges like a power of the IR cutoff at finite temperature because of the singularity of the Bose-Einstein distribution\footnote{If evaluated in Euclidean signature the more singular behaviour arises because the replacement of the frequency integral by a Matsubara sum means one integrates only over spatial momenta $\exd^3 k$.} $n_\ssB(k) = (e^{k/T}-1)^{-1} \simeq T/k$ for $k \ll T$ contributing a factor
\be \label{IRpower}
  \hbox{lower loop} \propto \frac{\lambda T}{\omega_{\IR}} \,,
\ee
where $\omega_{\IR}$ is an IR cutoff.

Notice that the zero-temperature limit of the entire graph of Fig.~\ref{figure:Cactus} (including the IR divergent part) precisely cancels with the graph where the upper loop is replaced by the mass counter-term once this counterterm is chosen to ensure $m^2 = 0$ at zero temperature. The same cancellation does not also occur at finite temperature because the counter-term only cancels the zero-temperature part of the top loop's contribution. 

Having Fig.~\ref{figure:Cactus} be proportional to \pref{IRpower} means that frequencies $\omega \le \lambda T$ contribute unsuppressed relative to Fig.~\ref{figure:Tadpole} because the large factor of $T/\omega_\IR$ compensates for the small factor $\lambda$. The same is true for graphs with multiple bubbles added on top of one another since each extra bubble contributes an additional factor of \pref{IRpower}. This signals a breakdown of perturbative methods since having $\lambda$ be small no longer ensures that graphs with additional divergent higher loops come with the penalty of a small loop-counting parameter, and suggests rethinking the split within the total Hamiltonian between $H_0$ and $H_{\rm int}$ to obtain a better-converging expansion.

The particular divergent contributions coming from IR divergent bubble graphs are indeed famously resummed by moving the temperature-dependent mass into the unperturbed Hamiltonian -- {\it i.e.}~by adding and subtracting the temperature-dependent mass shift $\delta m^2_\ssT$ and putting $m^2 + \delta m^2_\ssT$ into $H_0$. In this case all internal lines represent massive states and the new $\delta m_\ssT^2 = 0$ because the self-energy graphs systematically cancel with the corresponding graphs with the final bubble replaced by new mass counter-term, a well-known `hard-thermal-loop' resummation \cite{Gross:1980br,Altherr}.\footnote{Although controlled resummation can be possible for scalars whose mass vanishes at zero temperature, it need not be true in general that the perturbative breakdown associated with IR divergences can always be removed by resumming specific subsets of higher-order graphs. An example where this does not work arises when the total temperature-dependent mass $m^2 = m_0^2 + \delta m_\ssT^2$ vanishes for some nonzero temperature. This corresponds to arranging the theory to sit at a critical point, for which it is well-known that mean-field (perturbative) calculations are simply not a good approximation.}  This suggests that secular growth might similarly be resummed by perturbing around the full temperature-dependent mass.

\subsection{Influence Functionals}

An alternative approach to open systems uses the Feynman-Vernon influence functional \cite{Feynman:1963fq}, which is the path-integral version of the Hamiltonian evolution story told above for operators (see \cite{OpenSystems, FeynmanHibbs, Weiss, Calzetta:2008iqa} for textbook descriptions of this technique and \cite{IFs} for a non-exhaustive list of their early use in general non-equilibrium and gravitational settings). One of their advantages is they are easily adapted to focus directly on correlation functions -- as opposed to the reduced density matrix -- and so can often allow one to cut directly to the chase when computing observables.

To make the transition to path integrals we start with the standard expression for transition amplitudes $\langle \varphi_1 | U(t,t_0) | \varphi_2 \rangle$, where $U(t,t_0)$ is the unitary time evolution operator defined in (\ref{unitaryU}). These have the standard path integral representation
\begin{equation} \label{pathintegral}
\langle \varphi_2 | U(t,t_0) |  \varphi_1 \rangle \ = \ \int_{\varphi_1}^{\varphi_2} \mathcal{D}\phi\; e^{i S[\phi]}
\end{equation}
where $S[\phi]$ is the system's classical action 
\begin{equation} \label{pathintegralaction}
S[\phi] \ = \ \int_{t_0}^{t} \exd t' \; L [ \phi ; t' ] \,,
\end{equation}
and $|  \varphi_{j} \rangle $ are eigenstates of the Schr\"odinger-picture field operator $\hat{\phi}(\mathbf{x})$, so
\begin{equation}
\hat{\phi}(\mathbf{x}) |  \varphi_{j} \rangle = \varphi_j(\mathbf{x}) |  \varphi_{j} \rangle \,,
\end{equation}
and for notational clarity the $\mathbf{x}$-dependence of the eigenvalues in eq.~(\ref{pathintegral}) is suppressed when used as a label for a bra and ket. The limits of integration indicate that the integral sums over all configurations whose end points are chosen to be the specified initial and final eigenvalues: $\phi(\bfx,t_0) = \varphi_1(\bfx)$ and $\phi(\bfx,t) = \varphi_2(\mathbf{x})$. The path integral (\ref{pathintegral}) describes the probability amplitude that the system finds itself in the eigenstate $|\varphi_2\rangle$ at time $t$ given it began in the eigenstate $| \varphi_1 \rangle$ at the initial time $t_0$.

This can be turned into a path-integral representation for the density matrix by noting that the density matrix evolves as $\hat{\rho}(t) = U(t,t_0) \,\hat{\rho}_0 \,U^{\ast}(t,t_0)$ -- {\it c.f.}~eq.~\pref{rhoevolution} -- in the Schr\"odinger picture. This ensures its matrix elements can be written as
\begin{eqnarray}
\langle \varphi_2 | \hat{\rho}(t) | \varphi_1 \rangle & = & \langle \varphi_2 | U(t,t_0) \, \hat{\rho}_0 \,  U^{\ast}(t,t_0) | \varphi_1 \rangle  \\
& = & \sum_{\varphi_3 , \varphi_4} \langle \varphi_2 | U(t,t_0) | \varphi_4 \rangle \; \langle \varphi_{4} | \hat{\rho}_0 | \varphi_3 \rangle \; \langle \varphi_{1} | U(t,t_0) | \varphi_3 \rangle^{\ast}  \nn
\end{eqnarray}
which inserts two resolutions of the identity. Using (\ref{pathintegral}) one finds 
\begin{eqnarray} \label{densityaction}
\langle \varphi_2 | \hat{\rho}(t) | \varphi_1 \rangle & = &  \sum_{\varphi_3 , \varphi_4} \int_{\varphi_{4}}^{\varphi_2} \mathcal{D}\phi^{+}  \int_{\varphi_{3}}^{\varphi_1} \mathcal{D}\phi^{-}   \; e^{i S[\phi^{+}]- i S[\phi^{-}]}  \langle \varphi_{4} | \hat{\rho}_0 | \varphi_3 \rangle   \qquad
\end{eqnarray}
where the path integration takes place over two independent field variables, labelled $\phi^{+}$ and $\phi^{-}$, that satisfy the distinct boundary conditions $\phi^+(\bfx,t) = \varphi_2(\bfx)$, $\phi^+(\bfx,t_0) = \varphi_4(\bfx)$ and $\phi^-(\bfx,t) = \varphi_1(\bfx)$, $\phi^-(\bfx,t_0) = \varphi_3(\bfx)$.

\begin{figure}[h]
\centering
\includegraphics[scale=0.60]{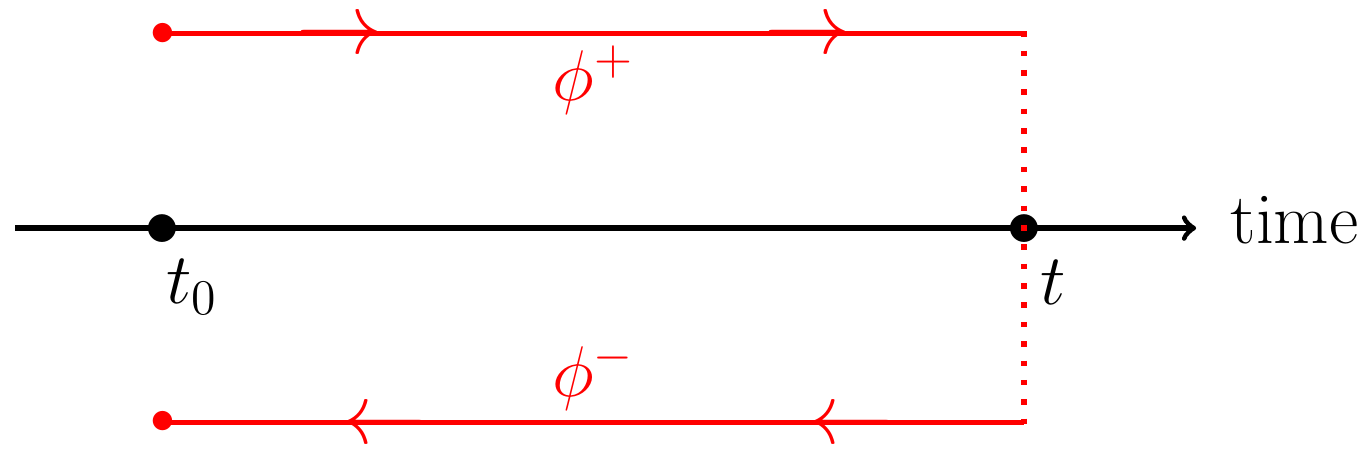} \\
\caption{Depiction of the closed-time path contour. Since only initial data is known (as opposed to the standard ``in-out'' situation in scattering calculations), averages like eq.~(\ref{corr_path}) computed via path integrals start at time $t_0$, flow out to $t$, and then back to $t_0$. The value of the field on the upper and lower branches are denoted by $\phi^{+}$ and $\phi^{-}$ and are treated as independent variables. In the literature, the contours are sometimes translated above and below the time axis by a tiny amount $\pm i \epsilon$ to help with convergence of the path integrations.}
\label{fig:Keldysh}
\end{figure}

Expressions such as (\ref{densityaction}) are the point of departure for the Schwinger-Keldysh formalism -- or  `in-in' or `closed-time path' formalism -- used in \S\ref{seculargrowth} to calculate field theoretic quantities like eq.~(\ref{Tsect}) for thermal correlators \cite{Bakshi:1962dv}. For a simple example of how this works, notice that one can use eq.~(\ref{densityaction}) to write equal-time correlation functions in terms of the diagonal components of $ \hat{\rho}$,
\begin{eqnarray} \label{corr_path}
& & \mathrm{Tr}\left[ \phi_{\ssH}(t,\mathbf{x}) \phi_{\ssH}(t,\mathbf{x}') \hat{\rho}_0 \right] = \mathrm{Tr}\left[ \hat{\phi}(\mathbf{x}) \hat{\phi}(\mathbf{x}') \hat{\rho}(t) \right] =  \sum_{\varphi} \varphi(\mathbf{x}) \varphi(\mathbf{x}')\; \langle \varphi | \hat{\rho}(t) | \varphi \rangle \notag \\
& \ & \ =  \sum_{\varphi, \varphi_3 , \varphi_4} \varphi(\mathbf{x}) \varphi(\mathbf{x}') \int_{\varphi_{4}}^{\varphi} \mathcal{D}\phi^{+}  \int_{\varphi_{3}}^{\varphi} \mathcal{D}\phi^{-}   \; e^{i S[\phi^{+}]- i S[\phi^{-}]}  \langle \varphi_{4} | \hat{\rho}_0 | \varphi_3 \rangle \ , \qquad
\end{eqnarray}
where Heisenberg picture operators are defined in terms of Schr\"odinger picture operators by
\begin{equation}
\phi_{\ssH}(t,\mathbf{x}) = U^{\ast}(t,t_0) \hat{\phi}(\mathbf{x})U(t,t_0) \ .
\end{equation}
In this formalism, one interprets path integrations such as the one in (\ref{corr_path}) as performed over a deformed time contour which starts at the initial time $t_0$, propagates out to the later time $t$ (where the boundary conditions in (\ref{corr_path}) are identified), and then {\it backwards} to the initial time $t_0$, as depicted in Figure \ref{fig:Keldysh}. The field variable living on either branch of the `closed-time path' are treated as independent variables ($\phi^{+}$ and $\phi^{-}$) as conveyed by earlier formulas.

The Schwinger-Keldysh framework is most useful when one only has information about the initial state ({\it i.e.} being ignorant of the `out' state) and so is well-equipped for dealing with non-equilibrium time evolution (generally the case in cosmology). It is also useful for time-evolving mixed states (including thermal states\footnote{When dealing with thermal states at temperature $1/\beta$, one usually further deforms the contour in Fig.~\ref{fig:Keldysh} to include a third piece that points in the imaginary time direction starting at time $t_0$, where one identifies $t_0$ with $t_0 + i \beta$ --- ultimately this is a manifestation of the Kubo-Martin-Schwinger (KMS) detailed-balance condition obeyed by thermal correlators \cite{KMS1,KMS2}.}) since quantum averages can be expressed in terms of expectation values over the `doubled' degrees of freedom --- what is useful about this is that it allows one to use standard techniques of QFT ({\it e.g.} Feynman diagrams) in this more complicated setting.
 
For the present purposes, formula (\ref{densityaction}) becomes most interesting in an open systems setting when the action $S[\phi_{\ssA}, \phi_{\ssB} ]$ is a function of system $\phi_{\ssA}$ and environment $\phi_{\ssB}$ degrees of freedom. Using the notation from earlier, this means that the components of the full density matrix (\ref{densityaction}) here become instead
\begin{eqnarray}
\langle a_2, b_2 | \hat{\rho}(t) | a_1, b_1 \rangle & = &  \sum_{a_3 , a_4, b_3, b_4} \int_{a_4}^{a_2} \mathcal{D}\phi^{+}_{\ssA}  \int_{b_4}^{b_2} \mathcal{D}\phi^{+}_{\ssB} \int_{a_3}^{a_1} \mathcal{D}\phi^{-}_{\ssA} \int_{b_3}^{b_1} \mathcal{D}\phi^{-}_{\ssB} \notag \\
& \ & \qquad\qquad \ \  \times \; e^{i S[\phi_{\ssA}^{+}, \phi_{\ssB}^{+} ] - i S[\phi_{\ssA}^{-}, \phi_{\ssB}^{-} ] }  \langle a_4, b_4 | \hat{\rho}_0 | a_3, b_3 \rangle  \,, 
\end{eqnarray}
where the lower (and upper) end points on the path integrals are fixed at time $t_0$ (and $t$) as in (\ref{densityaction}). When the system and environment interact through an action of the form
\begin{equation}
S[\phi_{\ssA}, \phi_{\ssB} ] \ = \ S_{\ssA}[\phi_{\ssA}] + S_{\ssB}[\phi_{\ssB}] + S_{\mathrm{int}}[\phi_{\ssA}, \phi_{\ssB}] \ .
\end{equation}
for some interaction $S_{\mathrm{int}}$, then one can trace over the environment to find the elements of the (Schr\"odinger-picture) reduced density matrix $\hat{\rho}_{\ssA}$ to get the path-integral representation
\begin{eqnarray}  \label{FV_reduced}
& \ & \langle a_2 | \hat{\rho}_{\ssA}(t) | a_1 \rangle =  \sum_{b} \langle a_2, b | \hat{\rho}(t) | a_1, b \rangle \\
& \ & \qquad = \sum_{a_3 , a_4} \int_{a_4}^{a_2} \mathcal{D}\phi_{\ssA}^{+} \int_{a_3}^{a_1} \mathcal{D}\phi^{-}_{\ssA} \; e^{i S_{\ssA}[\phi^{+}_{\ssA} ] - i S_{\ssA}[\phi_{\ssA}^{-} ]+ i S_{\ssI\ssF} [\phi_{\ssA}^{+}, \phi_{\ssA}^{-} ]}  \langle a_4 | \rho_{\ssA}(t_0) | a_3 \rangle \,.  \notag
\end{eqnarray}
This last expression packages the entire effect of the environment into $S_{\ssI\ssF} [\phi_{\ssA}^{+}, \phi_{\ssA}^{-} ]$, called the influence functional, defined by
\begin{eqnarray} \label{FeynmanVernon}
e^{iS_{\ssI\ssF} [\phi_{\ssA}^{+}, \phi_{\ssA}^{-}]} & := &  \sum_{b, b_3, b_4} \int_{b_4}^{b} \mathcal{D}\phi_{\ssB}^{+} \int_{b_3}^{b} \mathcal{D}\phi_{\ssB}^{-} \;   \\
&\ & \qquad \times\;  e^{i S_{\ssB}[\phi^{+}_{\ssB}]  + i S_{\mathrm{int}}[\phi^{+}_{\ssA}, \phi^{+}_{\ssB} ] - i S_{\ssB}[\phi^{-}_{\ssB}] - i S_{\mathrm{int}}[\phi^{-}_{\ssA}, \phi_{\ssB}^{-} ] } \; \langle b_4 | \rho_\ssB | b_3 \rangle  \ . \notag
\end{eqnarray}
which assumes an uncorrelated initial condition $\rho_0 = \rho_{\ssA}(t_0) \otimes \rho_{\ssB}$, as in (\ref{intial_uncorr}). 

A few comments are in order. In general $S_{\ssI\ssF}[\phi_{\ssA}^{+}, \phi_{\ssA}^{-}]$ is composed of interactions between $\phi_{\ssA}^{+}$ and $\phi_{\ssA}^{-}$ (as well as self-interactions for each). This is distinct from the usual situation in the Schwinger-Keldysh formalism -- see eq.~(\ref{densityaction}) -- where the actions for $\phi_{\ssA}^{+}$ and $\phi_{\ssA}^{-}$ split apart. There are also unitary and non-unitary contributions to $S_{\ssI\ssF}$, and finally the interactions are generally {\it non-local}.\footnote{Non-locality might be especially relevant in discussion of gravitation backgrounds with horizons, where other types of hypothetical non-local effects are sometimes considered.} All these ingredients further drive home the point that effective descriptions for open systems can be very non-Wilsonian. 

In \S\ref{ssec:Hotspot} we explore in some detail how this works for a toy model of a black hole that has been devised to be solvable and yet also to capture important open-system features. We there in particular use the influence functional to obtain an alternative derivation of a master equation and a stochastic Langevin equation.

\section{Applications to Rindler space}

We next turn to some simple illustrative applications of these techniques in spacetimes with horizons. Applications of Open EFT techniques are still relatively recent for gravity and we try to choose examples that illustrate current developments. 

We do so using applications to Rindler, de Sitter and simple black-hole geometries in turn, starting in this sector with the simplest -- Rindler -- case. We begin in each case with simple qubit examples that behave very much like the thermal system described above and for which the simplicity of the qubit sector allows calculations to be very explicit and assumptions to be robustly tested. We then describe illustrative examples of secular growth in more fully field-theoretic systems.

\subsection{Accelerated qubit thermalization}
\label{RindlerQubit}

We start by considering horizons generated by accelerated motion without a gravitational field. To explore this we use the same flat-space system as described above in \S\ref{ssec:ApprToEq} -- a qubit coupled to a massless scalar field -- but with two important differences: the scalar field is prepared in its (Minkowski) vacuum $\rho_\ssB = | \Omega \rangle \, \langle \Omega |$ (the $T \to 0$ limit of the above) and the qubit is uniformly accelerated rather than static. 

The qubit is a simple two-level accelerated DeWitt-Unruh detector \cite{Unruh:1976db, DeWitt:1980hx}, whose evolution in perturbation theory is well-studied in the literature (at least for early times). We describe how these early treatments can be extended to give reliable predictions at the late times relevant to thermalization to the Unruh temperature (which lies beyond the domain of validity of the earlier perturbative studies).

The system describing the detector and the quantum field is again described by the unperturbed Hamiltonian $H_0 = H_\ssA \otimes I_\ssB + I_\ssA \otimes H_\ssB$, where $H_\ssB$ is precisely as given in \pref{SampleLindblad} but with the qubit Hamiltonian generalized to include the time-dilation associated with its motion:
\be
 H_\ssA = \bfh \; \frac{\exd \tau}{\exd t} \quad \hbox{with} \quad \bfh := \frac{\omega}{2} \left(\begin{array}{cc} 1 & 0 \\ 0 & -1  \end{array} \right) 
\ee
where $\tau$ is the proper time $\exd \tau^2 = - \eta_{\mu\nu} \exd x^\mu \exd x^\nu$ evaluated along the qubit's world-line $x^\mu = y^\mu(\tau)$. This means that $\omega > 0$ is the splitting between qubit energy levels as measured in the rest-frame of the qubit. The complete Hamiltonian is $H_0 + H_{\rm int}$ where the qubit interaction in Schr\"odinger picture also contains a time-dilation factor, with Hamiltonian
\be \label{SampleLindbladIntR}
  H_{\rm int} = g \hat\alpha \otimes \hat\phi [y(\tau)] \; \frac{\exd \tau}{\exd t} 
  \quad \hbox{where} \quad
  \hat\alpha = \left(\begin{array}{cc} 0 & 1 \\ 1 & 0  \end{array} \right) \,.
\ee
where the `hat' again denotes a Schr\"odinger-picture operator and the dimensionless coupling $g \ll 1$ is chosen small enough to justify perturbative methods. 

With these choices the free evolution is given by the time-ordered expression
\begin{eqnarray}
U_0(t) \ = \ \mathcal{T}\exp\left( - i \int_0^t \exd s\ H_0 \right)  =  e^{- i\, \bfh \tau(t)} \otimes e^{- i H_\ssB t} 
\end{eqnarray}
and so the interaction-picture interaction Hamiltonian becomes
\begin{eqnarray}
V(t) \ = \ U^{\dagger}_0(t) H_{\mathrm{int}} U_0(t) \ = \ g \, \phi[y (\tau)] \otimes \alpha(\tau) \, \frac{\exd\tau}{\exd t} \,, \label{intintH}
\end{eqnarray}
where
\be
  \alpha(\tau) = e^{\mathi \, \bfh \tau} \hat \alpha \, e^{-\mathi \, \bfh \tau} = \left(\begin{array}{cc} 0 & 0 \\ 1 & 0  \end{array} \right) e^{-\mathi \omega \tau} +
   \left(\begin{array}{cc} 0 & 1 \\ 0 & 0  \end{array} \right) e^{\mathi \omega \tau} \,.
\ee

From here on we proceed precisely as in \S\ref{ssec:ApprToEq}, so we report only the parts of the calculation that change. The first change is to choose the field to be prepared in the Minkowski vacuum $|\Omega\rangle$ rather than a thermal state, so 
\begin{eqnarray} \label{uncorrelated}
\rho_\ssB = |{ \Omega }\rangle \langle{ \Omega }| \,. 
\end{eqnarray}
This is also the $T \to 0$ limit of the state considered in \S\ref{ssec:ApprToEq} and so for a static qubit situated at $\bfx = \bfx_0$ the Wightman function can be found by taking $\beta \to \infty$ in \pref{masslessWlims}: 
\be \label{masslessWlimsT0}
  W(s) =  \langle \Omega | \phi(\bfx_0,s) \phi(\bfx_0,0) |\Omega \rangle  =  -\, \frac{1 }{4\pi^2 (s - \mathi \epsilon )^2}   \,,
\ee
implying $R(\omega) = 0$ -- {\it c.f.}~eq.~\pref{QLRdef}. Unsurprisingly, stationary qubits that are initially in their ground state remain there despite coupling to the field if the field is prepared in its own ground state.

We instead move the qubit along a uniformly accelerated trajectory $x^\mu = y^\mu(\tau)$ 
\begin{eqnarray} \label{Rindler}
\hbox{with} \quad y^\mu (\tau) \ = \left[ \frac{1}{a} \sinh(a\tau), \frac{1}{a} \cosh(a\tau), 0, 0 \right] 
\end{eqnarray}
where $a>0$ denotes the qubit's proper acceleration and $\tau$ in this parameterization denotes proper time along the curve as measured using the Minkowski metric. A scalar field's Wightman function evaluated along this worldline is evaluated in closed-form in \cite{Takagi:1986kn,Langlois:2005nf}, and the massless limit of this result is\footnote{It can be tempting to rewrite $\sinh({a\tau}/{2}) - i \varepsilon$ as $\sinh[({a( \tau - i \varepsilon)}/{2}]$ with the reasoning that these are equivalent because infinitesimal $\varepsilon > 0$ is important only near $\tau =0$ \cite{Takagi:1986kn}. Although this reasoning is not false for real $\tau$, this replacement can be dangerous where $\tau$ is not real because it does {\it not} preserve important properties like the KMS condition mentioned in footnote 7.}
\be \label{WightmanDefMink}
  W(\tau) = \langle \Omega | \phi[  y (\tau) ] \phi[ y (0) ] |\Omega \rangle 
  = -\, \frac{a^2 }{16 \pi^2 \big[ \sinh({a\tau}/{2} ) - i \varepsilon \big]^2}\,,
\ee
which has the thermal form -- compare with \pref{masslessWlims} -- with $\beta = 2\pi/{a}$ corresponding to the usual Unruh temperature.

From here on the calculation follows along very much the same lines as in \S\ref{ssec:ApprToEq}. For qubits initially in their ground state and uncorrelated with the field, the leading rest-frame perturbative excitation rate agrees with earlier predictions \cite{Unruh:1976db, DeWitt:1980hx, Sciama:1981hr}:
\be  \label{NZQubit1aR}
\frac{\partial \rho_{\uparrow\uparrow}}{\partial \tau}  \simeq   g^2 R(\omega) \quad \hbox{where} \quad
R(\omega) = \frac{\omega}{2 \pi} \;\frac{1}{e^{2\pi \omega/a} - 1}   \,,
\ee
for proper times $2\pi/a \ll \tau \ll 2 \pi/(g^2 a)$. Although this perturbative result breaks down at large times, the arguments of \S\ref{ssec:ApprToEq} show that evolution is reliably well-approximated by a Markovian process within the parameter regime \pref{largeomegacond}, which in this instance can be written
\be \label{largeomegacondR}
  1 \gg \frac{2\pi \omega}{a}  \gg \frac{g^2}{2\pi}\,.
\ee

In this regime the proper-time evolution is given by
\be \label{markovianQubitR}
\frac{\partial \rho_{\uparrow\uparrow}}{\partial \tau}  \simeq  g^2 R(\omega) - 2 g^2  C(\omega)  \,\rho_{\uparrow\uparrow}(\tau) \,,
\ee
with
\be\label{QLCdefR}
  C(\omega) = \frac{\omega}{4\pi} \, \coth \left( \frac{\pi \omega}{a} \right) \,. 
\ee
The late-time solutions are given by \pref{rho11relax} and \pref{rho12relax} and describe thermalization to the Unruh temperature with relaxation times in the qubit rest frame given by
\be \label{xiDxiTvsCacc}
  \xi_c = 2\, \xi_d = \frac{1}{g^2 C(\omega)} = \frac{4\pi}{g^2 \omega} \, \tanh\left( \frac{\pi \omega}{a} \right)  \simeq \frac{4\pi^2}{g^2a} \left\{ 1 + O\left[ \left(\frac{\pi\omega}{a} \right)^2\right] \right\} \,.
\ee

\subsection{Secular growth and the Minkowski vacuum}
\label{ssec:Secular}

We next turn to a more fully field-theoretic example for which the interaction involves only quantum fields. In particular we show how loop corrections involving a scalar field prepared in its interacting vacuum in the presence of a self-interaction $\frac{1}{4!} \lambda \phi^4$ can in some circumstances have the same kinds of secular growth\footnote{See also \cite{Chaykov:2022zro}.} in its propagators as found above for the self-interacting thermal case. In doing so we will resolve a puzzle: if the Minkowski vacuum can describe thermal physics for some observers, then why doesn't bog-standard zero-temperature perturbation theory (with the field prepared in its Minkowski vacuum) also give rise to secular growth effects and late-time perturbative breakdown? 

To understand why, we re-evaluate the graph of Fig.~\ref{figure:Tadpole} at zero temperature. For these purposes it is useful to recall that the lowest-order position-space propagator at zero temperature is (for nonzero scalar mass)
\be \label{G0prop}
  G_0(x;y) = \frac{1}{4\pi^2} \;\frac{m}{\sqrt{(x-y)^2 + i\epsilon}} \; K_1\Bigl[ m \sqrt{(x-y)^2 + i \epsilon} \Bigr] \,,
\ee
where $(x-y)^2 = \eta_{\mu\nu} (x-y)^\mu (x-y)^\nu$ is negative for time-like separations and positive for space-like separations and $K_\nu(z)$ is a modified Bessel function of the second kind. The asymptotic form of $K_\nu(z)$ reproduces the usual massless limit:
\be
 G_0(x;y) = \frac{1}{4\pi^2} \; \frac{1}{(x-y)^2 + i\epsilon} \qquad (\hbox{$m=0$}) \,.
\ee

For a massive scalar Fig.~\ref{figure:Tadpole} evaluates to give
\be \label{Gtad1loop}
  G_{\rm tad}(x;y) = -i \Sigma(0) \int \frac{\exd^4p}{(2\pi)^4} \; \frac{ e^{i p \cdot (x-y)}}{(p^2 + m^2 - i \epsilon)^2} =  -\frac{\delta m^2}{8\pi^2}  \; K_0\Bigl[ m \sqrt{(x-y)^2 + i \epsilon} \Bigr] \,,
\ee
where the zero-momentum self-energy is
\be
  \Sigma(0) = -\delta m^2_{\rm ct} + 3i  \lambda \int \frac{\exd^4 k}{(2\pi)^4} \; \frac{1}{k^2+m^2 -i \epsilon} \,,
\ee
with $\delta m^2_{\rm ct}$ the mass counter-term that subtracts the UV-divergent part of the integral and $\delta m^2 = -\Sigma(0)$ is the UV finite mass shift after renormalization. 

In the massless limit this evaluates to 
\be \label{MinkTadxy}
  G_{\rm tad}(x;y) = \frac{\delta m^2}{8\pi^2} \; \ln \Bigl[\mu \sqrt{(x-y)^2 + i \epsilon} \Bigr] 
  \qquad\hbox{(when } m=0)\,.
\ee
up to a spacetime-independent IR-divergent constant. Here $\mu$ is the renormalization scale for the renormalized coupling $\lambda(\mu)$, whose precise value depends on how this IR divergence is regulated but plays no role when tracking the dependence of the result on $x-y$. 

We now evaluate the propagator using coordinates adapted to accelerating observers, for which the flat metric becomes
\be
  \exd s^2 =  \eta_{\mu\nu} \exd x^\mu \exd x^\nu = - (a\xi)^2 \exd \tau^2 + \exd \xi^2 + \exd y^2 + \exd z^2 \,.
\ee
Imagine now choosing both $x^\mu$ and $y^\mu$ to both lie along the specific accelerating world line described by 
\be
  x = \xi \cosh(a \tau) \qquad \hbox{and} \qquad t = \xi \sinh(a \tau) \,,
\ee
with the other two coordinates ($y$ and $z$) fixed. 
We evaluate $G(x,y)$ with $x^2 = y^2$ and $x^3 = y^3$ and choose a particular Rindler observer -- {\it i.e.}~fixed $\xi$ --- on whose accelerating world-line both $x^\mu$ and $y^\mu$ lie (and so are separated purely by a shift in Rindler time, $\tau$). We choose in particular the specific trajectory $\xi = 1/a$, for which Rindler time is also the proper time along the curve and the proper acceleration is $a$.

The invariant separation between two points separated by proper time $\tau$ is then given by
\be
 - (x-y)^2 = \frac{4}{a^2} \; \sinh^2 \left( \frac{a \tau}2 \right) \simeq \frac{1}{a^2}\, e^{a \tau} \Bigl[ 1 + O(e^{-2a\tau}) \Bigr] \,,
\ee
where the final approximate equality gives the asymptotic form when $a \tau \gg 1$. For such an observer (with $a \tau \gg 1$) eq.~\pref{MinkTadxy} implies an asymptotic time dependence of $G_{\rm tad}(\tau)$ of the form
\be \label{GRind1}
 G_{\rm tad}(\tau) =  \frac{\delta m^2}{8\pi^2} \; \ln \Bigl[\mu \sqrt{(x-y)^2 + i \epsilon} \Bigr] \simeq  \frac{\delta m^2 \, a \tau}{16\pi^2} + \hbox{subdominant} \,.
\ee

What value should be chosen for $\delta m^2$ in this last expression? As discussed earlier, the tadpole loop diverges in the UV and this divergence is cancelled by the mass counterterm, leaving a finite residual whose value depends on the renormalization scheme. 
A Minkowski observer would effectively choose $\delta m^2_\ssM =  - \Sigma_\ssM(0) = 0$ so that the unperturbed mass parameter is the physical mass. A punishment for not doing so would be to have IR divergences appear even at zero temperature in graphs like Fig.~\ref{figure:Cactus}. A Rindler observer would instead choose a counterterm that ensures the sum of the counterterm and tadpole graph vanishes if evaluated in the Rindler ground state\footnote{The Rindler state is the ground state of the Rindler Hamiltonian, defined as the Poincar\'e boost generator that generates translations in Rindler time.}  rather than the Minkowski one, for similar reasons. 

The Minkowski observer's choice sets $\delta m^2 = 0$ and so \pref{GRind1} gives zero. The Rinder observer's choice instead no longer completely cancels the tadpole graph when it is evaluated in the Minkowski vacuum, and so differs from the Minkowski choice by a finite amount. A standard calculation using the Rindler vacuum gives \cite{Takagi:1986kn, Davies:1974th, Boulware:1974dm, Troost:1978yk, Dowker:1978aza, Linet:1995mq}
\be \label{amass}
 \delta m^2_a  = \frac{ \lambda a^2}{16\pi^2} \,.
\ee
Using this in \pref{GRind1} then gives
\be
 G_{\rm tad}(\tau) =  \frac{\lambda a^3 \tau}{(16\pi^2)^2} + \hbox{subdominant} \,,
\ee
which precisely agrees with the thermal result \pref{Tsect} provided we identify temperature with acceleration in the usual way: $T = {a}/{2\pi}$.

If the Minkowski observer's counter-term choice had instead been made then secular growth would have instead appeared in the Rindler correlation function. Said differently, secular growth at late Rindler time cannot be avoided for {\it both} the Minkowski and Rindler vacua, and once it is excluded from one vacuum it necessarily appears for the other, and does so in precisely the way that would have been expected for a thermal state.

\section{Applications to de Sitter space}

We next turn to the simplest curved-space examples, which involve the de Sitter and near-de Sitter cosmologies likely to be associated with Dark-Energy dominated late-time cosmologies or inflationary cosmologies at very early times. de Sitter geometries are simple in the sense that de Sitter space shares the same number of isometries as does flat space, despite the presence of curvature. Because of this symmetry more explicit calculations are known for these geometries than for less symmetric ones that share the existence of horizons (like black holes). 

The geometry of interest is a spatially flat, homogeneous and isotropic metric of the Friedmann-LeMaitre-Robertson-Walker (FLRW) form
\be \label{FLRWmetric}
   \exd s^2 = - \exd t^2 + a^2(t) \, \exd \bfx \cdot \exd \bfx = a^2(\eta) \, \Bigl[ - \exd \eta^2 + \exd \bfx \cdot \exd \bfx \Bigr] 
\ee
where the geometry is specified in terms of a time-dependent scale factor $a(t)$, and for any given $a(t)$ conformal time $\eta$ and cosmic time $t$ are related by $\exd\, t = a \, \exd \eta$. The nonzero components of the Riemann tensor for this geometry are
\be \label{FLRWRiemann}
   {R^0}_{i\,0j} = -q \, H^2 g_{ij} \quad \hbox{and} \quad {R^i}_{jkl} = H^2 \Bigl( \delta^i_k g_{jl} - \delta^i_l g_{jk} \Bigr)
\ee
(plus those related to these by permuting indices), where 
\be \label{Hqdefs}
    H(t) := \frac{\dot a}{a} \quad \hbox{and} \quad q(t) := - \frac{a \, \ddot a}{\dot a^2} = -1 + \varepsilon_1 \quad
      \hbox{where} \quad  \varepsilon_1(t) := - \frac{\dot H}{H^2} 
\ee
and overdots (primes) denote $\exd/\exd t$ ($\exd/\exd \eta$). de Sitter space is the special case 
\be \label{dSdef}
   a = e^{Ht} =- \frac{1}{H\eta} \quad \hbox{where }  H \hbox{ is constant (and so } q = -1 \hbox{ and } \varepsilon_1 = 0) \,, 
\ee
and for this specific choice the Riemann tensor has the maximally symmetric form ${R^\mu}_{\nu\lambda\rho} = H^2 (\delta^\mu_\lambda g_{\nu\rho} - \delta^\mu_\rho g_{\nu\lambda})$.

The quantization of such systems is done semiclassically, with all fields (including the metric) split into a classical background plus a quantum fluctuation. This makes sense if we work within the spirit of effective field theories for gravity (GREFT), since these allow one to systematically ask when and why such semiclassical methods are justified. See the accompanying chapters in this review for more about these techniques (and see \cite{Burgess:2003jk, Adshead:2017srh}).   

A massless scalar field that only couples minimally to gravity within an FLRW universe satisfies 
\bea \label{KGeq}
  -\Box \phi = -\frac{1}{\sqrt{-g}} \, \partial_\mu \Bigl(\sqrt{-g}\;  g^{\mu\nu} \partial_\nu \phi \Bigr) 
 &=& \ddot \phi + 3H \dot \phi - \frac{1}{a^2} \nabla^2 \phi \nn\\
  &=&  \frac{1}{a^2} \Bigl( \phi'' + \frac{2a' }{a}\, \phi' - \nabla^2 \phi \Bigr) = 0 
\eea
where $\nabla^2 = \delta^{ij} \partial_i \partial_j$. Particle states for such a field can be labelled using momenta $\bfk$ because the spatial slices of the geometry are flat (and so translation invariant).

Specializing to de Sitter space and expanding the field in terms of the corresponding creation and annihilation operators
\be \label{dSphiExp}
  \phi(x) = \int \frac{\exd^3k}{(2\pi)^{3/2}} \Bigl[ v_\bfk(x) \, \mfa_\bfk + v^*_\bfk (x) \, \mfa_\bfk^* \Bigr] \quad\hbox{with} \quad v_\bfk(x) = \frac{u_\bfk(\eta)}{a} \, e^{i \bfk \cdot \bfx} \,,
\ee
implies 
\be
   u_\bfk'' + \left( \bfk^2 - \frac{2}{\eta^2} \right) u_\bfk = 0 \,,
\ee
for which the normalized solutions are
\bea \label{BDmodes}
    u_\bfk(\eta) &=& \frac{1}{\sqrt{2k}} \left(1 - \frac{i}{k \eta} \right) \, e^{-ik\eta} \\
    \hbox{and so} \quad
    v_\bfk(a,\bfx) &=& \frac{1}{\sqrt{2k^3}}\left(  \frac{k}{a}  +iH\right) \, e^{i[(k/aH)+\bfk\cdot \bfx]} \nn
\eea
(with $k := |\bfk|$) ensuring the standard commutation relations $[\mfa_\bfk, \mfa^*_\bfq] = \delta^3(\bfk-\bfq)$. The vacuum state $|\Omega\rangle$ defined by $\mfa_\bfk |\Omega \rangle = 0$ is called the Bunch-Davies vacuum.

For these modes physical momenta $\bfp(t) = \bfk/a(t)$ are time-dependent and fall monotonically and so for every mode there is a time $t_{\rm he}$ after which $|\bfp(t)| < H$, with crossover between these regimes (`horizon exit') occuring when $aH=k$. (Equivalently, the sweep from $-\infty < t < \infty$ corresponds to $- \infty < \eta < 0$ with $\eta \to 0$ in the far future and horizon exit occuring when $k\eta = -1$.) The modes \pref{BDmodes} are oscillatory in the remote past (when $k\eta \ll -1$) and are chosen to resemble standard flat-space modes in this regime. Their motion stops being adiabatic after horizon exit, with \pref{BDmodes} showing they stop oscillating (or `freeze') once $|k\eta| \ll 1$. 

Explicit expressions for massive mode functions are also known for de Sitter geometries, though we do not need them in what follows. One result we do use however is the expression (for a massive scalar field) for the renormalized expectation $\langle \phi^2(x) \rangle = \langle \Omega | \phi^2(x) | \Omega \rangle$ using the Bunch-Davies vacuum in de Sitter space. This is an ultraviolet divergent quantity and is independent of $x$ for massive fields. Renormalizing so that it vanishes when $H \to 0$ leaves a finite and nonzero value for de Sitter given by
\be\label{dSphi2m}
   \langle \phi^2(x) \rangle = \frac{3H^4}{8\pi^2 m^2} \,.
\ee

The divergence of this result as $m \to 0$ reflects the IR divergence that arises in this limit, as can be seen directly using the massless mode functions \pref{BDmodes}:
\be  \label{masslessphisqdS}
  \langle \phi^2(x) \rangle_{\rm massless} = \int \frac{\exd^3k}{(2\pi)^3} |v_\bfk(x)|^2 = \frac{1}{2\pi^2} \int_0^\infty \frac{\exd k}{k}\; \Bigl( k^3 |v_\bfk(x)|^2 \Bigr)
\ee 
where the second equality uses that $|v_\bfk(x)|^2$ is independent of the direction of $\bfk$. This expression diverges as $k \to 0$ due to the small-$k$ behaviour of $|v_\bfk|^2$ seen in \pref{BDmodes}. For later purposes we remark that if this integral is regulated in the UV by multiplying by a window function $\xi_\Lambda[k/a(t)]$ that discriminates against UV momenta -- such as if $\xi_\Lambda(z) =1$ for $z \ll \Lambda$ and $\xi_\Lambda(z) = 0$ for $z \gg \Lambda$ -- then
\bea \label{PhiSqDeriv}
   \partial_t \langle \phi^2(x) \rangle_{\rm massless} &=& H a \, \partial_a \int \frac{\exd^3k}{(2\pi)^3} |v_\bfk(x)|^2 \xi_\Lambda(k/a) \nn\\
   &=&
   -  \frac{H}{2\pi^2} \int_0^\infty \exd k \; \frac{\partial}{\partial k}  \Bigl( k^3 |v_\bfk(x)|^2 \xi_\Lambda \Bigr) =  \frac{H^3}{4\pi^2} \,,
\eea
and so is infrared finite. This uses that $a \, \partial_a = - k \, \partial_k$ when acting on any function of $k/a$ together with \pref{BDmodes} and the properties $\xi_\Lambda(\infty) = 0$ and $\xi_\Lambda(0) = 1$. Having $\langle \phi^2 \rangle$ be linear in $t$ makes many of its implications resemble those of a random walk.

The remainder of this section briefly sketches several applications of Open EFTs to de Sitter geometries. As in the previous section we first examine the simplest case of a qubit interacting with a quantum field, and then discuss several more field-theoretic situations both of which involve the need to resum secular growth. 

\subsection{Qubit thermalization}

One can probe the structure of de Sitter space by again considering a qubit coupled to a scalar field along the lines of \S\ref{RindlerQubit}. The presence of an event horizon --- in this case the de Sitter horizon caused by the ever-expanding nature of the universe --- gives rise to the so-called Gibbons-Hawking temperature $T_\GH = H/(2\pi)$. The qubit again thermalizes to this temperature in much the same way as in earlier sections, but this example also shows that horizons can capture other features also present in thermal baths: in this case the phenomenon of `critical slowing down' in which thermalization becomes very slow when the effective mass of the scalar is tuned to be very small. 

For later convenience we set up the problem in a way that is easy to generalize to other ({\it e.g.}~black hole) geometries. To this end we write metric in a slightly more general form
\begin{equation} \label{static}
\exd s^2 = - f(\mathbf{x}) \exd t^2 + \gamma_{\,ij}(\mathbf{x},t) \exd x^i \exd x^j  \ ,
\end{equation}
for which de Sitter space corresponds to the choices $f(\mathbf{x}) = 1$ and $\gamma_{\,ij} =e^{2 H t} \delta_{ij}$ (when written using cosmic time $t$). In these coordinates we assume that the qubit moves along a co-moving trajectory:
\begin{equation}
y^{\mu}(\tau) = [ t(\tau), \mathbf{x}(\tau) ] =  [ \tau, \mathbf{x}_0 ]  \,.
\end{equation}

We take the scalar quantum field describing the quantum environment to be governed by the Klein-Gordon Hamiltonian, which using the metric \pref{static} becomes
\begin{equation} \label{KG_H_static}
H_{\ssB} = \frac{1}{2} \int_{\Sigma_t} \exd^3 x \; \sqrt{ f \gamma } \left[ \frac{(\partial_t \phi)^2}{f} + \gamma^{\,ij} \partial_i \phi \partial_j \phi + ( m^2 - \xi R) \phi^2 \right]
\end{equation}
with $R = 12H^2$ the Ricci scalar computed using (\ref{FLRWRiemann}) and the integration is over a spacelike hypersurface $\Sigma_t$ of fixed $t$. We introduce here a non-minimal coupling to gravity parameterized by the dimensionless coupling $\xi$, and assume the scalar is prepared in the Bunch-Davies vacuum $|\Omega \rangle$.

With these choices the calculation of qubit response proceeds much as before. The autocorrelations of $\phi$ along the qubit worldline are
\begin{eqnarray} \label{dS_selfCorr}
W(\tau) & = & \langle \Omega |  \phi[y(\tau)] \phi[y(0)] | \Omega \rangle = \langle \Omega |  \phi(\tau,\bfx_0) \phi(0,\bfx_0) | \Omega \rangle \\
& = & \frac{H^2(\frac{1}{4} - \nu^2)}{16 \pi \cos(\pi \nu)} \; _2F_1\left( \frac{3}{2} + \nu , \frac{3}{2} -\nu ; 2 ; 1 + \left[ \sinh\left( \frac{H\tau}{2} \right) - i \varepsilon \right]^2 \right) \nn
\end{eqnarray}
where $_2F_1(a,b;c;z)$ is Gauss' hypergeometric function and 
\begin{equation}
\nu := \sqrt{ \frac{9}{4} - \frac{M^2}{H^2} } \quad \hbox{with effective mass} \quad M^2 := m^2 - 12 \xi H^2 \,.
\end{equation}
The late-time behaviour of this correlator is
\begin{eqnarray}   \label{WdSfalloff}
W(\tau) & \simeq & W_0 \; e^{ - \kappa \tau} \quad \hbox{when} \quad \kappa \tau \gg 1 \,,
\end{eqnarray}
where
\begin{equation}
W_0 := \frac{H^2}{4 \pi^{5/2} } i e^{i \pi \nu} \Gamma\left( \frac{3}{2} - \nu \right) \Gamma(\nu) \quad \mathrm{and} \quad \kappa := \left( \frac{3}{2} - \nu \right) H
\end{equation}
with $\Gamma(z)$ denoting Euler's gamma function.

If the qubit is prepared in its ground state then the purely perturbative rate with which it becomes excited gives the standard result for an Unruh-DeWitt detector: $\partial_{\tau} \rho_{\uparrow \uparrow} \simeq g^2 R(\omega)$ with $R(\omega)$ again defined by the first equality in \pref{QLRdef}, which in this case evaluates to
\begin{equation}
R(\omega) \ = \ \frac{H}{4 \pi^3} e^{ - {\pi\omega}/{H}} \left| \Gamma\left( \frac{3}{4} + \frac{\nu}{2} +  \frac{i\omega}{2H} \right) \Gamma\left( \frac{3}{4} - \frac{\nu}{2} + \frac{i\omega}{2H} \right)  \right|^2 \ .
\end{equation}
This perturbative result again breaks down at late times -- applying only when $g^2 C(\omega ) \tau \ll 1$ where $C(\omega)$ is defined by the first equality of \pref{QLCdef}. 

Late-time evolution can again be resummed, at least for time-scales long compared to the characteristic time $\kappa^{-1}$ set by the fall-off of environmental correlations \pref{WdSfalloff}. The domain of validity of the late-time Markovian regime turns out to be more delicate because of the dependence of $\kappa$ on the effective mass $M$. $\nu$ becomes imaginary if $M \gsim H$ and 
\begin{equation}
 \hbox{Re }\kappa = \frac{3H}{2} \,.
\end{equation}
The Markovian regime ends up being restricted to the parameter regime
\begin{equation}
  1 \gg \frac{ 2\pi\omega}{H} \gg \frac{g^2}{2\pi}
\end{equation}
for reasons similar to the ones given for the uniformly accelerated qubit in (\ref{largeomegacondR}). This results in the late-time solutions \pref{rho11relax} and \pref{rho12relax}, describing equilibration with the Gibbons-Hawking temperature over timescales
\be
  \xi_c = 2\, \xi_d = \frac{1}{g^2 C(\omega)} \simeq \frac{4\pi^3}{g^2 H} \left| \Gamma\left( \frac{3}{4} + \frac{\nu}{2} \right) \Gamma\left( \frac{3}{4} - \frac{\nu}{2} \right)  \right|^{-2} \,.
\ee

The behaviour is very different if one instead chooses $M/H \ll 1$ (and so $\nu \simeq \frac32$), since in this limit 
\begin{equation} \label{divergingW}
W_0 \simeq \frac{3H^4}{8 \pi^2 M^2} \qquad \mathrm{and} \qquad \kappa \simeq \frac{M^2}{3H}  \,.
\end{equation}
Both the amplitude and width of the correlation function $W(\tau)$ become parametrically large in the limit of small effective mass. Although Markovianity ultimately applies, it does so (for $M/H \ll 1$) only in the more restrictive regime  
\begin{equation} \label{criticalslowing}
1 \gg \frac{\omega}{H} \gg \frac{M^2}{H^2} \qquad \mathrm{and} \qquad \frac{M^6}{H^6} \gg g^2
\end{equation}
with relaxation timescales now taking the form
\be
  \xi_c = 2\, \xi_d = \frac{1}{g^2 C(\omega)} \simeq \frac{4\pi^2 M^4}{9 g^2 H^5} \,.
\ee

Notice that $\xi \gg 1/\kappa$ in this regime because the domain of validity (\ref{criticalslowing}) ensures the $1/g^2$ enhancement overwhelms the $M/H$ suppression. The Markovian approximation is so restrictive in this instance because it requires the width of $W(\tau)$ to be the shortest timescale in the problem, and this means that the enormous timescale $1/\kappa \simeq 3H/M^2$ must be smaller than any of the other scales associated with qubit evolution (like $\xi$ and $1/\omega$). 

We note in passing that it is also possible to solve explicitly for qubit evolution at late times even when this evolution is non-Markovian by returning to the Nakajima-Zwanzig equation (\ref{NZQubit1}) and (\ref{NZQubit2}). This more cumbersome calculation enlarges the domain of validity for which resummed late-time evolution can be obtained -- for instance applying in a regime where $\omega/H \ll 1$ can be either larger or smaller than $M/H \ll 1$, unlike in (\ref{criticalslowing}) above (see \cite{Kaplanek:2019vzj} for details).

\subsection{Coarse-grained fields}

We close this section with several examples of Open EFT calculations for which both system and environment are described by fields. Although this gives up the simplicity of the qubit examples, many of the lessons learned there continue to go through. In the examples considered we take the observed system to consist of super-Hubble modes, for which $k/a \ll H$, and seek the influence on these due to shorter-wavelength modes. 

We describe ongoing work aimed at two kinds of applications: one outlining the late-time evolution of the probability distribution for the amplitude of super-Hubble scalar field modes; the other computing the decoherence rate of field fluctuations during inflation, initially of a spectator scalar field but eventually for metric fluctuations more generally. 

Consider first a spectator scalar field $\Phi$ ({\it i.e.}~one whose energy density is negligible relative to the energy density responsible for the curvature of the de Sitter geometry). Expanding about a background configuration: $\Phi(t,\bfx) = \varphi(t) + \phi(t,\bfx)$ on a near-de Sitter metric, the system and environment are defined in terms of short- and long-wavelength modes -- {\it c.f.}~eq.~\pref{dSphiExp} -- so $\phi(x) = \phi_{\rm sys}(x) + \phi_{\rm env}(x)$ with
\bea \label{dSphiExpSys}
  \phi_{\rm sys}(x) &:=& \int  \frac{\exd^3k}{(2\pi)^{3/2}} \Bigl[ v_\bfk(x) \, \mfa_\bfk + v^*_\bfk(x) \, \mfa_\bfk^* \Bigr] f(k,k_*) \nn\\
  \hbox{and} \quad \phi_{\rm env}(x) &:=& \int  \frac{\exd^3k}{(2\pi)^{3/2}} \Bigl[ v_\bfk(x) \, \mfa_\bfk + v^*_\bfk(x) \, \mfa_\bfk^* \Bigr] \Bigl[ 1 - f(k,k_*) \Bigr]\,,
\eea
and $0 \leq f(k,k_*) \leq 1$ a window function that distinguishes `short' from `long` wavelengths relative to a reference $k_*$. 

For instance if $f(k,k_*) = \Theta(k_* - k)$ is a Heaviside step function then the system consists of those modes whose comoving momentum satisfies $k < k_*$. (More generally, smoother choices for $f$ that transition from 0 to 1 over a region $k_* - \delta < k < k_*+\delta$ could also be entertained.) All modes with $k < k_*$ are super-Hubble after some time $t_0$ where $p_*(t_0) := k_*/a(t_0) < H$. Alternatively, it can also be convenient for some purposes to allow $f(k,k_*)$ also to depend on $t$, such as if the system/environment split is defined in terms of physical wavelengths. For instance, if the system is defined by $p(t) < \Lambda$ for all $t$ where $\Lambda$ is a fixed physical scale then $f(k,\Lambda,t) = \Theta[\Lambda a(t) - k]$. In such circumstances the derivation given in \S\ref{ssec:OpenQS} for the time-evolution of the system density matrix $\rho_\ssA(t)$ must be revisited to allow for a time-dependent division between system and environment.

\subsubsection{Stochastic inflation}
\label{sec:stoch}

The evidence for the existence of secular effects is clearer for de Sitter geometries because explicit calculations of subleading perturbative effects have been performed. This section summarizes an example, together with the preliminary evidence that the secular growth visible in it can be resummed \cite{SIref} using the formalism of Stochastic Inflation \cite{StochInf}. The story of how stochastic inflation itself is now becoming understood as the leading part of a more systematic approximation is the topic Daniel Green's chapter \cite{Green:2022ovz} in this review. 

To this end consider again massless $\lambda \phi^4$ theory, with action as given by \pref{lambdaphi4}, specialized to the FLRW metric \pref{FLRWmetric} with de Sitter scale factor \pref{dSdef}. For simplicity take the stress energy associated with the scalar field to be much smaller than the value of the cosmological constant responsible for the de Sitter curvature -- {\it i.e.}~a `spectator' field. The consistency of this assumption can be verified {\it ex post facto} by checking that the scalar stress energy in the state of interest is order $H^4$. See however \cite{Vennin:2015hra, Vennin:2020kng} for the treatment of non-spectator scalars whose stress energy drives inflation, and \cite{Burgess:2015ajz} for the extension of spectator scalars to include scalar masses and slow-roll corrections.

$O(\lambda)$ corrections have been computed explicitly for $\langle \phi^2(x) \rangle$ evaluated in the adiabatic (Bunch-Davies) vacuum whose mode functions are given in \pref{BDmodes}. For a massive scalar field the symmetries of de Sitter space (and the Bunch-Davies vacuum) ensure $\langle \phi^2(x) \rangle$ is independent of $x$ and given by \pref{dSphi2m} (so is singular\footnote{Notice the same is not true for the energy density since $\langle m^2 \phi^2 \rangle \sim H^4$.} as $m \to 0$). As discussed around eq.~\pref{masslessphisqdS}, this singularity shows up in the massless limit as an IR divergence in $\langle \phi^2(x) \rangle$ and depending on how it is regulated\footnote{Introducing a small nonzero mass is an example of a time-independent IR regularization.} this can introduce a time-dependence to $\langle \phi^2 \rangle$ in the massless limit. For instance \pref{PhiSqDeriv} gives $\langle \phi^2 \rangle \propto t = H^{-1} \ln a$ and including $O(\lambda)$ corrections turns out to give \cite{SIref}
\be \label{phi2dSsec}
    \langle \phi^2(x) \rangle = \frac{H^2}{4\pi^2} \, \ln a \left[ 1 - \frac{\lambda}{36\pi^2} \, \ln^2 a + \cdots \right] \,,
\ee
up to an additive constant whose value is regularization dependent but irrelevant for the time-dependence of the right-hand side. The factor of $\ln^2 a = (Ht)^2$ multiplying $\lambda$ is the secular growth that undermines trust in perturbative methods at late time. 

Besides showing the existence of secular growth, this same calculation also provides evidence that this growth can be controllably resummed. The evidence comes from comparing the (IR finite) derivative $\partial_t \langle \phi^2(x) \rangle$ computed from \pref{phi2dSsec} with the predictions of Stochastic Inflation \cite{StochInf}. Stochastic Inflation starts with the similarity between the leading prediction $\langle \phi^2(t) \rangle \propto t$ and the variance of the distance travelled in a random walk. The proposal is to compute the time evolution of field correlations on super-Hubble scales by building on this random-walk analogy; by regarding the approximately position-independent value $\varphi$ taken by the super-Hubble part of the field to be a random variable in the region of size $H^{-1}$ surrounding $x$ (a `Hubble patch'). If $P(\varphi,t)$ denotes the probability of it taking the value $\varphi$ at time $t$ then correlators can be computed using formulae like
\be 
   \langle \phi^{2n}(t) \rangle = \int \exd \varphi \; \varphi^{2n} \, P(\varphi,t) \,.
\ee

The random-walk part of the picture enters when computing the time-dependence of $\langle \phi^{2n}(t) \rangle$, with the time-evolution of $P(\varphi,t)$ taken to be governed by a Fokker-Planck equation 
\be\label{FokkerPlanck}
   \partial_t P = \frac{H^3}{8\pi^2} \, \frac{\partial^2 P}{\partial\varphi^2} + \frac{1}{3H} \, \frac{\partial}{\partial \varphi} \left( \frac{\partial V}{\partial \varphi} \; P \right) \,,
\ee
as one would expect for a random walk in the presence of a potential $V(\varphi)$. The second term of the right-hand side of this equation is designed to properly evolve the mean
\be
  \partial_t \langle \phi(t) \rangle = \int \exd \varphi \; \varphi \, \partial_t P(\varphi,t) = - \frac{1}{3H} \langle V'(\phi) \rangle
\ee
corresponding to the evolution $3H \dot\phi + V'(\phi) = 0$. This is the evolution equation for $\phi$ that would be obtained from \pref{KGeq} if a potential $V(\phi)$ were added and if restricted to the `slow-roll' regime where $\ddot \phi$ can be neglected. 

The first term on the right-hand side of \pref{FokkerPlanck} is similarly chosen to reproduce the leading $H^3/(4\pi^2)$ contribution -- {\it c.f.}~eq.~\pref{PhiSqDeriv} -- to the rate of change of the variance. For instance, specializing \pref{FokkerPlanck} to $V = \frac{1}{4!} \lambda \phi^4$ leads to the prediction
\be
  \partial_t \langle \phi^2(t) \rangle  = \int \exd \varphi \; \varphi^2 \, \partial_t P(\varphi,t) = \frac{H^3}{4\pi^2} - \frac{\lambda}{9H} \langle \phi^4 \rangle \,.
\ee
Ref.~\cite{SIref} sets up the hierarchy of evolution equations for $\partial_t \langle \phi^{2n}(t) \rangle$ implied by \pref{FokkerPlanck} by continuing as above for the specific case $V = \frac{1}{4!} \lambda \phi^4$, and then solves the resulting recursion relations. This leads to the more explicit predictions
\bea
 \langle \phi^{2n}(t) \rangle &=& (2n-1)!! \left( \frac{H^2}{4\pi^2} \, \ln a \right)^n \left[ 1 - \frac{n(n+1)}{2} \left( \frac{\lambda}{36\pi^2} \right) \ln^2 a + \right. \\
 && \quad \left. + \frac{n}{280}(35n^3 + 170 n^2 + 225 n + 74) \left( \frac{\lambda}{36\pi^2} \, \ln^2 a \right)^2 + \cdots \right] \,.\nn
\eea

This expression agrees -- including $O(\lambda)$ corrections -- with results like \pref{phi2dSsec} for the evolution of these quantities predicted by explicit field theory calculations. This suggests that the stochastic formulation captures the long-wavelength part of the perturbative result, potentially giving insight into how small secular effects evolve. Furthermore, it does so in a way that seems to give access to the late-time future towards which the secular evolution ultimately leads. In the stochastic formulation the evolution describes relaxation towards a static state, whose form can be found by solving \pref{FokkerPlanck} for $P_\infty(\varphi)$ under the assumption that $\partial_t P_\infty = 0$. This leads to 
\be\label{FokkerPlancktIndep}
    \frac{H^3}{8\pi^2} \, \frac{\partial^2 P_\infty}{\partial\varphi^2} + \frac{1}{3H} \, \frac{\partial}{\partial \varphi} \left( \frac{\partial V}{\partial \varphi} \; P_\infty \right) = 0\,,
\ee
with late-time solution
\be\label{PinftyStoch}
   P_\infty(\varphi) = C \, \exp \left[ -  \frac{8\pi^2V(\varphi)}{3H^4} \right] \,.
\ee
In the case of a free massive field $V = \frac12 m^2 \varphi^2$ eq.~\pref{PinftyStoch} describes a Gaussian distribution around mean $\langle \phi \rangle = 0$ with the correct variance $\langle \phi^2 \rangle = 3H^4/(8\pi^2m^2)$. But for an interacting potential $V = \frac{1}{4!} \lambda \varphi^4$ eq.~\pref{PinftyStoch} instead predicts evolution towards a very non-Gaussian distribution. 

Considerable effort has been devoted to proving that late-time evolution of quantum fields on de Sitter space is well-described by Stochastic Inflation \cite{StochInf,Starobinsky:1994bd,Tsamis:2005hd}, and in particular how it might emerge more systematically as the leading approximation for long-wavelength modes within an open-system approach along the lines used here \cite{StochasticPrelim}. Recent efforts have developed diagrammatic arguments to identify more systematically both how the stochastic limit arises and what its leading corrections are \cite{StochasticGraphs}. We defer to Daniel Green's chapter of this review \cite{Green:2022ovz} for a more expert description of these developments. 

\subsubsection{Primordial decoherence}

We close this section with a variation on the above themes that describes a more practical field-theoretic Open EFT calculation. We apply this formalism to compute the decoherence rate of primordial fluctuations during inflation. More specifically, we describe the speed with which gravitational self-interactions can allow unseen short-wavelength metric fluctuations to decohere their observed longer wavelength cousins that are believed to seed primordial fluctuations within inflationary cosmologies (see also \cite{Grishchuk:1989ss, Brandenberger:1990bx, Calzetta:1995ys, Kiefer:1998qe, Hollowood:2017bil, Martin:2018lin}).

To this end consider the following single-field inflationary model with inflaton $\Phi$ coupled to gravity with action
\begin{equation} \label{actionstart}
S = \int \exd^4 x\; \sqrt{ - g} \left[ \frac{M_p^2}{2} \, R
    - \frac{1}{2} g^{\mu\nu} \, \partial_{\mu} \Phi \,
    \partial_{\nu} \Phi  - V(\Phi) \right]
\end{equation}
where $V(\Phi)$ is designed so the classical homogeneous solutions $\varphi(t)$ describe slow-roll inflation, and so in particular ensure $\varepsilon_1(\varphi) := - \dot H/H^2 \simeq \frac12(M_p\, \partial_\phi V/V)^2 \ll 1$. The scalar is {\it not} assumed to be a spectator and as a result fluctuations about the background mix scalar and metric modes. Writing $\Phi(\bfx,t) = \varphi(t) + \phi(\bfx,t)$ and expanding the metric
\begin{eqnarray} \label{ADMmetric}
  \exd s^2 = - N^2 \exd t^2 + h_{ij} \big( \exd x^i + N^i \exd t \big)
  \big( \exd x^j + N^j \exd t \big) \,,
\end{eqnarray}
standard arguments allows us to write the metric fluctuation to second order as
\begin{equation} \label{metricsplit}
     h_{ij} = a^2 e^{2\zeta} \hat h_{ij} \quad\hbox{with}\quad
     \hat h_{ij} = \delta_{ij} + \gamma_{ij} + \frac12 \,
     \delta^{kl} \gamma_{ik} \gamma_{lj} + \cdots \,,
\end{equation}
where $a(t)$ is the near-de Sitter scale factor for the background FLRW metric, $\det \hat h_{ij} = 1$ and $\delta^{ij} \partial_i \gamma_{jk} = \delta^{ij} \gamma_{ij} = 0$. 

$\gamma_{ij}$ describes the metric's tensor perturbations (gravitational waves) while one combination of $\phi$ and $\zeta$ describes the metric's scalar perturbations and the other combination represents a gauge freedom corresponding to different ways to foliate the spacetime into time slices. Two convenient gauge conditions are the choices $\phi = 0$ (co-moving gauge) or $\zeta = 0$ (spatially-flat gauge). 

In co-moving gauge the leading (quadratic) part of the action governing fluctuations has the form \cite{Kodama:1984ziu, Mukhanov:1990me}  
\begin{equation}
\label{freescalaraction}
  ^{(2)}S = \int \exd  t \; \exd^3 x \left\{
  \frac{\dot{\varphi}^2}{2H^2}\bigg[ a^3 \dot{\zeta}^2
    - a (\partial \zeta )^2 \bigg]  +   
\frac{\Mp^2}{8} \bigg[ a^3 \dot{\gamma}^{\,ij} \dot{\gamma}_{\,ij}
- a (\partial^{k} \gamma^{\,ij} )(\partial_{k} \gamma_{\,ij} ) \bigg] \right\} \,,
\end{equation}
where spatial indices are raised and lowered using $\delta_{ij}$, so $(\partial \zeta)^2 = \delta^{ij} \partial_i \zeta \, \partial_j \zeta$. This has the canonical form $\frac12 \, \int \exd \eta [(v')^2 + (v_{ij}')^2 + \cdots]$ 
%
%
once rewritten in terms of the Mukhanov-Sasaki variables 
\begin{equation}
\label{eq:defzeta}
v(\eta, \bfx) = a \,\Mp \sqrt{2\slrl } \, \zeta(\eta,\bfx ) 
\quad \hbox{and} \quad
v_{ij}(\eta, \bfx) = \frac{1}{2} \, a \,\Mp \gamma_{ij}(\eta, \bfx)\,,
\end{equation}
where the slow-roll parameter enters because of the background classical slow-roll relation $\dot{\varphi}^2=2H^2\Mp^2\slrl$. 

Interaction terms are obtained by expanding the action \pref{actionstart} to cubic and higher order in the fluctuations, and every extra power of $v$ or $v_{ij}$ costs a power of $1/M_p$. For instance, at cubic order one finds the interactions \cite{Maldacena:2002vr}
\bea  \label{allgravitonscalarcubics}
{}^{(3)}S &=& \int \exd t \, \exd^3x \, \Mp^2
\biggl[ \slrl ^2  a\,  \zeta \left(\partial \zeta\right)^2 +  \slrl  a \gamma^{\,ij}  \partial_i \zeta  \partial_j \zeta 
 +  \frac{ \slrl }{8} \, a \, \zeta \, \partial^l\gamma^{\,ij} \partial_l
\gamma_{ij}+ \cdots \biggr] \nn\\
 &=& \int \frac{\exd \eta \exd^3x}{aM_p} 
\left\{ \frac{\sqrt{\slrl}}{2\sqrt2}  \; v\Bigl[ \left(\partial v\right)^2 + \partial^lv^{\,ij} \partial_l
v_{ij} \Bigr] +  v^{\,ij}  \partial_i v  \partial_j v 
  + \cdots \right\} \,,
\eea
where the ellipses involve numerous other interactions, all of which either do not involve the scalar mode $v$, or are suppressed by more slow-roll parameters or involve more time derivatives than the ones explicitly shown. Further terms involving quartic and higher powers of $v$ and $v_{ij}$ are also present, suppressed by at least two powers of $1/M_p$, and so on. 

From here the goal is to split the fields into system and environment components, where the system consists of those modes whose wavelengths are visible to observers in the late universe. We therefore follow \cite{GravDecohere} and split the fields $v$ and $v_{ij}$ as in \pref{dSphiExpSys}, with the environment/system split occuring at a scale $k_*$ chosen as the shortest modes currently accessible in late-time cosmology. We compute how these modes evolve while outside the Hubble scale during the tail end of inflation, focussing on how the system modes are decohered by the shorter-wavelength environment. This involves tracking the evolution of the off-diagonal components $\langle \varphi_1 | \varrho | \varphi_2 \rangle$ of the reduced density matrix rather than the diagonal ones $P(\varphi) = \langle \varphi | \varrho | \varphi \rangle$ whose evolution is relevant to the validity of Stochastic Inflation discussed above.

Inspection of the general evolution equation \pref{NakaZwanExplicit} shows that decoherence first arises at second order in the interaction that couples system to environment, and at lowest order in $1/M_p$ the relevant interaction are cubic in the fields, with the fields split into system and environment parts: $v = v_{\rm sys} + v_{\rm env}$. Because only primordial scalar fluctuations have been observed we focus on how these decohere and so can ignore those cubic interactions not involving $v$. Interactions involving more than the minimal slow-roll suppression can also be dropped as subdominant. Finally, the freezing of super-Hubble modes implies that time-derivatives can also be dropped relative to spatial derivatives in all interactions provided we focus on super-Hubble environmental modes. What remains are then only the interactions given by \pref{allgravitonscalarcubics}. Not all of these interactions are even required because derivatives acting on system fields are always suppressed relative to their shorter-wavelength cousins in the environment. Momentum conservation also precludes having only one environment field since one large momentum cannot sum with two small ones to give zero. 

In the end the only interactions that matter to leading order are the first two appearing in  \pref{allgravitonscalarcubics}, where the differentiated fields are environmental modes and the undifferentiated fields belong to the system. Of these, the first interaction mediates decoherence of long-wavelength scalar fluctuations by the short-wavelength scalar environment and the second interaction describes decoherence of long-wavelength scalar fluctuations by the short-wavelength tensor environment. The interaction Hamiltonian to be used in \pref{NakaZwanExplicit} therefore becomes
\begin{equation}
\label{HintRdef}
      {\cH}_{\mathrm{int}}(\eta) =  G(\eta) \int\exd^3x\;
      {v}_{\rm sys} (\eta, \bfx) \otimes \Bigl[ B^{\ssS}(\eta,\bfx) +  B^{\ssT}(\eta,\bfx) \Bigr] \,,
\end{equation}
where the effective coupling and the scalar and tensor environmental interaction operators are
\begin{equation}
\label{eq:B:def}
G(\eta) := - \frac{\sqrt{\slrl }}{2\sqrt{2}\;  \Mp \, a  }
\,, \quad
      B^{\ssS}   : =  \partial^{i} {v}_{\rm env} \partial_{i} {v}_{\rm env} 
      \quad \hbox{and} \quad
      B^{\ssT}   : =  \partial^{l} {v}^{ij} \partial_{l} {v}_{ij} \,,
\end{equation}

Because this interaction is linear in $v_{\rm sys}$ its use at second order in \pref{NakaZwanExplicit} gives an evolution equation for the system state that is at most quadratic in $v_{\rm sys}$. Together with momentum conservation this implies that the field state for each mode $\bfk$ evolves independent of the others. Starting in the remote past with each mode uncorrelated (as is true in particular for the Bunch-Davies state) then ensures they remain uncorrelated and allows the system's reduced density matrix to be written
\be \label{prodvac}
    \varrho_{\rm sys} (\eta) = \bigotimes_{k < k_*} \varrho_\bfk(\eta) \,,
\ee
with each $\rho_\bfk(\eta)$ evolving independently. 

Using \pref{HintRdef} and \pref{prodvac} in \pref{NakaZwanExplicit} reveals that the environmental correlation relevant to primordial scalar fluctuations is $C_{\rm env}(\eta,\eta'; \bfy) = C^\ssS(\eta, \eta' ; \bfy) + C^\ssT(\eta, \eta' ; \bfy)$ where
\be 
\label{2pt_text}
C^{\ssS}(\eta,\eta' ; \bfx - \bfx')  :=   \left\langle 
\left[ {B}^{\ssS}(\eta,\bfx) - \mathfrak{B}^{\ssS}(\eta) \right]
\left[ {B}^{\ssS}(\eta',\bfx') - \mathfrak{B}^{\ssS}(\eta')  \right]   \right\rangle \,,
\ee
with $\mathfrak{B}^{\ssS}(\eta) := \langle {B}^{\ssS}(\eta,\bfx)  \rangle$. The result for $C^{\ssT}$ is identical with the replacement ${B}^{\ssS} \to {B}^{\ssT}$. In these expressions expectation values for all environmental modes are taken in the Bunch-Davies state. The combination that controls the evolution of the state $\rho_\bfk(\eta)$ is then $C_\bfk(\eta,\eta')$ where
\begin{equation}
\label{CR_FT}
C_{\rm env}(\eta, \eta' ; \bfy)   = \int \frac{\exd^{3} k}{(2\pi)^{3/2}}  \;
 {C}_{\bfk}(\eta,\eta')   \, e^{i \bfk \cdot \bfy } \,.
\end{equation}
These correlation functions are evaluated explicitly in \cite{GravDecohere} where it is also shown that they are peaked in a way that allows approximate Markovian evolution in the super-Hubble regime $|k\eta| \ll 1$. 

The Markovian evolution equation to which one is led in this way is
\bea 
\label{Lindblad_1}
\frac{\mathcal{V}}{(2\pi)^3}
\frac{\partial  {\varrho}_{\bfk}}{\partial \eta}
&\simeq&   - \, \mathrm{Re}[\mathfrak{F}_{\bfk}(\eta,\eta_{\mathrm{in}})]
\, \Bigl[  v_{\bfk}(\eta) , \big[  v_{\bfk}(\eta),
{\varrho}_{\bfk}(\eta) \big] \Bigr] \\
&& \qquad\qquad\qquad
- i \, \mathrm{Im}[\mathfrak{F}_{\bfk}(\eta,\eta_{\mathrm{in}})]
\, \left[ \left[v_{\bfk}(\eta) \right]^2 ,
{\varrho}_{\bfk}(\eta) \right]  \ ,\nn
\eea
where $\mathcal{V}$ denotes the volume of space and enters due to the way we normalize momentum modes. The coefficient function is defined by 
\be 
\label{Ffrakdef}
\mathfrak{F}_{\bfk}(\eta,\eta_{\mathrm{in}})  : =
(2\pi)^{3/2} \int_{ \eta_{\rm in} }^{\eta} \exd \eta' \,
G(\eta) G(\eta')  {C}_{\bfk}(\eta,\eta') \,.
\ee
Explicit expressions for $\mathfrak{F}_\bfk$ are given in \cite{GravDecohere}, which shows in particular that Re $\mathfrak{F}_\bfk$ is UV finite. UV divergences do appear in Im $\mathfrak{F}_\bfk$ and do so in a way that can be renormalized into parameters in the effective lagrangian in the usual EFT way (as reviewed, for example, in \cite{Burgess:2003jk}). Because the second line of \pref{Lindblad_1} describes Liouville evolution it cannot contribute to decoherence, which therefore depends only on Re $\mathfrak{F}_\bfk$ and as a consequence is UV finite. 

In the super-Hubble regime ($k\eta \to 0$) Re $\mathfrak{F}_\bfk$ is given by
\begin{equation}
\label{ReFSHresult}
\mathrm{Re}\; \mathfrak{F}_{\bfk}(\eta, \eta_{\mathrm{in}})  
\simeq  \frac{3\slrl  H^2 k^2}{1024 \pi^2 \Mp^2 }
\left\{ \frac{20 \pi}{( - k \eta)^2}
+\frac{g\left(k_*, - k \eta_{\mathrm{in}} \right)}
{( - k \eta)} + \mathcal{O}\left[ (- k \eta)^0 \right]\right\} \,.
\end{equation}
where the overall factor of 3 arises as 2 + 1 where the 2 comes from the tensor environment $C^\ssT$ and the 1 from the scalar environment $C^\ssS$. This form is universal in the sense that all details like the precise position $k_*$ of the system/environment split and the initial time $\eta_{\rm in}$ where system and environment start off uncorrelated appear only in subdominant terms, such as the known function $g(k_*,-k\eta_{\rm in})$ in \pref{ReFSHresult}. Notice that the universal leading term grows strongly at late times.  

Eq.~\pref{Lindblad_1} can be solved explicitly \cite{GravDecohere} and because its right-hand side is quadratic in $v_{\rm sys}$ it returns a gaussian state whose time evolution can be solved in great detail. We confine ourselves to exploring one consequence of \pref{Lindblad_1}: its implications for the `purity' of the observed system, defined by
\begin{equation}
\label{purity}
\mfp_{\bfk}(\eta) :=
\mathrm{Tr}\left[ {\varrho}^2_{ \bfk} (\eta)\right] =: \frac{1}{\sqrt{1 + \Xi_\bfk(\eta)}} \,.
\end{equation} 
Purity is measure of the state's decoherence because it satisfies $0 \leq \mfp_\bfk \leq 1$, with $\mfp_\bfp  = 1$ if and only if $\varrho_\bfk$ is a pure state and so $\varrho_\bfk$ is also pure if and only if $\Xi_\bfk =0$. Decoherence is said to be effective when $\mfp_{\bfk} \ll 1$. Eq.~\pref{Lindblad_1} implies 
\be 
\label{eq:purity:exact}
\Xi_\bfk(\eta) = 8 \int_{\eta_{\rm in}}^\eta \exd\eta'\; 
\mathrm{Re}[\mathfrak{F}_{\bfk}(\eta',\eta_{\mathrm{in}})]P_{vv}(k,\eta') \,,
\ee 
where $P_{vv}$ is the power spectrum, given for the Bunch-Davies vacuum by $|u_\bfk(\eta)|^2$ with mode functions as given in \pref{BDmodes}. See \cite{GravDecohere} for more details. 

The strong growth of $\mathfrak{F}_\bfk(\eta',\eta_{\rm in})$ for $k \eta' \to 0$ implies the integral is dominated by the super-Hubble limit $-k\eta \leq - k\eta_{\mathrm{in}}\ll 1$, where the universal form seen in \pref{ReFSHresult} applies. Using this leads to the late-time prediction 
\be 
\label{eq:purity:exactr}
\Xi_\bfk(\eta) \simeq    \frac{5 \slrl  }{64 \pi^2} \bigg( \frac{ H^2 }{ \Mp^2 } \bigg) \frac{1}{(-k\eta)^3}  =   \frac{5 \slrl  }{64 \pi^2} \left( \frac{ H^2 }{ \Mp^2 } \right) \left( \frac{aH}{k} \right)^3  \,.
\ee
The dependence on $\slrl$ and $H/M_p$ found here follows directly from the couplings appearing in the underlying interaction \pref{allgravitonscalarcubics}, and if linearized in $\Xi_\bfk$ agrees (up to normalization) with the perturbative result found in \cite{Nelson:2016kjm}. But linearization of \pref{eq:purity:exactr} is not required because in the small $k\eta$ limit the universal form of $\mathfrak{F}_\bfk$ implies it is independent of $\eta_{\rm in}$, and this allows \pref{Lindblad_1} to be used to resum late time behaviour. So although use of perturbative methods requires $\slrl$ and $H/M_p$ to be small, the solutions to the evolution equation \pref{Lindblad_1} can be trusted even for times late enough that  $\Xi_\bfk$ is not small because the factor of $a^3$ is large enough to compensate for the small perturbative prefactors.  

Notice that because these arguments explicitly use proximity to de Sitter (by perturbing in $\slrl$) the prediction \pref{eq:purity:exactr} applies at the end of inflation, and not at the much later epoch when the observed modes re-enter the Hubble scale and become observed. At this writing it is an open question how the purity evolves during the post-inflationary universe, but it is intuitive that a very classical state is not expected to be recohered by the later evolution of the universe.

\section{Black holes}

We finally consider preliminary applications of Open EFT techniques to black holes, whose static properties are captured (for non-rotating black holes) for many purposes by the metric 
\begin{equation}
\exd s^2 = - f(r) \, \exd t^2 + \frac{ \exd r^2}{f(r)} + r^2 \exd\theta^2 + r^2 \sin^2 \theta \exd \varphi^2  \,,
\end{equation}
where $f(r) = 1 - r_s/r$ for the simplest case: a Schwarzschild black hole with Schwarzschild radius $r_{s} = 2 G M$. This metric has the same form as in \pref{static}, with $\gamma_{ij} = \mathrm{diag}[f^{-1},r^2 ,r^2 \sin^2\theta ]$. 

The difficulty doing explicit calculations for black hole backgrounds means much less is known about open-system behaviour for these geometries. We therefore content ourselves to briefly describing some simple examples -- such as qubit evolution -- for which explicit calculations can in some circumstances be done. We also explore the properties of a toy model of a black hole for which the observed system is a field, using the technique of influence functionals. 

\subsection{Qubit thermalization}

The qubit set-up proceeds similar to previous sections and we pick the qubit to hover at a fixed position in space (in Schwarzschild coordinates), so
\begin{equation} \label{BHtrajectory}
y^{\mu}(\tau) = \Bigl[ t(\tau), r(\tau), \theta(\tau), \varphi(\tau) \Bigr] = \left[ \frac{\tau}{\sqrt{1 - {r_s}/{r_0}}}, r_0, \theta_0, \varphi_0 \right] 
\end{equation}
where $\tau$ gives proper time along the curve and $r_0,\theta_0,\varphi_0$ are all constants. This trajectory is {\it not} a geodesic and so the qubit shares many properties with the uniformly accelerated Rindler example considered previously. Gravitational redshift implies the energy splitting seen at infinity between the two qubit levels is
\be
  \omega_\infty = \omega \sqrt{1 - \frac{r_s}{r_0}} \,,
\ee    
where $\omega$ is the splitting in the qubit's rest frame at the qubit's position.

We take our quantum field environment again to be a massless real scalar field, with action as in \pref{lambdaphi4} but with $\lambda = 0$ since we ignore field self-interactions. There are a variety of `vacuum' states in which such a field could be prepared, including the Boulware \cite{Boulware}, Hartle-Hawking \cite{HartleHawking} or Unruh \cite{Unruh:1976db} vacua, and in principle qubit response requires calculating the autocorrelation function $W(x,x') = \langle \Omega | \phi(x) \phi(x') |\Omega \rangle$ for two points along the qubit trajectory using the state of interest. 

This evaluation is particularly simple if the invariant separation between the two field points is sufficiently small that the geodesic distance, $s(x,x')$, between them is much smaller than the local curvature scale, in which case it is dominated by the universal {\it Hadamard} form \cite{Hadamard1,Hadamard2,Hadamard3} that dominates the coincident limit:
\begin{equation} \label{WightmanHad}
\langle \Omega|\phi(x) \phi(x')| \Omega\rangle \simeq \frac{1}{8 \pi^2 \sigma(x,x') + i \varepsilon [ \mathcal{T}(x) - \mathcal{T}(x') ]} \qquad ( x \to x' ) \ ,
\end{equation}
where $\sigma(x,x') = \frac12 s^2(x,x')$ is the so-called Synge world function and $\mathcal{T}$ is any future-increasing function of time which gets multiplied by the regulator $\varepsilon$ so that the singularity structure of (\ref{WightmanHad}) is that of the Wightman function. This limit applies to any state of Hadamard form and reflects the intuitive fact that physical vacuum states on curved spacetimes should be indistinguishable from their flat space counterparts so long as one probes wavelengths much shorter than the local radius of curvature. The Hartle-Hawking and Unruh vacua in particular are Hadamard states. 
 
Evaluating \pref{WightmanHad} along the trajectory \pref{BHtrajectory} for the specific case of a Schwarzschild geometry leads to the expression for $W(\tau) = \langle \Omega | \phi[  y (\tau) ] \phi[ y (0) ] |\Omega \rangle$:
\be  \label{Wightman_BH_traj}
  W(\tau)  \simeq - \, \frac{1}{64 \pi^2 r_{s}^2 \left( 1 - \frac{r_s}{r_0} \right) \left[ \sinh\left(\frac{t(\tau)}{4 r_s} \right) - i \varepsilon \right]^2}\,,
\ee
which holds in the regime $\big|\sigma\big[ y(\tau), y(0)\big]\big| \ll r_{s}^2$, or equivalently
\begin{equation} \label{BH_Condition}
 \left( 1 - \frac{r_s}{r_0} \right) \sinh^2\left[\frac{t(\tau)}{4 r_s} \right]   \ll 1 \,.
\end{equation}
Here $t(\tau) = \tau (1-r_s/r_0)^{-1/2}$ denotes the Schwarzschild time as measured along the qubit trajectory \pref{BHtrajectory}.

Comparing \pref{Wightman_BH_traj} with (\ref{WightmanDefMink}) shows that $W(\tau)$ falls off in a suggestively thermal way so one might hope to deploy the Open EFT formalism to capture late-time Markovian behaviour over time-scales much longer than the falloff time. The trick is to do so while remaining within the regime (\ref{BH_Condition}) for which \pref{Wightman_BH_traj} is valid. Happily $|\sigma(x,x')| \ll r_s^2$ can be consistent with late times $t(\tau) \gg r_{s}$ provided $r_0$ is chosen close enough to the horizon to ensure \cite{Kaplanek:2020iay}
\be 
 1 \ll  \frac{t(\tau)}{r_{s}} \ll \left| 2 \log\left( \frac{1 - {r_s}/{r_0}}{4} \right) \right| \,.  
\ee
Having control of the approximation (\ref{Wightman_BH_traj}) for $W(\tau)$ in this manner, one proceeds as for the uniformly accelerated qubit of \S\ref{RindlerQubit}. The same arguments as given above show that Markovian evolution is valid in the limit where
\begin{equation}
1 \gg  4 \pi r_s \omega_{\infty} \gg \frac{g^2}{2\pi} \,,
\end{equation}
and when this is true the resummed Markovian evolution at late times describes qubit thermalization to its local Hawking temperature 
\be
   T_\ssH(r_0) = \frac{(4 \pi r_s)^{-1}}{\sqrt{1 - r_s/r_0}} \,,
\ee
for which $\beta(r_0) \omega = 4\pi r_s \omega \sqrt{1-r_s/r_0} = 4\pi r_s \omega_\infty$. 

The thermalization time-scales as seen by the observer at infinity are found to be
\begin{equation}
\xi_{c\,\infty} = 2 \xi_{d\,\infty} = \frac{4 \pi \tanh( 2 \pi r_{s} \omega_{\infty} )}{ g^2 \omega_{\infty} }   \simeq \frac{8 \pi^2 r_{s}}{g^2}
\end{equation}
which correspond to the blue-shifted time-scales $\xi_{c,d}(r_0) = \xi_{c,d\,\infty} \sqrt{1-r_s/r_0}$ in the qubit's frame. Because all scales redshift in the same way we have $\xi_\infty T_\ssH = \xi(r_0) \, T_\ssH(r_0)$ and so the same hierarchy of time-scales seen by the qubit is also seen at infinity.

\subsection{Hotspots and Influence Functionals}
\label{ssec:Hotspot}

A hindrance when studying interacting quantum fields near black holes is the relative lack of explicit calculations that include self-interactions in addition to the interaction with the classical gravitational field (see however {\it e.g.}~\cite{Akhmedov:2015xwa,Emelyanov:2016tws}). Although qubit calculations in a black hole background can be informative, they are also exceedingly simple and so might not capture features that arise when more complicated systems involving fields are observed. It is for these more complicated systems that influence functionals also come into their own. 

In this section we illustrate the use of many of the tools described above in a slightly more complicated field-theoretical system. Because we do not yet have good black hole examples to hand we instead illustrate their use using a simpler solvable system, called a `hotspot'  \cite{Kaplanek:2021sbo,Kaplanek:2021fnl,Burgess:2021luo}, that captures some of the features of a localized thermal source.

\subsubsection{Hotspots: small hot sources}
\label{sssec:HotspotToyBH}

A hotspot is a simple field-theoretic system for which environment degrees of freedom are a collection of $N$ massless scalar fields $\chi^a$ localized to an infinite spatial region $\cR_\ssB$ of spacetime and prepared in a thermal state. This is meant to emulate a black hole whose interior is hidden behind an event horizon. $\cR_\ssB$ is taken to be infinite so as to ensure that the fields do not have gaps in the spacing of their particle states. The observable sector representing the exterior of a black hole is a single massless real scalar field $\phi$ living in a different spatial region $\mathcal{R}_{\ssA}$ that is also taken to be infinitely large. This field can be prepared in any convenient state for the purposes of study, but we choose here the free-field vacuum state.

\begin{figure}[t]
\begin{center}
\includegraphics[width=30mm,height=30mm]{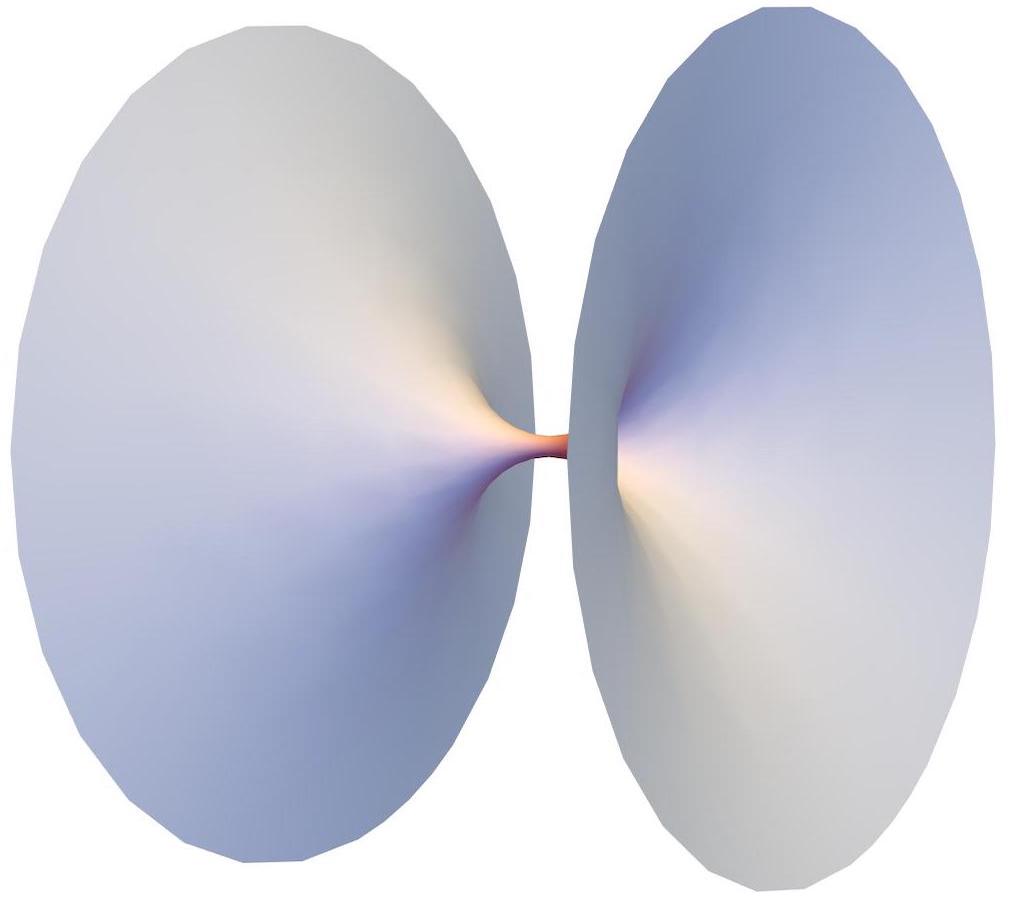} 
\caption{A cartoon of the two spatial branches, $\cR_\ssA$ and $\cR_\ssB$, in which the system field $\phi$ and the $N$ environmental fields $\chi^a$ repsectively live. Mixing between these fields occurs only in the localized throat region, which can be taken to be a small sphere of radius $r_h$ (or effectively a point in the limit that $r_h$ is much smaller than all other scales of interest).} \label{fig:FunnelFig} 
\end{center}
\end{figure}

The two kinds of fields interact with one another only on the intersection of the regions, $\cR_\ssA \cap \cR_\ssB$, which is taken to be the surface of a sphere of radius $r_h$ (meant as a proxy for the event horizon, see Fig.~\ref{fig:FunnelFig}). The system is solvable if the interactions are limited to a Caldeira-Leggett style bilinear mixing of these two kinds of fields \cite{Caldeira:1981rx}, since in this case the entire problem remains gaussian. It is often of interest to focus on the far-field case where $r_h \to 0$ corresponding to measurements that are performed in the observable sector at distances much larger than $r_h$. 

The free actions for these two fields are taken to be
\begin{equation}\label{freeSAB}
S_{\ssA0}[\phi] := - \frac{1}{2} \int_{\mathcal{R}_{\ssA}} \exd^4 x \; \partial_\mu \phi \partial^\mu \phi \quad \mathrm{and} \quad S_{\ssB}[\chi] := - \frac{1}{2} \int_{\mathcal{R}_{\ssB}} \exd^4 x \; \delta_{ab} \partial_\mu\chi^a  \partial^\mu\chi^b \,.
\end{equation}
Although the gravitational field of the environment can be included by assigning a Schwarzschild geometry to region $\mathcal{R}_{\ssA}$ (with $r_h > 2GM$ unless we want this to actually be a black hole), we here take both $\cR_\ssA$ and $\mathcal{R}_{\ssB}$ to be flat for simplicity. While this does {\it not} mimic the perfect infall near a black hole horizon, it does capture a localized interaction with a thermal object, and so can act as a benchmark against which quantum black hole calculations can be compared.

In the limit $r_h \to 0$ the Caldeira-Leggett style interaction between the two sectors is given by a mixing term of the form
\begin{equation} \label{hotspot_int}
S_{\mathrm{int}} = -   g_{a}\int \exd t \;  \chi^a(t,\mathbf{0}) \phi(t,\mathbf{0})  
\end{equation}
so that the fields interact only at a single hotspot point $\mathbf{x}=\mathbf{0}$. There is an implied sum over the $N$ couplings $g_{a}$ in (\ref{hotspot_int}), and it is assumed that the couplings are of the same size: $g_{a} = {\tilde{g}}/{\sqrt{N}}$ for all $a$ (with the factor $\sqrt{N}$ extracted for later convenience). 

This interaction also implies the existence of another localized self-interaction,
\begin{equation} \label{ct_int}
S_{\mathrm{ct}} = - \frac{\lambda}{2}  \int \exd t \;  \phi^2(t,\mathbf{0})   \,,
\end{equation}
which must be present in order to absorb some of the UV divergences that \pref{hotspot_int} generates. The treatment of these divergences associated with couplings to localized sources we handle using the general formalism developed in \cite{PPEFT}. Because $S_{\rm ct}$ does not couple sectors $A$ and $B$ it is often convenient to combine it with $S_{\ssA0}$ from \pref{freeSAB} and write
\be \label{SAForm}
   S_\ssA = S_{\ssA0} + S_{\rm ct} = -   \frac{1}{2} \int_{\mathcal{R}_{\ssA}} \exd^4 x  \Bigl[  \partial_\mu \phi \partial^\mu \phi  +  \lambda \, \delta^3(\bfx)\,  \phi^2(t,\mathbf{0}) \Bigr] \,.
\ee

The initial conditions at $t=0$ for the full density matrix are assumed to be uncorrelated -- {\it c.f.}~eq.~(\ref{intial_uncorr}):
\begin{equation} \label{factor_hotspot}
\rho_0 = \rho(t=0) = \rho_{\ssA0} \otimes \rho_{\ssB}
\end{equation}
with $\phi$ prepared in its vacuum $\rho_{\ssA0} \ = \ | \Omega \rangle \langle \Omega |$ and the environment fields $\chi^a$ prepared in a thermal state at temperature $1/\beta$:
\begin{equation}
\rho_{\ssB} = \frac{e^{- \beta H_{\ssB}}}{\mathrm{Tr}[ e^{- \beta H_{\ssB}}  ]} \quad \mathrm{with} \quad H_{\ssB} = \frac{1}{2} \int_{\mathcal{R}_{\ssB}}\exd^3x \;  \delta_{ab} \left[  \partial_t \chi^a \partial_t \chi^b  + \nabla \chi^a \cdot \nabla \chi^b   \right] \,.
\end{equation}
With this choice $\langle \varphi_1|\rho_{\ssA0}|\varphi_2 \rangle = \Psi[\varphi_1] \Psi^*[\varphi_2]$ where $\Psi[\varphi] := \langle \varphi | \Omega \rangle$ is the wave-functional of the vacuum. 

In general, interactions introduce correlations between $\phi$ and $\chi^a$ spoiling the factorized form of (\ref{factor_hotspot}) and for this reason we assume that the interaction $S_{\mathrm{int}}$ is suddenly turned on at $t=0$ --- this choice allows one to prepare the initially uncorrelated state and then observe how the combined system reacts to the onset of couplings (sometimes called a quench). One can think of this sudden approximation as a cartoon of the formation of a black hole, with outgoing transient waves radiating out from the spacetime event where the hotspot is formed.

Because the model is Gaussian the 2-point correlators contain all of the information. In the $N \to \infty$ limit back-reaction of the external field $\phi$ onto the environment can be neglected, so the environment fields satisfy\footnote{Time and spatial translation symmetry of the thermal bath is used here. Note also that the $i \epsilon$-prescription used here only works for real $t$.} 
%
%
%
\begin{equation}
\mathrm{Tr}_{\ssB}\left[ \chi^{a}(t,\mathbf{x}) \chi^b(0,\mathbf{0}) \rho_{\ssB} \right] = \delta^{ab}\frac{ \coth\left[ \frac{ \pi}{\beta} ( t + |\mathbf{x}| - i \epsilon)\right] - \coth\left[ \frac{\pi}{\beta}( t - |\mathbf{x}| - i \epsilon) \right]}{8 \pi \beta |\mathbf{x}|} \,.
\end{equation}
Of particular importance later on is the special case
\begin{equation} \label{Wightman_hotspot}
\tilde{g}^2 \mathcal{W}_{\ssB}(t) :=  g_{a} g_{b} \mathrm{Tr}_{\ssB}\left[ \chi^{a}(t,\mathbf{0}) \chi^b(0,\mathbf{0}) \rho_{\ssB} \right] = - \frac{\tilde{g}^2}{4 \beta^2 \sinh^2\left[ \frac{\pi }{\beta} ( t - i \epsilon) \right]}
\end{equation}
since this appears explicitly in the influence functional. 

More interesting is the response of the external field $\phi$, which feels the effects of the environment even as $N \to \infty$. Because the system is gaussian the $\phi$ response function can be computed exactly as a function of $\tilde g$, $\lambda$ and spacetime position, with the result given explicitly in \cite{Kaplanek:2021sbo}. It suffices here to quote the result in the equal-time limit expanded out to order $\tilde g^2$, for which (after a renormalization of the self-coupling $\lambda$) the result is
\bea \label{hotspot_corr}
&\ & \mathrm{Tr}_{\ssB}\left[ \phi_{\ssH}(t,\mathbf{x}) \phi_{\ssH}(t,\mathbf{x}') \rho_{\ssA} \right]  \simeq \frac{1}{4\pi^2 \big[ - (t - t' - i \epsilon)^2 + |\bfx - \bfx'|^2 \big]}  \\ 
&\ & \qquad + \frac{\lambda}{16\pi^3} \left[ \frac{\Theta(t-|\bfx|)}{|\bfx|} \frac{1}{(|\bfx| + i \epsilon)^2 - |\bfx'|^2} + \frac{\Theta(t-|\bfx'|)}{|\bfx'|} \frac{1}{(|\bfx'| - i \epsilon)^2 - |\bfx|^2} \right] \nn \\
& \ & \qquad\quad - \frac{\tilde{g}^2 \Theta(t-|\bfx|) \Theta(t - |\bfx'|)}{64 \pi^2 \beta^2 |\bfx| |\bfx'| \sinh^2 \left[ \frac{\pi}{\beta}( |\bfx| - |\bfx'| + i \epsilon ) \right]} \nn \\
&\ & \qquad\quad\quad + \frac{\tilde{g}^2}{32 \pi^4} \left[ \frac{\Theta(t-|\bfx|)}{\big[ (|\bfx| + i \epsilon)^2 - |\bfx'|^2 \big]^2} +  \frac{\Theta(t-|\bfx'|)}{\big[ (|\bfx'| - i \epsilon)^2 - |\bfx|^2 \big]^2} \right] \nn \\
&\ & \qquad\quad\quad\quad + \frac{\tilde{g}^2}{64\pi^4} \left[ \frac{\delta(t - |\bfx|)}{|\bfx| \big[ (t + i \epsilon)^2 + |\bfx'|^2 \big]} + \frac{\delta(t - |\bfx'|)}{|\bfx'| \big[ (t - i \epsilon)^2 + |\bfx|^2 \big]} \right] \,, \nn
\eea
The delta-function terms reveal an outgoing transient wave radiating out from the system's shock at $\mathbf{x} = t = 0$. The step functions show how the external system's properties change after this wave passes. Translation symmetry is clearly broken while invariance under rotations about the hotspot position is preserved. 

A  connection to the previous qubit calculations can also be made by coupling a qubit to the $\phi$-field alone at a nonzero distance from the hotspot. Doing so shows that this qubit thermalizes to the temperature of the hotspot in certain regimes of parameter space --- see \cite{Kaplanek:2021fnl} for further details.

\subsubsection{Hotspot influence functional}

We now take the hotspot model and use it to illustrate some of the features of the influence functional formalism, and how it overlaps with the other approaches encountered above. As in eq.~(\ref{FV_reduced}), the components of the reduced Schr\"odinger-picture density matrix for the $\phi$ field have the path-integral representation
\begin{equation} \label{hotspot_comp}
\langle \varphi_2 | \hat{\rho}_{\ssA}(t) | \varphi_1 \rangle = \sum_{\varphi_3 , \varphi_4} \int_{\varphi_4}^{\varphi_2} \mathcal{D}\phi^{+} \int_{\varphi_3}^{\varphi_1} \mathcal{D}\phi^{-} \; e^{i S_{\ssA}[\phi^+] - i S_{\ssA}[\phi^{-}]+ i S_{\ssI\ssF} [\phi^{+}, \phi^{-} ]}  \langle \varphi_4 | \rho_{\ssA0} | \varphi_3 \rangle 
\end{equation}
where the influence functional is given by (\ref{FeynmanVernon}), which in the present instance is
\begin{eqnarray} \label{SIF_chi}
e^{iS_{\ssI\ssF} [\phi^{+}, \phi^{-}]} & := &  \sum_{\chi, \chi_3, \chi_4} \int_{\chi_4}^{\chi} \mathcal{D}\chi^{+} \int_{\chi_3}^{\chi} \mathcal{D}\chi^{-} \;   \\
&\ & \qquad \times\;  e^{i S_{\ssB}[\chi^{+}] + i S_{\mathrm{int}}[\phi^+, \chi^{+} ] - i S_{\ssB}[\chi^{-}] - i S_{\mathrm{int}}[\phi^{-}, \chi^{-} ] } \; \langle \chi_4 | \rho_\ssB | \chi_3 \rangle  \ . \notag
\end{eqnarray}
We emphasize that the actions in (\ref{SIF_chi}) are integrated from time $t=0$ (where the initial conditions are applied) up to time $t$ (where we evaluate the reduced density matrix on the left-hand side), and so the boundary conditions in the path integrations (for the field eigenstates) are applied at these times as well. 

The leading influence functional contributions computed perturbatively in $\tilde g$ are
\begin{eqnarray} \label{IF_hospot_1}
S_{\ssI\ssF} [\phi^{+}, \phi^{-}] & \simeq & \frac{i}{2} \left\langle S^2_{\mathrm{int}}[\phi^+,\chi^+] \right\rangle_{\ssB} + \frac{i}{2}  \left\langle S^2_{\mathrm{int}}[\phi^-,\chi^-] \right\rangle_{\ssB} \\
& \ & \qquad  - i \left\langle S_{\mathrm{int}}[\phi^+,\chi^+] S_{\mathrm{int}}[\phi^+,\chi^+] \right\rangle_{\ssB}  + \ldots \notag
\end{eqnarray}
where we neglect terms $\mathcal{O}(\tilde g^3)$ and use the notation
\begin{equation}
\langle O \rangle_{\ssB} := \sum_{\chi, \chi_3, \chi_4} \int_{\chi_4}^{\chi} \mathcal{D}\chi^{+} \int_{\chi_3}^{\chi} \mathcal{D}\chi^{-} \; O \; e^{i S_{\ssB}[\chi^{+}] - i S_{\ssB}[\chi^{-}] } \; \langle \chi_4 | \rho_\ssB | \chi_3 \rangle 
\end{equation}
to denote averaging over environment fields. Notice that the average can be interpreted in terms of the Fig.~\ref{fig:Keldysh} (with $\phi^{\pm}$ replaced by $\chi^{\pm}$), where the boundary condition at time $t$ identifies $\chi^{+}(t,\mathbf{x}) = \chi^{-}(t,\mathbf{x})= \chi(\mathbf{x})$. This means that time integration can be thought of as being from zero to $t$ along the upper `$+$' branch, and then back again along the lower `$-$' branch. 

It remains to compute the three averages appearing in eq.~(\ref{IF_hospot_1}). Writing out the first term more explicitly using (\ref{hotspot_int}) one finds
\begin{eqnarray} \label{Sint2_av}
& \ & \left\langle S_{\mathrm{int}}[\phi^+,\chi^+]^2 \right\rangle_{\ssB} = g_a g_b \int_0^t \exd t' \int_0^t \exd t'' \; \phi^{+}(t',\mathbf{0}) \; \phi^{+}(t'',\mathbf{0}) \\
&& \times \sum_{\chi, \chi_3, \chi_4} \int_{\chi_4}^{\chi} \mathcal{D}\chi^{+} \int_{\chi_3}^{\chi} \mathcal{D}\chi^{-} \; \chi^{+a}(t',\mathbf{0}) \chi^{+b}(t'',\mathbf{0})  \; e^{i S_{\ssB}[\chi^{+}] - i S_{\ssB}[\chi^{-}] } \; \langle \chi_4 | \rho_\ssB | \chi_3 \rangle \,. \notag
\end{eqnarray}
A standard path integration exercise \cite{Weinberg:1995mt} shows that the lower line of \pref{Sint2_av} is equivalent to the following environmental propagator in the state $\rho_\ssB$: 
\be  \label{first_average}
 \mathrm{Tr}_{\ssB}\left[ \mathcal{T}_+\left\{ \chi^{a}(t',\mathbf{0}) \chi^{b}(t'',\mathbf{0}) \right\} \rho_{\ssB} \right]   = \delta^{ab} \; \mathcal{F}_{\ssB}(t' - t'')
\ee
where $\mathcal{T}_+$ denotes time ordering of the field operators on the upper `$+$' branch of the closed time path depicted in Fig.~\ref{fig:Keldysh}, and $\mathcal{F}_{\ssB}$ is the Feynman propagator for a single field in the state $\rho_{\ssB}$. Note that $\mathcal{F}_{\ssB}$ is related to the  Wightman function $\mathcal{W}_{\ssB}$ -- given in eq.~(\ref{Wightman_hotspot}) -- in the standard way: 
\begin{eqnarray} \label{FeynmanWightman}
  \mathcal{F}_{\ssB}(t' - t'')   =   \mathcal{W}_{\ssB}(t' - t'') \Theta(t' - t'') +   \mathcal{W}_{\ssB}(t'' - t') \Theta(t'' - t') \,,
\end{eqnarray}
and (using $g_a g_b \delta^{ab} = \tilde g^2$) this allows (\ref{Sint2_av}) to be written
\begin{eqnarray}  \label{first_av_W}
 \left\langle S_{\mathrm{int}}[\phi^+,\chi^+]^2 \right\rangle_{\ssB} & = & \tilde{g}^2 \int_0^t \exd t' \int_0^t \exd t'' \; \phi^{+}(t',\mathbf{0}) \phi^{+}(t'',\mathbf{0}) \mathcal{F}_{\ssB}(t' - t'') \nn\\
 & = & 2 \tilde{g}^2 \int_0^t \exd t' \int_0^{t'} \exd t'' \; \phi^{+}(t',\mathbf{0}) \phi^{+}(t'',\mathbf{0}) \mathcal{W}_{\ssB}(t' - t'') \ .
\end{eqnarray}

A similar exercise shows the second average in (\ref{IF_hospot_1}) is given by
\begin{eqnarray} \label{second_ave}
\left\langle S_{\mathrm{int}}[\phi^-,\chi^-]^2 \right\rangle_{\ssB} & = & \tilde{g}^2 \int_0^t \exd t' \int_0^t \exd t'' \; \phi^{-}(t',\mathbf{0}) \phi^{-}(t'',\mathbf{0}) \mathcal{F}^{\ast}_{\ssB}(t' - t'')\nn\\ 
& = & 2 \tilde{g}^2 \int_0^t \exd t' \int_0^{t'} \exd t'' \; \phi^{-}(t',\mathbf{0}) \phi^{-}(t'',\mathbf{0}) \mathcal{W}^{\ast}_{\ssB}(t' - t'')\,,
\end{eqnarray}
which again uses \pref{FeynmanWightman} together with the property $ \mathcal{W}_{\ssB}(-t) =  \mathcal{W}_{\ssB}^{\ast}(t)$ that is always valid for Hermitian fields. These imply that the path integration here {\it anti}-time orders the two fields on the lower `$-$' branch (as might have been expected given the interpretation of the lower branch in Fig.~\ref{fig:Keldysh} as going backwards in time relative to the upper branch). 

The third term in (\ref{IF_hospot_1}) involves a field on both the upper and lower branch,  
\begin{eqnarray} \label{third_ave}
\left\langle S_{\mathrm{int}}[\phi^+,\chi^+]S_{\mathrm{int}}[\phi^-,\chi^-] \right\rangle_{\ssB} & = & \tilde{g}^2 \int_0^t \exd t' \int_0^t \exd t'' \; \phi^{+}(t',\mathbf{0}) \phi^{-}(t'',\mathbf{0}) \; \mathcal{W}^{\ast}_{\ssB}(t' - t'') \nn\\
& = & \tilde{g}^2 \int_0^t \exd t' \int_0^{t'} \exd t'' \; \bigg[ \phi^{-}(t',\mathbf{0}) \phi^{+}(t'',\mathbf{0}) \; \mathcal{W}_{\ssB}(t' - t'') \notag \\
&& \qquad + \phi^{+}(t',\mathbf{0}) \phi^{-}(t'',\mathbf{0}) \; \mathcal{W}^{\ast}_{\ssB}(t' - t'')  \bigg] 
\end{eqnarray}
and so  -- recalling that $\tilde{g}^2 \mathcal{W}^{\ast}_{\ssB}(t' - t'') = g_a g_b \mathrm{Tr}_{\ssB}[ \chi^{a}(t'',\mathbf{0}) \chi^{b}(t',\mathbf{0}) \rho_{\ssB} ]$ -- time-ordering is {\it not} enforced by the path integration. The integration is nevertheless `path-ordered', however, inasmuch as the field on the lower branch (with time $t''$) is to the left of the field on the upper branch (with time $t'$) in the correlator. The final equality arranges the integration range here to match the ones in \pref{first_av_W} and \pref{second_ave}.

Combining everything allows the $\cO(\tilde g^2)$ terms of the influence functional (\ref{IF_hospot_1}) to be explicitly written as
\begin{eqnarray} \label{SIF_answer}
S_{\ssI\ssF}[\phi^{+}, \phi^{-}] & \simeq & i \tilde{g}^2 \int_0^t \exd t' \int_0^{t'} \exd t'' \; \bigg[ \ \phi^{+}(t',\mathbf{0}) \phi^{+}(t'',\mathbf{0}) \; \mathcal{W}_{\ssB}(t' - t'')  \\
& \ & + \phi^{-}(t',\mathbf{0}) \phi^{-}(t'',\mathbf{0}) \; \mathcal{W}^{\ast}_{\ssB}(t' - t'') - \phi^{-}(t',\mathbf{0}) \phi^{+}(t'',\mathbf{0}) \; \mathcal{W}_{\ssB}(t' - t'') \notag \\
& \ & \qquad \qquad \qquad \ \ - \phi^{+}(t',\mathbf{0}) \phi^{-}(t'',\mathbf{0}) \; \mathcal{W}^{\ast}_{\ssB}(t' - t'')  \bigg] \,, \notag
\end{eqnarray}
with $\cW_\ssB(t)$ given explicitly for the thermal environment by \pref{Wightman_hotspot}. Anything that can be computed using the reduced density matrix can be computed given $S_{\ssI\ssF}$ through the connection \pref{hotspot_comp}. For instance $\phi$ equal-time correlation functions are straightforwardly evaluated and lead again to the result (\ref{hotspot_corr}).

In the remainder of this subsection we explore some of the other physical implications of the above influence functional. 

\subsubsection{Hotspot master equation}

Expression \pref{hotspot_comp} ensures that the influence functional encodes all of the information contained in the reduced density matrix. This includes the ability to derive a master equation like \pref{NakaZwanExplicit} for its evolution (which was also truncated at second order in the interactions).

To derive the master equation from the influence functional one evaluates
\begin{equation} \label{IF_master}
\frac{\partial \langle \varphi_{2} | \hat{\rho}_{\ssA}(t) | \varphi_{1} \rangle }{ \partial t } \simeq \frac{ \langle \varphi_{2} | \hat{\rho}_{\ssA}(t + \Delta t) | \varphi_{1} \rangle  - \langle \varphi_{2} | \hat{\rho}_{\ssA}(t) | \varphi_{1} \rangle }{ \Delta t } 
\end{equation}
for small enough $\Delta t$, using (\ref{hotspot_comp}) to compute the reduced density matrix elements. Expanding for $\Delta t \ll t$ yields
\begin{eqnarray}
 \partial_t \langle \varphi_{2} |  \hat{\rho}_{\ssA}(t) | \varphi_{1} \rangle& \simeq & \sum_{\varphi_3 , \varphi_4} \int_{\varphi_4}^{\varphi_2} \mathcal{D}\phi^{+} \int_{\varphi_3}^{\varphi_1} \mathcal{D}\phi^{-} \; i \partial_t \Bigl\{ S_{\ssA}[\phi^+] - S_{\ssA}[\phi^{-}]+ S_{\ssI\ssF} [\phi^{+}, \phi^{-} ] \Bigr\} \notag \\
&& \qquad\qquad \times e^{i S_{\ssA}[\phi^+] - i S_{\ssA}[\phi^{-}]+ i S_{\ssI\ssF} [\phi^{+}, \phi^{-} ]}  \langle \varphi_4 | \rho_{\ssA0} | \varphi_3 \rangle 
\end{eqnarray}
up to terms $\mathcal{O}(\Delta t)$. The two terms depending on $\partial_t S_{\ssA}$ can be re-expressed in operator language in terms of the Hamiltonian corresponding to $S_{\ssA}$, which given \pref{SAForm} is 
\begin{equation}
H_{\ssA} = \frac{1}{2} \int \exd^3 \mathbf{x}\; \bigg[ (\partial_t \phi )^2 + |\nabla \phi|^2 \bigg] \ + \frac{\lambda}{2} \phi^2(t,\mathbf{0}) \,,
\end{equation}
giving $\partial_t \langle \varphi_{2} | \hat{\rho}_{\ssA}(t) | \varphi_{1} \rangle   \simeq   - i  \langle \varphi_{2} | \bigl[ H_{\ssA} , \hat{\rho}_{\ssA} \bigr] | \varphi_1 \rangle + $ ($\partial_t S_\IF$ term).
%

To simplify the final term, we evaluate $\partial_t S_\IF$ to leading nontrivial order in $\tilde g$, using (\ref{SIF_answer}) to write 
\begin{eqnarray}
i \partial_t S_{\ssI\ssF} & = & - \tilde{g}^2 \int_0^{t} \exd s \; \left\{   \Bigl[ \phi^{+}(t,\mathbf{0}) \phi^{+}(s,\mathbf{0}) - \phi^{-}(t,\mathbf{0}) \phi^{+}(s,\mathbf{0}) \Bigr] \; \mathcal{W}_{\ssB}(t - s) \right. \\
& \ & \qquad \qquad \qquad \left. + \Bigl[ \phi^{-}(t,\mathbf{0}) \phi^{-}(s,\mathbf{0}) - \phi^{+}(t,\mathbf{0}) \phi^{-}(s,\mathbf{0}) \Bigr] \; \mathcal{W}^{\ast}_{\ssB}(t - s) \right\} \,. \notag
\end{eqnarray}
Using $\phi^{+}(t,\mathbf{0}) = \varphi_2(\mathbf{0})$ and $\phi^{-}(t,\mathbf{0}) = \varphi_1(\mathbf{0})$, this becomes
\begin{eqnarray} \label{masterIF_3}
  \partial_t \langle \varphi_{2} | \hat{\rho}_{\ssA}(t) | \varphi_{1} \rangle  & \simeq & - i  \langle \varphi_{2} | \bigl[ H_{\ssA} , \hat{\rho}_{\ssA} \bigr] | \varphi_1 \rangle  - \big[ \varphi_2 (\mathbf{0} ) - \varphi_1(\mathbf{0}) \big] \int_0^t \exd s\; \mathcal{W}_{\ssB}(t - s) F^{+}(t,s) \notag \\
 & \ & \qquad - \big[ \varphi_1(\mathbf{0} ) - \varphi_2(\mathbf{0}) \big] \int_0^t \exd s\; \mathcal{W}^{\ast}_{\ssB}(t - s) F^{-}(t,s) \,,
\end{eqnarray}
which introduces the shorthand
\begin{eqnarray} \label{Fpm_approx}
F^{\pm}(t,s) &:=& \tilde{g}^2 \sum_{\varphi_3 , \varphi_4} \int_{\varphi_4}^{\varphi_2} \mathcal{D}\phi^{+} \int_{\varphi_3}^{\varphi_1} \mathcal{D}\phi^{-} \; \phi^{\pm}(s,\mathbf{0}) \\
 && \qquad\qquad\qquad \times \,  e^{i S_{\ssA}[\phi^+] - i S_{\ssA}[\phi^{-}]+ i S_{\ssI\ssF} [\phi^{+}, \phi^{-} ]}  \langle \varphi_4 | \rho_{\ssA 0} | \varphi_3 \rangle  \,.\nn
\end{eqnarray}
This satisfies $F^{-}(t,s) = F^{+\ast}(t,s)$, as can be seen using the identity $S_\IF^*[\phi^+,\phi^-] = S_\IF[\phi^-,\phi^+]$ that follows from eq.~\pref{SIF_chi}. The function $F^\pm(t,s)$ has a simple operator interpretation if we drop terms beyond leading order in $\tilde{g}^2$, since then
\begin{eqnarray} \label{Fpm_approx2}
F^{\pm}(t,s)  &\simeq& \tilde{g}^2 \sum_{\varphi_3 , \varphi_4} \int_{\varphi_4}^{\varphi_2} \mathcal{D}\phi^{+} \int_{\varphi_3}^{\varphi_1} \mathcal{D}\phi^{-} \; \phi^{\pm}(s,\mathbf{0}) \, e^{i S_{\ssA}[\phi^+] - i S_{\ssA}[\phi^{-}]}  \langle \varphi_4 | \rho_{\ssA 0} | \varphi_3 \rangle \,,\nn\\
&=&  \tilde{g}^2 \langle \varphi_2 | e^{- i H_\ssA (t - s)} \hat{\phi}(\mathbf{0}) e^{- i H_\ssA s} \rho_{\ssA0} e^{+ i H_\ssA t} | \varphi_1 \rangle \,.
\end{eqnarray}

Using this in the above expressions we obtain the evolution equation,
\begin{eqnarray} \label{masterIF_4}
 \partial_t \hat{\rho}_{\ssA}(t)  & \simeq  & - i \Bigl[ H_{\ssA} , \hat{\rho}_{\ssA}(t) \Bigr]  \\
&\  & \quad - \tilde{g}^2 \int_0^t \exd s\; \left\{ \mathcal{W}_{\ssB}(t - s)\; \Bigl[ \hat{\phi}(\mathbf{0}) \,, \; e^{- i H_\ssA (t - s)} \hat{\phi}(\mathbf{0}) e^{- i H_\ssA s} \rho_{\ssA0} e^{+ i H_\ssA t} \Bigr] \ + \mathrm{h.c.} \right\} \notag
\end{eqnarray}
where we peel off the eigenstates to yield an operator equation, and where `h.c.'~means Hermitian conjugate of the previous terms in the curly bracket. This is more illuminating in the interaction picture, defined using $H_\ssA$ as the unperturbed Hamiltonian so that $\rho_{\ssA}(t) := e^{+ i H_\ssA t} \hat{\rho}_{\ssA}(t) e^{- i H_\ssA t}$, since the above becomes
\bea  \label{masterIF_5}
 \partial_t \rho_{\ssA}(t)  & \simeq  & -  \frac{i\lambda}{2} \; \Bigl[ \phi^2(t,\mathbf{0}) \,, \; \rho_{\ssA}(t) \Bigr]  \\
&& \qquad \qquad   - \tilde{g}^2 \int_0^t \exd s\; \left\{  \mathcal{W}_{\ssB}(t - s) \, \Bigl[ \phi(t, \mathbf{0}) \,, \phi(s, \mathbf{0}) \rho_{\ssA0} \Bigr] \ + \mathrm{h.c.} \right\} \ . \notag
\eea
This is the leading perturbative result since dropping additional powers of $\tilde g$ is what justifies dropping $S_\IF$ in going from \pref{Fpm_approx} to \pref{Fpm_approx2}. Notice that because $\rho_\ssA(s) = \rho_{\ssA0} + \cO(\tilde g^2)$ in the interaction picture, eq.~\pref{masterIF_5} agrees with the perturbative limit of the Nakajima-Zwanzig equation given in \pref{NakaZwanExplicit}, which in this example would be 
\bea \label{masterIF_6}
 \partial_t \rho_{\ssA}(t) & \simeq  & -  \frac{i\lambda}{2} \; \Bigl[ \phi^2(t,\mathbf{0}) \,, \; \rho_{\ssA}(t) \Bigr]  \\
&& \qquad \qquad - \tilde{g}^2 \int_0^t \exd s\; \left\{  \mathcal{W}_{\ssB}(t - s)\, \Bigl[ \phi(t, \mathbf{0}) \,, \phi(s, \mathbf{0}) \,\rho_{\ssA}(s) \Bigr] \ + \mathrm{h.c.} \right\} \,. \notag
\eea

It is fairly common practice in the literature to justify master equations like \pref{masterIF_6} by deriving \pref{masterIF_5} in perturbation theory and then simply replacing $\rho_{\ssA0} \to \rho_\ssA(s)$ with the justification that these agree at lowest order in $\tilde g^2$. Of course this argument is actually ambiguous, as we could have equivalently taken any number of other combinations of operators that agree with $\rho_{\ssA0}$ as $\tilde g \to 0$ by the same argument, and all of these choices can disagree\footnote{An example instead replaces $\rho_{\ssA0} \simeq \rho_{\ssA}(t)$ and so predicts an evolution equation -- the Redfield equation -- that is time-local (as opposed to involving convolutions with $\rho_\ssA(s)$ as in \pref{masterIF_6}).} on the predicted evolution beyond order $\tilde g^2$.  From this point of view the derivation that passes through the Nakajima-Zwanzig equation \pref{NakaZwanExplicit} is preferable, because this derives the higher-order dependence on couplings like $\tilde g$ by explicitly tracing out the environment order-by-order in $\tilde g$.  

For the hotspot these arguments allow a test of the validity of the Markovian limit of the Nakajima-Zwanzig equation. To this end one again notices that the correlaton function $\mathcal{W}_{\ssB}(t - s)$ given in \pref{Wightman_hotspot} is sharply peaked in time for an intervel set by the inverse temperature $\beta$. This allows the remainder of the integrand to be Taylor expanded about $s = t$ if the remainder of the integrand varies over time scales much longer than $\beta$. When this is true the leading contribution behaves as if $\mathcal{W}_{\ssB}(t - s)$ were proportional to $\delta(t-s)$ and the upper integration limit can be taken to infinity, resulting in the approximate Markovian evolution
\begin{eqnarray} \label{masterIF_Markovian}
 \partial_t \rho_{\ssA}(t)   & \simeq  & -  \frac{i\lambda}{2} \; \Bigl[ \phi^2(t,\mathbf{0}) \,, \; \rho_{\ssA}(t) \Bigr]  \\
&& \qquad \qquad - \tilde{g}^2 \int_0^\infty \exd s \; \left\{ \mathcal{W}_{\ssB}(s)\, \Bigl[ \phi(t, \mathbf{0}) \,, \phi(t, \mathbf{0}) \, \rho_{\ssA}(t) \Bigr] \ + \mathrm{h.c.} \right\} \,. \notag
\end{eqnarray}

In the present instance a sufficient condition for ensuring both $\phi(t, \mathbf{0})$ and $\rho_\ssA(t)$ vary slowly compared to $\beta$ is when the external field theory's UV cutoff $\Lambda$ obeys $\beta \Lambda \ll 1$. This makes the hotspot temperature a UV scale (which for black holes also would require the event horizon size $r_h$ to be negligibly small, as assumed above). In this limit integrating the Markovian master equation (\ref{masterIF_Markovian}) to compute the correlation functions for the system gives (after a tedious computation)
\bea
\mathrm{Tr}_{\ssB}\left[ \phi_{\ssH}(t,\mathbf{x}) \phi_{\ssH}(t,\mathbf{x}') \rho_{\ssA} \right] & \simeq & \frac{1}{4\pi^2 |\bfx - \bfx'|^2} - \frac{\lambda}{16\pi^3 |\mathbf{x}| |\mathbf{x}'| (|\mathbf{x}| + |\mathbf{x}'|)} \\
& \ & \qquad + \frac{\tilde{g}^2}{32 \pi^3 \beta |\bfx| |\bfx'|} \delta( |\mathbf{x}| - |\mathbf{x}' | ) + \frac{\tilde{g}^2}{16 \pi^4 ( |\mathbf{x}|^2 - |\mathbf{x}'|^2 )^2} \nn
\eea
which specializes to $t>|\mathbf{x}|,|\mathbf{x}'|$ (after transients of the sudden approximation have passed). This is precisely the $|\mathbf{x}|, |\mathbf{x}'| \ll \beta$ limit of the correlator given in (\ref{hotspot_corr}) -- see \cite{Burgess:2021luo} for more details.

\subsubsection{Langevin equations and Stochastic evolution}

Another use for the path-integral influence functional formulation is to derive a stochastic evolution equation for the fields in which the environment is reduced to stochastic noise variable appearing in the equations of motion for the observed field. This type of formulation is useful when computing the time evolution of correlation functions.  

To see how, relabel the two fields $\phi^\pm$ using
\begin{equation}
\phi^{\mathrm{cl}} :=  \phi^{+} + \phi^{-}   \qquad \mathrm{and}   \qquad \phi^{\mathrm{q}} :=  \phi^{+} - \phi^{-}  \,,
\end{equation}
The notation `cl' stands for classical and `q' stands for quantum, with the logic that $\phi^{\rm cl}$ emerges as a macroscopic mean field in a regime where $\phi^{\mathrm{q}}$ averages to zero.  Fluctuations of $\phi^{\rm q}$ about zero will be the source of the noise mentioned above with which $\phi^{\rm cl}$ interacts. 

In terms of these variables the unperturbed actions become
\begin{equation}
   S_\ssA[\phi^+] - S_\ssA[\phi^-] = - \frac12 \int \exd^4 x \left[ \partial_\mu \phi^{\rm q} \partial^\mu \phi^{\rm cl} + \lambda \, \delta^3(\bfx) \, \phi^{\rm q} \phi^{\rm cl} \right]\,,
\end{equation}
while eq.~(\ref{SIF_answer}) for $S_{\ssI\ssF}$ is
\begin{eqnarray} \label{SIF_answer_2}
S_{\ssI\ssF}[\phi^{+}, \phi^{-}] & \simeq & \frac{i \tilde{g}^2}{2}  \int_0^t \exd t' \int_0^{t} \exd t'' \; \phi^{\mathrm{q}}(t',\mathbf{0}) \, \mathrm{Re}\left[ \mathcal{W}_{\ssB}(t' - t'') \right] \phi^{\mathrm{q}}(t'',\mathbf{0}) \\
& \ & \qquad \qquad -  i \tilde{g}^2 \int_0^t \exd t' \int_0^{t'} \exd t'' \; \phi^{\mathrm{q}}(t',\mathbf{0}) \, \mathrm{Im}\left[ \mathcal{W}_{\ssB}(t' - t'') \right] \phi^{\mathrm{cl}}(t'',\mathbf{0}) \,. \notag
\end{eqnarray}
Notice that the integration limits of the first line of (\ref{SIF_answer_2}) are not time-ordered while those of the second line are. 

The only contribution within $S_\ssA[\phi^+] - S_\ssA[\phi^-] + S_\IF[\phi^+,\phi^-]$ that is quadratic in $\phi^{\rm q}$ comes from the first line of \pref{SIF_answer_2}. Furthermore the factor of $i$ ensures that $e^{i S_\IF}$ becomes a {\it real} gaussian exponential in $\phi^{\rm q}$, as would have appeared for a statistical average rather than a quantum one. This makes it suggestive to use the Hubbard-Stratonovich identity \cite{Calzetta:2008iqa}
\begin{equation}
e^{ - \frac12 \int \mathrm{d} t' \int \mathrm{d} t'' \; \phi^{\rm q}(t') N(t',t'') \phi^{\rm q}(t'')} = \int \mathcal{D} \nu  \; P[\nu] \; e^{ i \int \exd t'\; \nu(t') \phi^{\rm q}(t')} \ ,
\end{equation}
which expresses the left-hand side in terms of a Gaussian integral over a stochastic dummy field $\nu$ which is defined to have zero mean and correlation functions given by the kernel in the exponent on the left-hand side:
\begin{equation}
\left\langle \nu(t) \right\rangle_{\ssP} = 0 \qquad \mathrm{and} \qquad \left\langle \nu(t) \nu(t') \right\rangle_{\ssP}  = N(t,t')
\end{equation}
where the averages here denote
\begin{equation}
\left\langle O \right\rangle_{\ssP} \ := \ \int \mathcal{D}\nu \; P[\nu] \; O \ .
\end{equation}

Armed with this formula the influence functional (\ref{SIF_answer_2}) appearing in the path integral can be rewritten in terms of this stochastic noise field as
\begin{equation}
e^{i S_{\ssI\ssF}} = \left\langle \exp\left( i \int_0^t \exd t'\; \psi(t') \phi^{\mathrm{q}}(t',\mathbf{0}) \right) \right\rangle_{P}
\end{equation}
with the definitions
\begin{equation}
\psi(t) \ := \ \nu(t) - \tilde{g}^2 \int_{0}^{t} \exd t' \; \mathrm{Im}[\mathcal{W}_{\ssB}(t-t')] \phi^{\mathrm{cl}}(t',\mathbf{0}) \  ,
\end{equation}
and in the present case the noise kernel is
\be\label{NdefW}
  N(t,t') = \tilde{g}^2\; \mathrm{Re}\, \mathcal{W}_{\ssB}(t,t')   \,.
\ee

The point of this exercise is that $\phi^{\rm q}$ now appears only linearly in the argument $S_{\mathrm{eff}}[\phi^{+},\phi^{-}] := S_{\ssA}[\phi^{+}] - S_{\ssA}[\phi^{-}] +  S_{\ssI\ssF}[\phi^{+},\phi^{-}] $ of the exponential appearing in the path integral. Its integration therefore gives a functional delta function that can be used to perform the $\phi^{\rm cl}$ path integral, leading to the constraint 
\begin{eqnarray}
  &&0= \frac{\delta S_{\mathrm{eff}}}{\delta \phi^{\mathrm{q}} } = \Box \phi(x) \\
  && \qquad\qquad +  \delta^3(\mathbf{x})  \bigg[ -  \lambda \phi(t,\mathbf{0}) - \nu(t) + \tilde{g}^2 \int_0^t \exd s \; \mathrm{Im}[\mathcal{W}_{\ssB}(t-s)] \phi(s,\mathbf{0}) \bigg]    \,,\nn
\end{eqnarray}
where we rescale $\phi := \frac12 \, \phi^{\rm cl}$. 

The upshot is that correlation functions of $\phi = \frac12 \phi^{\rm cl}$ are to be computed as a stochastic average over $\nu(t)$ with correlator $N(t,t')$ given in terms of Re $\cW_\ssB(t,t')$ by \pref{NdefW} and with the spacetime dependence of $\phi$ determined by the {\it Langevin} equation 
\begin{equation}
\Box \phi(x) +  \delta^3(\mathbf{x})  \bigg[ -  \lambda \phi(t,\mathbf{0}) + \tilde{g}^2 \int_0^t \exd s \; \mathrm{Im}[\mathcal{W}_{\ssB}(t-s)] \phi(s,\mathbf{0}) \bigg]   =  \delta^3(\mathbf{x}) \nu(t) \ ,
\end{equation}
with the stochastic field $\nu$ appearing as a source.  Notice in the particular example of the hotspot we have $ \mathrm{Im}\;\mathcal{W}_{\ssB}(t-s) = \delta'(t-s)/4\pi$ and so the equation is really local in time,
\begin{equation} \label{DeltaExample}
\Box \phi(x) +  \delta^3(\mathbf{x})  \bigg[ \left( -  \lambda + \frac{\tilde{g}^2}{4\pi} \delta(0) \right) \phi(t,\mathbf{0}) + \tilde{g}^2 \partial_t \phi(t,\mathbf{0}) \bigg] \ =\ \delta^3(\mathbf{x}) \nu(t) \,.
\end{equation}

The $\delta(0)$ divergence appearing in \pref{DeltaExample} can be absorbed into a renormalization of $\lambda$. The $ \mathrm{Im}[\mathcal{W}_{\ssB}(t-t')]$ term is called the dissipation kernel because it ends up introducing single time derivatives into the equation of motion for $\phi$ --- what turns out to be a generic feature of influence functionals. It is the noise kernel $\mathrm{Re}\, \mathcal{W}_{\ssB}(t-t')$ that is responsible for the decoherence seen in earlier sections. 

As a final note, the Langevin equation encountered here provides the same information as does a Fokker-Planck equation -- like eq.~(\ref{FokkerPlanck}) encountered in \S\ref{sec:stoch} -- for the evolution of the probability distribution $P[\phi]$ that $\phi$ inherits from the distribution $P[\nu]$. Doing so brings us full circle and back to the density matrix since $P[\phi] = \langle \phi | \rho_\ssA | \phi \rangle$. For more details we refer the reader to the literature \cite{IFs,Starobinsky:1994bd,FokkerLangevin}.

\section{Summary}

Our focus in this chapter is on open quantum systems, which we argue carry important lessons for quantum studies in gravitational fields. We focus in particular on situations where hierarchies of scale simplify calculations for these systems -- what can be called Open EFTs -- as is appropriate in this section devoted to effective field theories. Such simplifications are widely used throughout physics, but their use for gravitating open quantum systems still remains young. 

The hope is that the gravity community can profit from the decades of study of quantum fields interacting with open systems, particularly for problems involving event horizons. The potential benefits are many and include an improved ability to understand evolution at the very late times that are both of great interest for problems like information loss, and are places where simpler traditional perturbative methods are known always to fail. Differences between Open EFTs and traditional Wilsonian ones -- such as the appearance of nonlocality in time for open systems -- might yet bring surprises.

\section*{Acknowledgements}

We thank Thomas Colas, Archie Cable, Rich Holman, Jerome Martin and Vincent Vennin for helpful conversations. CB's research was partially supported by funds from the Natural Sciences and Engineering Research Council (NSERC) of Canada. Research at the Perimeter Institute is supported in part by the Government of Canada through NSERC and by the Province of Ontario through MRI. GK is supported by
the Simons Foundation award ID 555326 under the Simons Foundation Origins of the Universe
initiative, Cosmology Beyond Einstein's Theory as well as by the European Union Horizon 2020
Research Council grant 724659 MassiveCosmo ERC2016COG.


\begin{thebibliography}{99}




\bibitem{EFTBook}
C.P.~Burgess,
``Introduction to Effective Field Theory,''
Cambridge University Press, 2020,
ISBN 978-1-139-04804-0, 978-0-521-19547-8.

\bibitem{OpenSystems}
H.-P.~Breuer and F.~Petruccione, 
``The Theory of Open Quantum Systems,''
Oxford University Press, 2007,
ISBN 978-019921390.

\bibitem{OpenPP}

J.~R.~Ellis, N.~E.~Mavromatos and D.~V.~Nanopoulos,
``Testing quantum mechanics in the neutral kaon system,''
Phys. Lett. B \textbf{293} (1992), 142-148.

C.~P.~Burgess and D.~Michaud,
``Neutrino propagation in a fluctuating sun,''
Annals Phys. \textbf{256} (1997), 1-38
[arXiv:hep-ph/9606295 [hep-ph]];

C.~H.~Chang, W.~S.~Dai, X.~Q.~Li, Y.~Liu, F.~C.~Ma and Z.~j.~Tao,
``Possible effects of quantum mechanics violation induced by certain quantum gravity on neutrino oscillations,''
Phys. Rev. D \textbf{60} (1999), 033006.

F.~Benatti and R.~Floreanini,
``Open system approach to neutrino oscillations,''
JHEP \textbf{02} (2000), 032.

G.~Barenboim, N.~E.~Mavromatos, S.~Sarkar and A.~Waldron-Lauda,
``Quantum decoherence and neutrino data,''
Nucl. Phys. B \textbf{758} (2006), 90-111.

E.~Braaten, H.~W.~Hammer and G.~P.~Lepage,
``Open Effective Field Theories from Deeply Inelastic Reactions,''
Phys. Rev. D \textbf{94} (2016) no.5, 056006.

D.~Hellmann, H.~P\"as and E.~Rani,
``Quantum Gravitational Decoherence in the 3 Neutrino Flavor Scheme,''
[arXiv:2208.11754 [hep-ph]].

S.~Cao and D.~Boyanovsky,
``Non-equilibrium dynamics of Axion-like particles: the quantum master equation,''
[arXiv:2212.05161 [astro-ph.CO]].


\bibitem{OpenQG}

D.~Boyanovsky,
``Effective Field Theory out of Equilibrium: Brownian quantum fields,''
New J. Phys. \textbf{17} (2015) no.6, 063017.

D.~Boyanovsky,
``Effective field theory during inflation: Reduced density matrix and its quantum master equation,''
Phys. Rev. D \textbf{92} (2015) no.2, 023527.

T.~J.~Hollowood and J.~I.~McDonald,
``Decoherence, discord and the quantum master equation for cosmological perturbations,''
Phys. Rev. D \textbf{95} (2017) no.10, 103521.

A.~Baidya, C.~Jana, R.~Loganayagam and A.~Rudra,
``Renormalization in open quantum field theory. Part I. Scalar field theory,''
JHEP \textbf{11} (2017), 204.

C.~Agon, V.~Balasubramanian, S.~Kasko and A.~Lawrence,
``Coarse Grained Quantum Dynamics,''
Phys. Rev. D \textbf{98} (2018) no.2, 025019.

S.~Shandera, N.~Agarwal and A.~Kamal,
``Open quantum cosmological system,''
Phys. Rev. D \textbf{98} (2018) no.8, 083535.

C.~Ag\'on and A.~Lawrence,
``Divergences in open quantum systems,''
JHEP \textbf{04} (2018), 008.

J.~Martin and V.~Vennin,
``Observational constraints on quantum decoherence during inflation,''
JCAP \textbf{05} (2018), 063
[arXiv:1801.09949 [astro-ph.CO]].

J.~Martin and V.~Vennin,
``Non Gaussianities from Quantum Decoherence during Inflation,''
JCAP \textbf{06} (2018), 037.

S.~Choudhury, A.~Mukherjee, P.~Chauhan and S.~Bhattacherjee,
``Quantum Out-of-Equilibrium Cosmology,''
Eur. Phys. J. C \textbf{79} (2019) no.4, 320
[arXiv:1809.02732 [hep-th]].

C.~Burrage, C.~K\"ading, P.~Millington and J.~Min\'a\v{r},
``Open quantum dynamics induced by light scalar fields,''
Phys. Rev. D \textbf{100} (2019) no.7, 076003.

M.~Parikh, F.~Wilczek and G.~Zahariade,
``The Noise of Gravitons,''
Int. J. Mod. Phys. D \textbf{29} (2020) no.14, 2042001
[arXiv:2005.07211 [hep-th]].

M.~Parikh, F.~Wilczek and G.~Zahariade,
``Signatures of the quantization of gravity at gravitational wave detectors,''
Phys. Rev. D \textbf{104} (2021) no.4, 046021
[arXiv:2010.08208 [hep-th]].

S.~Banerjee, S.~Choudhury, S.~Chowdhury, J.~Knaute, S.~Panda and K.~Shirish,
``Thermalization in Quenched De Sitter Space,''
[arXiv:2104.10692 [hep-th]].

S.~Brahma, A.~Berera and J.~Calder\'on-Figueroa,
``Universal signature of quantum entanglement across cosmological distances,''
[arXiv:2107.06910 [hep-th]].

T.~Colas, J.~Grain and V.~Vennin,
``Benchmarking the cosmological master equations,''
Eur. Phys. J. C \textbf{82} (2022) no.12, 1085
[arXiv:2209.01929 [hep-th]].

A.~Daddi Hammou and N.~Bartolo,
``Cosmic decoherence: primordial power spectra and non-Gaussianities,''
[arXiv:2211.07598 [astro-ph.CO]].

R.~Loganayagam, M.~Rangamani and J.~Virrueta,
``Holographic open quantum systems: Toy models and analytic properties of thermal correlators,''
[arXiv:2211.07683 [hep-th]].



\bibitem{Nakajima}
S. Nakajima, 
``On Quantum Theory of Transport Phenomena,'' 
Prog.\ Theor.\ Phys.\ {\bf 20} 948 (1958). 

\bibitem{Zwanzig}
 R. Zwanzig, 
 ``Ensemble Method in the Theory of Irreversibility,'' 
  J.\ Chem.\ Phys.\ {\bf 33} 1338 (1960).
  
\bibitem{Lindblad}
  G.~Lindblad,
  ``On the Generators of Quantum Dynamical Semigroups,''
  Commun.\ Math.\ Phys.\  {\bf 48} (1976) 119.

\bibitem{Gorini}
  V.~Gorini, A.~Frigerio, M.~Verri, A.~Kossakowski and E.C.G.~Sudarshan,
  ``Properties of Quantum Markovian Master Equations,''
  Rept.\ Math.\ Phys.\  {\bf 13} (1978) 149.
  
\bibitem{Kaplanek:2019dqu}
G.~Kaplanek and C.~P.~Burgess,
``Hot Accelerated Qubits: Decoherence, Thermalization, Secular Growth and Reliable Late-time Predictions,''
JHEP \textbf{03} (2020), 008.

\bibitem{Burgess:2018sou}
C.~P.~Burgess, J.~Hainge, G.~Kaplanek and M.~Rummel,
``Failure of Perturbation Theory Near Horizons: the Rindler Example,''
JHEP \textbf{10} (2018), 122
[arXiv:1806.11415 [hep-th]].

\bibitem{Schwinger:1960qe}
  J.~S.~Schwinger,
  ``Brownian motion of a quantum oscillator,''
  J.\ Math.\ Phys.\  {\bf 2} (1961) 407.

\bibitem{Keldysh:1964ud}
  L.~V.~Keldysh,
  ``Diagram technique for nonequilibrium processes,''
  Zh.\ Eksp.\ Teor.\ Fiz.\  {\bf 47} (1964) 1515
   [Sov.\ Phys.\ JETP {\bf 20} (1965) 1018].

\bibitem{Gross:1980br}
D.~J.~Gross, R.~D.~Pisarski and L.~G.~Yaffe,
``QCD and Instantons at Finite Temperature,''
Rev. Mod. Phys. \textbf{53} (1981), 43.

\bibitem{Altherr}
T.~Altherr, ``Infrared problem in g?4 theory at finite temperature,''
Physics Letters B,
Volume 238, Issues 2-4, 1990, 360-366.

\bibitem{Feynman:1963fq}
R.~P.~Feynman and F.~L.~Vernon, Jr.,
``The Theory of a general quantum system interacting with a linear dissipative system,''
Annals Phys. \textbf{24} (1963), 118-173

\bibitem{FeynmanHibbs}
R.~P.~Feynman and A.~R.~Hibbs, 
``Quantum Mechanics and Path Integrals,''
McGraw-Hill, New York, 1965.

\bibitem{Weiss}
U.~Weiss, ``Quantum Dissipative Systems,'' World Scientific (2000).

\bibitem{Calzetta:2008iqa}
E.~A.~Calzetta and B.~L.~B.~Hu,
``Nonequilibrium Quantum Field Theory,''
Cambridge University Press, 2022.



\bibitem{IFs}

A.~O.~Caldeira and A.~J.~Leggett,
``Path integral approach to quantum Brownian motion,''
Physica A \textbf{121} (1983), 587-616.

V.~Hakim and V.~Ambegaokar,
``Quantum theory of a free particle interacting with a linearly dissipative environment,''
Phys. Rev. A \textbf{32} (1985), 423-434.

C.~M.~Smith and A.~O.~Caldeira,
``Generalized Feynman-Vernon approach to dissipative quantum systems,''
Phys. Rev. A \textbf{36} (1987), 3509-3511.

H.~Grabert, P.~Schramm and G.~L.~Ingold,
``Quantum Brownian motion: The Functional inegral approach,''
Phys. Rept. \textbf{168} (1988), 115-207.

B.~L.~Hu, J.~P.~Paz and Y.~h.~Zhang,
``Quantum Brownian motion in a general environment: 1. Exact master equation with nonlocal dissipation and colored noise,''
Phys. Rev. D \textbf{45} (1992), 2843-2861.

B.~L.~Hu and A.~Matacz,
``Quantum Brownian motion in a bath of parametric oscillators: A Model for system - field interactions,''
Phys. Rev. D \textbf{49} (1994), 6612-6635
[arXiv:gr-qc/9312035 [gr-qc]].

B.~L.~Hu and A.~Matacz,
``Back reaction in semiclassical cosmology: The Einstein-Langevin equation,''
Phys. Rev. D \textbf{51} (1995), 1577-1586
[arXiv:gr-qc/9403043 [gr-qc]].

E.~Calzetta and B.~L.~Hu,
Phys. Rev. D \textbf{49} (1994), 6636-6655
[arXiv:gr-qc/9312036 [gr-qc]].

D.~Boyanovsky, H.~J.~de Vega, R.~Holman, D.~S.~Lee and A.~Singh,
``Dissipation via particle production in scalar field theories,''
Phys. Rev. D \textbf{51} (1995), 4419-4444
[arXiv:hep-ph/9408214 [hep-ph]].

F.~Lombardo and F.~D.~Mazzitelli,
``Coarse graining and decoherence in quantum field theory,''
Phys. Rev. D \textbf{53} (1996), 2001-2011
[arXiv:hep-th/9508052 [hep-th]].





\bibitem{Bakshi:1962dv}
P.~M.~Bakshi and K.~T.~Mahanthappa,
``Expectation value formalism in quantum field theory. 1.,''
J. Math. Phys. \textbf{4} (1963), 1-11.

\bibitem{KMS1}
R.~Kubo,
``Statistical mechanical theory of irreversible processes. 1. General theory and simple applications in magnetic and conduction problems,''
J. Phys. Soc. Jap. \textbf{12} (1957), 570-586.

\bibitem{KMS2}
P.~C.~Martin and J.~S.~Schwinger,
``Theory of many particle systems. 1.,''
Phys. Rev. \textbf{115} (1959), 1342-1373.
  


\bibitem{Unruh:1976db}
  W.G.~Unruh,
  ``Notes on black hole evaporation,''
  Phys.\ Rev.\ D {\bf 14} (1976) 870.

\bibitem{DeWitt:1980hx}
  B.S.~DeWitt,
  ``Quantum Gravity: The New Synthesis'' in ``General Relativity, An Einstein Centenary Survey,'' edited by S.~W.~Hawking and W.~Israel, Cambrdige University Press (1979).

\bibitem{Sciama:1981hr}
  D.W.~Sciama, P.~Candelas and D.~Deutsch,
  ``Quantum Field Theory, Horizons and Thermodynamics,''
  Adv.\ Phys.\  {\bf 30} (1981) 327.
  
\bibitem{Chaykov:2022zro}
S.~Chaykov, N.~Agarwal, S.~Bahrami and R.~Holman,
``Loop corrections in Minkowski spacetime away from equilibrium 1: Late-time resummations,''
[arXiv:2206.11288 [hep-th]].
 
\bibitem{Takagi:1986kn}
  S.~Takagi,
  ``Vacuum noise and stress induced by uniform accelerator: Hawking-Unruh effect in Rindler manifold of arbitrary dimensions,''
  Prog.\ Theor.\ Phys.\ Suppl.\  {\bf 88} (1986) 1.
  
\bibitem{Langlois:2005nf}
  P.~Langlois,
  ``Causal particle detectors and topology,''
  Annals Phys.\  {\bf 321} (2006) 2027 
  [gr-qc/0510049].

\bibitem{Davies:1974th}
  P.~C.~W.~Davies,
  ``Scalar particle production in Schwarzschild and Rindler metrics,'' 
  J.\ Phys.\ A {\bf 8} (1975) 609.

\bibitem{Boulware:1974dm}
  D.~G.~Boulware,
  ``Quantum Field Theory in Schwarzschild and Rindler Spaces,'' 
  Phys.\ Rev.\ D {\bf 11} (1975) 1404.
  
\bibitem{Troost:1978yk}
  W.~Troost and H.~van Dam,
  ``Thermal Propagators and Accelerated Frames of Reference,'' 
  Nucl.\ Phys.\ B {\bf 152} (1979) 442.

\bibitem{Dowker:1978aza}
  J.~S.~Dowker,
  ``Thermal properties of Green's functions in Rindler, de Sitter, and Schwarzschild spaces,'' 
  Phys.\ Rev.\ D {\bf 18} (1978) no.6, 1856.

\bibitem{Linet:1995mq}
  B.~Linet,
  ``Euclidean scalar and spinor Green's functions in Rindler space,'' 
  gr-qc/9505033.
 

\bibitem{Burgess:2003jk}
C.~P.~Burgess,
``Quantum gravity in everyday life: General relativity as an effective field theory,''
Living Rev. Rel. \textbf{7} (2004), 5-56.

\bibitem{Adshead:2017srh}
P.~Adshead, C.~P.~Burgess, R.~Holman and S.~Shandera,
``Power-counting during single-field slow-roll inflation,''
JCAP \textbf{02} (2018), 016.

\bibitem{Kaplanek:2019vzj}
G.~Kaplanek and C.~P.~Burgess,
``Hot Cosmic Qubits: Late-Time de Sitter Evolution and Critical Slowing Down,''
JHEP \textbf{02} (2020), 053.

\bibitem{SIref}
N.~C.~Tsamis and R.~P.~Woodard,
``Matter contributions to the expansion rate of the universe,''
Phys. Lett. B \textbf{426} (1998), 21-28.

\bibitem{StochInf}
A.~A. Starobinsky, 
``Stochastic de Sitter (inflationary) stage in the
  early Universe,''
  {Lect. Notes Phys.}
  {\bf 246} (1986) 107--126.
  
\bibitem{Green:2022ovz}
D.~Green,
``EFT for de Sitter Space,''
[arXiv:2210.05820 [hep-th]].

\bibitem{Vennin:2015hra}
V.~Vennin and A.~A.~Starobinsky,
``Correlation Functions in Stochastic Inflation,''
Eur. Phys. J. C \textbf{75} (2015), 413.

\bibitem{Vennin:2020kng}
V.~Vennin,
``Stochastic inflation and primordial black holes,''
[arXiv:2009.08715 [astro-ph.CO]].

\bibitem{Burgess:2015ajz}
C.~P.~Burgess, R.~Holman and G.~Tasinato,
``Open EFTs, IR effects \& late-time resummations: systematic corrections in stochastic inflation,''
JHEP \textbf{01} (2016), 153.

\bibitem{Starobinsky:1994bd}
A.~A. Starobinsky and J.~Yokoyama, 
``Equilibrium state of a self-interacting scalar field in the De Sitter background,''
  {Phys. Rev. D} {\bf
  50} (1994) 6357--6368.

\bibitem{Tsamis:2005hd}
N.~C.~Tsamis and R.~P.~Woodard,
``Stochastic quantum gravitational inflation,''
Nucl. Phys. B \textbf{724} (2005), 295-328
[arXiv:gr-qc/0505115 [gr-qc]].

\bibitem{StochasticPrelim}
M.~Mijic,  {Stochastic dynamics of coarse grained quantum fields in the
  inflationary universe},
   {Phys. Rev. D} {\bf
  49} (1994) 6434--6441, [\href{http://arxiv.org/abs/gr-qc/9401030}{{\tt
  gr-qc/9401030}}].

D.~Seery, {Infrared effects in inflationary correlation functions},
  {Class. Quant.
  Grav.} {\bf 27} (2010) 124005, [\href{http://arxiv.org/abs/1005.1649}{{\tt
  1005.1649}}].


C.~P.~Burgess, R.~Holman, G.~Tasinato and M.~Williams,
``EFT Beyond the Horizon: Stochastic Inflation and How Primordial Quantum Fluctuations Go Classical,''
JHEP \textbf{03} (2015), 090.

H.~Collins, R.~Holman and T.~Vardanyan,
``The quantum Fokker-Planck equation of stochastic inflation,''
JHEP \textbf{11} (2017), 065.

\bibitem{StochasticGraphs}
V.~Gorbenko and L.~Senatore,
``$\lambda \phi^4$ in dS,''
[arXiv:1911.00022 [hep-th]].

M.~Mirbabayi,
``Infrared dynamics of a light scalar field in de Sitter,''
JCAP \textbf{12} (2020), 006.

M.~Baumgart and R.~Sundrum,
``De Sitter Diagrammar and the Resummation of Time,''
JHEP \textbf{07} (2020), 119.

T.~Cohen and D.~Green,
``Soft de Sitter Effective Theory,''
JHEP \textbf{12} (2020), 041.

M.~Mirbabayi,
``Markovian dynamics in de Sitter,''
JCAP \textbf{09} (2021), 038.

M.~Baumgart and R.~Sundrum,
``Manifestly Causal In-In Perturbation Theory about the Interacting Vacuum,''
JHEP \textbf{03} (2021), 080.

T.~Cohen, D.~Green and A.~Premkumar,
``A Tail of Eternal Inflation,''
[arXiv:2111.09332 [hep-th]];
%
``Large Deviations in the Early Universe,''
[arXiv:2212.02535 [hep-th]].

\bibitem{Grishchuk:1989ss}
L.~P.~Grishchuk and Y.~V.~Sidorov,
``On the Quantum State of Relic Gravitons,''
Class. Quant. Grav. \textbf{6} (1989), L161-L165.

\bibitem{Brandenberger:1990bx}
R.~H.~Brandenberger, R.~Laflamme and M.~Mijic,
``Classical Perturbations From Decoherence of Quantum Fluctuations in the Inflationary Universe,''
Mod. Phys. Lett. A \textbf{5} (1990), 2311-2318.

\bibitem{Calzetta:1995ys}
E.~Calzetta and B.~L.~Hu,
``Quantum fluctuations, decoherence of the mean field, and structure formation in the early universe,''
Phys. Rev. D \textbf{52} (1995), 6770-6788.

\bibitem{Kiefer:1998qe}
C.~Kiefer, D.~Polarski and A.~A.~Starobinsky,
``Quantum to classical transition for fluctuations in the early universe,''
Int. J. Mod. Phys. D \textbf{7} (1998), 455-462
[arXiv:gr-qc/9802003 [gr-qc]].

\bibitem{Hollowood:2017bil}
T.~J.~Hollowood and J.~I.~McDonald,
``Decoherence, discord and the quantum master equation for cosmological perturbations,''
Phys. Rev. D \textbf{95} (2017) no.10, 103521.

\bibitem{Martin:2018lin}
J.~Martin and V.~Vennin,
``Non Gaussianities from Quantum Decoherence during Inflation,''
JCAP \textbf{06} (2018), 037.

\bibitem{Kodama:1984ziu}
H.~Kodama and M.~Sasaki,
``Cosmological Perturbation Theory,''
Prog. Theor. Phys. Suppl. \textbf{78} (1984), 1-166.

\bibitem{Mukhanov:1990me}
V.~F.~Mukhanov, H.~A.~Feldman and R.~H.~Brandenberger,
``Theory of cosmological perturbations. Part 1. Classical perturbations. Part 2. Quantum theory of perturbations. Part 3. Extensions,''
Phys. Rept. \textbf{215} (1992), 203-333.

\bibitem{Maldacena:2002vr}
J.~M.~Maldacena,
``Non-Gaussian features of primordial fluctuations in single field inflationary models,''
JHEP \textbf{05} (2003), 013.

\bibitem{GravDecohere}
C.P.~Burgess, R.~Holman, G.~Kaplanek, J.~Martin and V.~Vennin,
``Minimal decoherence from inflation,''
[arXiv:2211.11046].

\bibitem{Nelson:2016kjm}
E.~Nelson, 
``{{Quantum Decoherence During Inflation from Gravitational
  Nonlinearities}},''
 JCAP  {\bf 1603}
  (2016) 022 , [arXiv:1601.03734].


\bibitem{Boulware}
D.G.~Boulware,
``Hawking Radiation and Thin Shells,''
Phys. Rev. D 13 (1976) 2169.

\bibitem{HartleHawking}
J.B.~Hartle and S.W.~Hawking, 
``Path Integral Derivation of Black Hole Radiance,''
Phys. Rev. D 13 (1976) 2188.

\bibitem{Hadamard1}
J~ Hadamard, ``Lectures on Cauchy?s problem in linear partial differential equations'', Yale University Press, New Haven, U.S.A.

\bibitem{Hadamard2}
B.~S.~DeWitt and R.~W.~Brehme,
``Radiation damping in a gravitational field,''
Annals Phys. \textbf{9} (1960), 220-259.

\bibitem{Hadamard3}
S.~A.~Fulling, M.~Sweeny and R.~M.~Wald,
``Singularity Structure of the Two Point Function in Quantum Field Theory in Curved Space-Time,''
Commun. Math. Phys. \textbf{63} (1978), 257-264.

\bibitem{Kaplanek:2020iay}
G.~Kaplanek and C.~P.~Burgess,
``Qubits on the Horizon: Decoherence and Thermalization near Black Holes,''
JHEP \textbf{01} (2021), 098.

\bibitem{Akhmedov:2015xwa}
E.~T.~Akhmedov, H.~Godazgar and F.~K.~Popov,
``Hawking radiation and secularly growing loop corrections,''
Phys. Rev. D \textbf{93} (2016) no.2, 024029
[arXiv:1508.07500 [hep-th]].

\bibitem{Emelyanov:2016tws}
S.~Emelyanov,
``Near-horizon physics of an evaporating black hole: One-loop effects in the $\lambda\Phi^4$-theory,''
[arXiv:1608.05318 [hep-th]].

\bibitem{Kaplanek:2021sbo}
G.~Kaplanek, C.~P.~Burgess and R.~Holman,
``Influence through mixing: hotspots as benchmarks for basic black-hole behaviour,''
JHEP \textbf{09} (2021), 006.

\bibitem{Kaplanek:2021fnl}
G.~Kaplanek, C.~P.~Burgess and R.~Holman,
``Qubit heating near a hotspot,''
JHEP \textbf{08} (2021), 132.

\bibitem{Burgess:2021luo}
C.~P.~Burgess, R.~Holman and G.~Kaplanek,
``Quantum Hotspots: Mean Fields, Open EFTs, Nonlocality and Decoherence Near Black Holes,''
Fortsch. Phys. \textbf{70} (2022) no.4, 2200019.

\bibitem{Caldeira:1981rx}
A.~O.~Caldeira and A.~J.~Leggett,
``Influence of dissipation on quantum tunneling in macroscopic systems,''
Phys. Rev. Lett. \textbf{46} (1981), 211.

\bibitem{PPEFT}
C.~P.~Burgess, P.~Hayman, M.~Williams and L.~Zalavari,
``Point-Particle Effective Field Theory I: Classical Renormalization and the Inverse-Square Potential,''
JHEP \textbf{04} (2017), 106.

C.~P.~Burgess, P.~Hayman, M.~Rummel, M.~Williams and L.~Zalavari,
``Point-Particle Effective Field Theory II: Relativistic Effects and Coulomb/Inverse-Square Competition,''
JHEP \textbf{07} (2017), 072.

R.~Plestid, C.~P.~Burgess and D.~H.~J.~O'Dell,
``Fall to the Centre in Atom Traps and Point-Particle EFT for Absorptive Systems,''
JHEP \textbf{18} (2020), 059.

\bibitem{Weinberg:1995mt}
S.~Weinberg,
``The Quantum theory of fields. Vol. 1: Foundations,''
Cambridge University Press (2005).

\bibitem{FokkerLangevin}
G.~A.~Pavliotis, ``Stochastic processes and applications: diffusion processes, the Fokker-Planck and Langevin equations,'' Vol.~60, Springer (2014).

\end{thebibliography}
\end{document}